\documentstyle[aps,prd,preprint,floats,tighten,eqsecnum,amssymb,epsf]{revtex}
\newcommand{\fslash}[1]{\mbox{$\!\not\!#1$}}

\newcommand{\beq}{\begin{equation}}
\newcommand{\eeq}{\end{equation}}
\newcommand{\bea}{\begin{eqnarray}}
\newcommand{\eea}{\end{eqnarray}}
\newcommand{\ba}{\begin{array}}
\newcommand{\ea}{\end{array}}

\newcommand{\bef}{\begin{figure}}
\newcommand{\eef}{\end{figure}}
\newcommand{\bce}{\begin{center}}
\newcommand{\ece}{\end{center}}

\newcommand{\gsim}{\mathrel{\hbox{\rlap{\lower.55ex \hbox {$\sim$}}
                   \kern-.3em \raise.4ex \hbox{$>$}}}}
\newcommand{\lsim}{\mathrel{\hbox{\rlap{\lower.55ex \hbox {$\sim$}}
                   \kern-.3em \raise.4ex \hbox{$<$}}}}

\begin{document}
\tightenlines

\preprint{\vtop{\hbox{}
\vskip24pt}}

\title{Color Superconductivity at Moderate Baryon Density}

\author{Mei Huang
\footnote{E-mail: huang@th.physik.uni-frankfurt.de}}

\address{1. Institute for Theoretical Physics, J.W.Goethe University,
Frankfurt am Main, Germany \\
2. Physics Department, Tsinghua University, Beijing, China}

\maketitle

\begin{abstract}

This article focuses on the two-flavor color superconducting phase 
at moderate baryon density. In order to simultaneously investigate the 
chiral phase transition and the color superconducting phase transition,
the Nambu-Gorkov formalism is extended to treat the quark-antiquark 
and diquark condensates on an equal footing. The competition between the 
chiral condensate and the diquark condensate is analyzed. The cold 
dense charge neutral two-flavor quark system is investigated in detail. 
Under the local charge neutrality condition, the ground state of two-flavor
quark matter is sensitive to the coupling strength in the diquark channel.
When the diquark coupling strength is around the value obtained from the 
Fierz transformation or from fitting the vacuum bayron mass, the ground 
state of charge neutral two-flavor quark matter is in a thermal stable 
gapless 2SC (g2SC) phase. The unusual properties at zero as well as nonzero
temperatures and the chromomagnetic properties of the g2SC phase 
are reviewed. Under the global charge neutrality condition, 
assuming the surface tension is negligible, the mixed phase composed of 
the regular 2SC phase and normal quark matter is more favorable than the 
g2SC phase. A hybrid nonstrange neutron star is constructed.

\end{abstract}

\pacs{}
\pagenumbering{empty} 

\tableofcontents

\pagenumbering{arabic}

\newpage

\section{Introduction}
\label{sec:Intro}


\subsection{QCD phase structure}

Quantum Chromodynamics (QCD) is regarded as the fundamental theory of 
quarks and gluons. Its ultimate goal is to explain all 
strong interaction experiments at all energies, high and low. QCD is 
an asymptotically free theory \cite{asymptotic}. At very high energies,  
interaction forces become weak, thus perturbation calculations can be used. 
The perturbative QCD predictions have been extensively confirmed by experiments, 
while QCD in the non-perturbative regime is still a challenge to theorists. 
The fundamental quarks and gluons of QCD have not been seen as free particles, 
but are always confined within hadrons. It is still difficult 
to construct the hadrons in terms of nearly massless quarks and gluons. The observed 
baryon spectrum indicates that the (approximate) chiral symmetry is spontaneously 
broken in the vacuum. As a result, the eight pseudoscalar mesons $\pi$, $K$ and $\eta$
are light pseudo-Nambu-Goldstone bosons, and the constituent quark obtains dynamical 
mass, which contributes to the baryon mass. Besides conventional mesons and baryons, 
QCD itself does not exclude the existence of the non-conventional states such as 
glueballs, hybrid mesons and multi-quark states \cite{Klempt:2004yz}. Last year,
pentaquarks, e.g., $\Theta^+ (u^2d^2{\bar s})$ 
\cite{Theta-pentaquark} with the mass $M=1540\pm 10 {\rm MeV}$ and $\Xi^{--} 
(s^2d^2{\bar u})$ \cite{Xi-pentaquark} with $M=1862 \pm 18 {\rm MeV}$, were 
discovered in experiments. 
This has stirred a great interest of theorists to understand the structure of
the pentaquarks \cite{Diakonov-1997,Jaffe-Wilczek,Shuryak:2003zi,Zhu} as well 
as other fundamental problems of QCD in the non-perturbative regime.

Since 1970s, people have been interested in QCD at extreme conditions. It is 
expected that the chiral symmetry can be restored, and quarks and gluons 
will become deconfined at high temperatures and/or densities
\cite{Lee:1974ma,Collins:1974ky,Baym:1976yu,qgpT}. Results from lattice show
that the quark-gluon plasma (QGP) does exist. For the system with zero net baryon
density, the deconfinement and chiral symmetry restoration phase transitions
happen at the same critical temperature \cite{Karsch}. 
At asymptotically high temperatures, e.g., during the first microseconds of the 
``Big Bang", the many-body system composed of quarks and gluons 
can be regarded as an ideal Fermi and Boson gas. 
It is believed that the ``little Bang" can be produced in RHIC and LHC. 
Recently, it was shown that the new state of matter produced in RHIC is far away 
from the asymptotically hot QGP, but in a strongly 
coupled regime. This state is called strongly coupled quark-gluon plasma (sQGP). 
In this strong coupling system, the meson bound states still play an important role 
\cite{Strong-Hatsuda,Shuryak-SQGP}. For most recent reviews about QGP, 
e.g., see Ref.~\cite{Rischke-r,Gyulassy-Mclerran,Jacobs-Wang,Heinz-review}.

Studying QCD at finite baryon density is the traditional subject of nuclear 
physics. The behaviour of QCD at finite baryon density and low temperature is 
central for astrophysics to understand the structure of compact stars, and 
conditions near the core of collapsing stars (supernovae, hypernovae).
Cold nuclear matter, such as in the interior of a Pb nucleus, is
at $T=0$ and $\mu_B\simeq m_N = 940 {\rm MeV}$. Emerging from this point,
there is a first-order nuclear liquid-gas phase transition, 
which terminates in a critical endpoint at a temperature 
$\sim 10 {\rm MeV}$ \cite{liquid-gas-Tc}. If one squeezes matter further 
and further, nuleons will overlap. Quarks and gluons in one nucleon can feel 
quarks and gluons in other nucleons. Eventually, deconfinement phase 
transition will happen. Unfortunately, at the moment, lattice QCD is facing 
the ``sign problem" at nonzero net baryon densities. Our understanding at 
finite baryon densities has to rely on effective QCD models. Phenomenological 
models indicated that, at nonzero baryon density, the QGP phase and the hadron 
gas are separated by a critical line of roughly a constant energy density 
$\epsilon_{cr} \simeq 1 {\rm GeV}/{\rm fm}^3$ \cite{Heinz-Stocker}. 

In the case of asymptotically high baryon density, the system is a color 
superconducor. This was proposed 25 years ago by Frautschi \cite{Frautschi} and 
Barrois \cite{Barrois}. Reaching this conclusion does not need any other knowledge, 
if one knows that one-gluon exchange between two quarks is attractive in the 
color antitriplet channel,  and if one is also familiar with the theory of 
superconductivity, i.e., the Bardeen, Cooper, and Schrieffer (BCS) theory 
\cite{BCS}. Based on the BCS scenario, if there is a weak 
attractive interaction in a cold Fermi sea, the system is unstable with respect 
to the formation of particle-particle Cooper-pair condensate in the momentum space.  
Detailed numerical calculations of color superconducting gaps were firstly
carried out by Bailin and Love \cite{Bailin-Love}. They concluded that the 
one-gluon exchange induces gaps on the order of 1 ${\rm MeV}$ at several times 
of nuclear matter density. This small gap has little effect on cold dense quark 
matter, thus the investigation of cold quark matter lay dormant for several decades.  
It was only revived recently when it was found that the color superconducting gap 
can be of the order of $100~{\rm MeV}$ \cite{CS-1997}, which is two orders larger 
than early perturbative estimates in Ref.~\cite{Bailin-Love}. For this reason, 
the topic of color superconductivity stirred a lot of interest in recent years. 
For review articles on the color superconductivity, see for example,
Refs.~\cite{Rischke-r,RW-r,Hong-r,Alford-r,Nardulli-r,Schafer-r,Buballa-r,Ren-r}.

In this paper, I will focus on the color superconducting phase structure at 
intermediate baryon density regime based on my own experience. It is worth to 
mention that there is another good review article Ref. ~\cite{Buballa-r} in 
this regime from another point of view. 
The outline of this article is as follows: A brief overview 
of color superconducting phases is given in Sec. \ref{cs-phases}.
In Sec. \ref{NG}, the generalized Nambu-Gorkov formalism is introduced to 
treat the chiral and diquark condensates on an equal footing, and the 
competition between the chiral and color superconducting phase transitions 
is investigated. Sec.~\ref{g2sc} focuses on homogeneous neutral two-flavor 
quark matter, the gapless 2SC (g2SC) phase and its properties are reviewed. 
In Sec.~\ref{mixed}, a neutral mixed phase composed of the 2SC phase and 
normal quark matter is discussed. At last, a brief outlook is given in 
Sec.~\ref{outlook}.

\subsection{Color superconducting phases}
\label{cs-phases}

Let us start with the system of free fermion gas. Fermions obey the Pauli 
exclusion principle, which means no two identical fermions can occupy 
the same quantum state. The energy distribution for fermions (with mass $m$) 
has the form of
\begin{eqnarray}
f(E_p)=\frac{1}{e^{\beta(E_p-\mu)}+1}, ~~ \beta=1/T,
\end{eqnarray}
here $E_p=\sqrt{p^2+m^2}$, $\mu$ is the chemical potential and $T$ is the 
temperature. At zero temperature, $f(E_p)=\theta(\mu-E_p)$. The ground state 
of the free fermion gas is a filled Fermi sea, i.e., all states with the 
momenta less than the Fermi momentum $p_F=\sqrt{\mu^2-m^2}$ are occupied, 
and the states with the momenta greater than the Fermi momentum $p_F$ are 
empty. Adding or removing a single fermion costs no free energy at the 
Fermi surface. 

For the degenerate Fermi gas, the only relevant fermion degrees of freedom 
are those near the Fermi surface. Considering two fermions near the Fermi 
surface, if there is a net attraction between them, it turns out that they
can form a bound state, i.e., Cooper pair \cite{Cooper}. The binding energy
of the Cooper pair $\Delta(K)$ ($K$ the total momentum of the pair),
is very sensitive to $K$, being a maxium where $K=0$. There is an infinite 
degeneracy among pairs of fermions with equal and 
opposite momenta at the Fermi surface. Because Cooper pairs are composite 
bosons, they will occupy the same lowest energy quantum state at zero 
temperature and produce a Bose-Einstein condensation. Thus the ground state 
of the Fermi system with a weak attractive interaction is a complicated 
coherent state of particle-particle Cooper pairs near the Fermi 
surface \cite{BCS}. Exciting a quasiparticle and a hole which interact with the 
condensate requires at least the energy of $2 \Delta$. 

In QED case in condensed matter, the interaction between two electrons by
exchanging a photon is repulsive. The attractive interaction to form 
electron-electron Cooper pairs is by exchanging a phonon, which is a collective 
excitation of the positive ion background. The Cooper pairing of the electrons 
breaks the electromagnetic gauge symmetry, and the photon obtains an effective 
mass. This indicates the Meissner effect \cite{Meissner}, i.e., a superconductor 
expels the magnetic fields.

In QCD case at asymptotically high baryon density, the 
dominant interaction between two quarks is due to the one-gluon exchange. 
This naturally provides an attractive interaction between two quarks.  
The scattering amplitude for single-gluon exchange in an $SU(N_c)$ gauge theory 
is proportional to 
\begin{eqnarray}
(T_a)_{ki}(T_a)_{lj}=-\frac{N_c+1}{4N_c} (\delta_{jk}\delta_{il} - \delta_{ik}\delta_{jl})
+\frac{N_c-1}{4N_c} (\delta_{jk}\delta_{il} + \delta_{ik}\delta_{jl}).
\label{T-OGE}
\end{eqnarray}
Where $T_a$ is the generator of the gauge group, and $i,j$ and $k,l$ are the fundamental 
colors of the two quarks in the incoming and outgoing channels, respectively. Under
the exchange of the color indices of either the incoming or the outgoing quarks, 
the first term is antisymmetric, while the second term is symmetric. For $N_c=3$, 
Eq.~(\ref{T-OGE}) represents that the tensor product of two fundamental colors 
decomposes into an (antisymmetric) color antitriplet and a (symmetric) color sextet,
\begin{eqnarray}
[{\bf 3}]^c \otimes [{\bf 3}]^c = [\bar{\bf 3}]_a^c \oplus [{\bf 6}]_s^c.
\end{eqnarray}
In Eq.~(\ref{T-OGE}), the minus sign in front of the antisymmetric contribution indicates
that the interaction in this antitriplet channel is attractive, while the interaction 
in the symmetric sextet channel is repulsive. 

For cold dense quark matter, the attractive interaction in the color
antitriplet channel induces the condensate of the quark-quark Cooper pairs, and
the ground state is called the ``color superconductivity". Since the diquark cannot
be color singlet, the diquark condensate breaks the local color $SU(3)_c$ symmetry,
and the gauge bosons connected with the broken generators obtain masses. 
Comparing with the Higgs mechanism of dynamical gauge symmetry breaking in the 
Standard Model, here the diquark Cooper pair can be regarded as a composite Higgs 
particle. The calculation of the energy gap and the critical temperature from the 
first principles has been derived systematically in Refs. 
\cite{weak-Son,weak-Schafer,weak-Dirk,weak-Tc-Dirk,weak-Hong,weak-HongShovkovy,weak-Ren,weak-Manuel}.  

In reality, we are more interested in cold dense quark matter at moderate baryon density
regime, i.e., $\mu_q \sim 500 MeV$, which may exist in the interior of neutron stars. 
It is likely that cold dense quark droplet might be created in the laboratory 
through heavy ion collisions in GSI-SPS energy scale. 
At these densities, an extrapolation of the asymptotic arguments becomes unreliable, we 
have to rely on effective models. Calculations in the framework of pointlike 
four-fermion interactions based on the instanton vertex 
\cite{CS-1997,Instanton-Schafer,Berges-Rajagopal,Carter-Diakonov},
as well as in the Nambu--Jona-Lasinio (NJL) model 
\cite{NJL-Klevansky,NJL-Aichelin,NJL-Buballa,NJL-Ebert,NJL-Huang} show that color 
superconductivity does occur at moderate densities, and the magnitude of diquark gap 
is around $100~{\rm MeV}$. 

Even though the antisymmetry in the attractive channel signifies that only quarks 
with different colors can form Cooper pairs, color superconductivity has very rich 
phase structure because of its flavor, spin and other degrees of freedom. In the 
following, I list some of the known color superconducting phases.

\vskip 0.3cm
{\bf The 2SC phase} 
\vskip 0.3cm

Firstly we consider a system with only massless $u$ and $d$ quarks, 
assuming that the strange quark is much heavier than the up and down quarks. The color 
superconducting phase with only two flavors is normally called the 2SC phase.

Renormalization group arguments \cite{weak-Son,RG-Schafer,RG-Evans} suggest that 
possible quark pairs always condense in the $s-$wave. This means that the spin wave 
function of 
the pair is anti-symmetric. Since the diquark condenses in the color antitriplet 
$\bar{\bf 3}_c$ channel, the color wave function of the pair is also 
anti-symmetric. The Pauli principle requires that the total wave function 
of the Cooper pair has to be antisymmetric under the exchange of the two 
quarks forming the pair. Thus the flavor wave function has to be 
anti-symmetric, too. This determines the structure of the order parameter
\begin{eqnarray}
\label{order2sc}
\Delta_{ij}^{\alpha\beta}=\Delta \epsilon_{ij}\epsilon^{\alpha\beta b},
\end{eqnarray}
where color indices $\alpha, \beta \in (r,g,b)$ and flavor indices 
$i, j \in(u,d)$. From the order parameter Eq. (\ref{order2sc}), we can see 
that the condensate picks a color direction (here the $blue$ direction, which is
arbitrarily selected). The ground state is invariant under 
an $SU(2)_c$ subgroup of the color rotations that mixes the red and green colors, but 
the blue quarks are singled out as different. Thus the color $SU(3)_c$ is broken down 
to its subgroup $SU(2)_c$, and five of the gluons obtain masses, which indicates the
Meissner effect \cite{Meissner2SC}.  

In the 2SC phase, the Cooper pairs are $ud-du$ singlets and the global flavor 
symmetry $SU(2)_L \otimes SU(2)_R$ is intact, i.e., the chiral symmetry is not broken. 
There is also an unbroken global symmetry which plays the role of $U(1)_B$. Thus 
no global symmetry are broken in the 2SC phase.

\vskip 0.3cm    
{\bf The CFL phase}
\vskip 0.3cm 

In the case when the chemical potential is much larger than the strange 
quark mass, we can assume $m_u=m_d=m_s=0$, and there are three degenerate massless 
flavors in the system. The spin-0 order parameter should be color and flavor 
anti-symmetric, which has the form of
\begin{eqnarray}
\Delta_{ij}^{\alpha\beta}= \Delta \sum_I \epsilon_{ijI} \epsilon^{\alpha\beta I},
\label{CFL-gap}
\end{eqnarray}  
where color indices $\alpha, \beta \in (r,g,b)$ and flavor indices $i, j \in(u,d,s)$. 
Writing $\sum_I\epsilon_{ijI} \epsilon^{\alpha\beta I}=\delta_i^{\alpha} \delta_j^{\beta} 
-\delta_j^{\alpha} \delta_i^{\beta} $, we can see that the order parameter
\begin{eqnarray}
\Delta_{ij}^{\alpha\beta}= \Delta (\delta_i^{\alpha} \delta_j^{\beta} 
-\delta_j^{\alpha} \delta_i^{\beta})
\end{eqnarray}  
describes the color-flavor locked (CFL) phase proposed in Ref.~\cite{CFL-ARW}. Many  
other different treatments 
\cite{CFL-Chiral-Schaefer,CFL-Evans,CFL-Shovkovy}
agreed that a condensate of the form (\ref{CFL-gap}) is the dominant condensate in 
three-flavor QCD. 

In the CFL phase, all quark colors and flavors participate in the pairing.
The color gauge group is completely broken, and all eight gluons become
massive \cite{CFL-ARW,Meissner-CFL}, which ensures that there are no infrared 
divergences associated with gluon propagators. Electromagnetism is no 
longer a separate symmetry, but corresponds to gauging one of the flavor 
generators. A rotated electromagnetism (``$\tilde{Q}$") remains unbroken. 

Two global symmetries, the chiral symmetry and the baryon number, are broken in 
the CFL phase, too. In zero-density QCD, the spontaneous breaking of chiral symmetry 
is due to the condensation of left-handed quarks with right-handed quarks. 
Here, at high baryon density, the chiral symmetry breaking occurs due to a 
rather different mechanism: locking of the flavor rotations to color. 
In the CFL phase, there is only pairing of left-handed quarks with left-handed 
quarks, and right-handed quarks with right-handed quarks, i.e.,
\begin{eqnarray}
< \psi_{Li}^{\alpha}\psi_{Lj}^{\beta}> 
= - < \psi_{Ri}^{\alpha}\psi_{Rj}^{\beta}>.
\end{eqnarray}
Where $L,R$ indicate left- and right-handed, respectively, $\alpha,\beta$ are 
color indices and $i,j$ are flavor indices. A gauge invariant form
\cite{RW-r,CFL-effective,CFL-Son}
\begin{eqnarray}
< \psi_{Li}^{\alpha}\psi_{Lj}^{\beta}  
   {\bar \psi}_{R\alpha}^k {\bar \psi}_{R\beta}^l> 
\sim  < \psi_{Li}^{\alpha}\psi_{Lj}^{\beta}> 
   <{\bar \psi}_{R\alpha}^k {\bar \psi}_{R\beta}^l>
\sim \Delta^2 \epsilon_{ijm}\epsilon^{klm}
\end{eqnarray}
captures the chiral symmetry breaking. The spectrum of excitations in the CFL phase 
contains an octet of Goldstone bosons associated with the chiral symmetry breaking. 
This looks remarkably like those at low density. In the excitation spectrum 
of the CFL phase, there is another singlet $U(1)$ Goldstone boson related to the 
baryon number symmetry breaking, which can be described using the order parameter
\begin{eqnarray}
<udsuds>\sim<\Lambda \Lambda>.
\end{eqnarray}  
In QCD with three degenerate light flavors, the spectrum in the CFL phase looks 
similar to that in the hyper-nuclear phase at 
low-density. It is suggested that the low density hyper-nuclear phase and the high 
density quark phase might be continuously connected \cite{Continuity-Schafer}. 

\vskip 0.3cm
{\bf Spin-1 color superconductivity} 
\vskip 0.3cm
In the case of only one-flavor quark system, due to the antisymmetry in 
the color space, the Pauli principle requires that the Cooper pair has to 
occur in a symmetric spin channel. 
Therefore, in the simplest case, the Cooper pairs carry total spin one. 
Spin-1 color 
superconductivity was firstly studied in Ref.~\cite{Bailin-Love}, for more 
recent and detailed
discussions about the spin-1 gap, its critical temperature and Meissner 
effect, see 
Refs.~ \cite{weak-Dirk,weak-Tc-Dirk,Spin1-Schafer,Spin1-Buballa,Spin1-Tc-Schmitt,Spin1-Meissner}.
For a review, see Ref. ~\cite{Schmitt-PhD}.

\vskip 0.3cm
{\bf Pairing with mismatch: LOFF, CFL-K, g2SC and gCFL}
\vskip 0.3 cm

To form the Cooper pair, the ideal case is when the two pairing quarks have the same 
Fermi momenta, i.e., $p_{F,i}=p_{F,j}$ with $p_{F,i}=\sqrt{\mu_{F,i}^2-m_i^2}$, like 
in the ideal 2SC, CFL, and spin-1 color superconducting phases. However, in reality, 
the nonzero strange quark mass or the requirement of charge neutrality induces a 
mismatch between the Fermi momenta of the two pairing quarks. When the mismatch is 
very small, it has little effect on the Cooper pairing. While if the mismatch is very 
large, the Cooper pair will be destroyed. The most interesting situation happens when 
the mismatch is neither very small nor very large.  

{\bf LOFF:} In the regime just on the edge of decoupling of the two pairing quarks 
(due to the nonzero strange quark mass for the $qs$ Cooper pair with $q\in (u,d)$ or 
the chemical potential difference for the $ud$ Cooper pair), a ``LOFF" 
(Larkin-Ovchinnikov-Fulde-Ferrell) state may be formed. The LOFF state was firstly 
investigated in the context of electron superconductivity in the presence of 
magnetic impurities \cite{LO,FF}. It was found that near the unpairing transition, 
it is favorable to form a state in which the Cooper pairs have nonzero momentum. 
This is favored because it gives rise to a regime of phase space where each of the 
two quarks in a pair can be close to its Fermi surface, and such pairs can be 
created at low cost in free energy. This sort of condensates spontaneously break 
translational and rotational invariance, leading to gaps which vary periodically 
in a crystalline pattern. The crystalline color superconductivity has been 
investigated in a series of papers, e.g., see 
Refs. ~\cite{LOFF-1,LOFF-2,LOFF-3,LOFF-4,LOFF-5,LOFF-Bowers,LOFF-Ren,LOFF-Nardulli}.

{\bf CFL-K:} The strange quark mass $m_s$ induces an effective chemical potential 
$\mu_s=m_s^2/(2p_F)$, and the effects of the strange quark mass can be quite dramatic. 
In the CFL phase, the $K^+$ and $K^0$ modes may be unstable for large values of the 
strange quark mass to form a kaon condensation 
\cite{CFLK-Schafer,CFLK-Reddy,Schafer-Goldstone,Igor-Goldstone}.
In the framework of effective theory 
\cite{CFL-effective,CFL-Son,CFL-mesons-Rho,CFL-mesons-Hong,CFL-mesons-Manuel}, 
the masses of the Goldstone bosons can be determined as 
\begin{eqnarray}
\label{mgb}
 m_{\pi^\pm} &=& \mp\frac{m_d^2-m_u^2}{2p_F}+\left [
       \frac{4A}{f_{\pi}^2}(m_u+m_d)m_s\right ]^{1/2}, \nonumber \\
 m_{K^\pm} &=& \mp\frac{m_s^2-m_u^2}{2p_F}+\left [
       \frac{4A}{f_{\pi}^2}(m_u+m_s)m_d\right ]^{1/2}, \nonumber \\
 m_{K^0,{\bar K}^0} &=& \mp\frac{m_s^2-m_d^2}{2p_F}+\left [
       \frac{4A}{f_{\pi}^2}(m_s+m_d)m_u\right ]^{1/2}, \nonumber \\
\end{eqnarray}
with $A=3\Delta^2/(4\pi^2)$ \cite{CFL-Son,smass-Schafer}.  
It was found that the kaon masses are substantially affected 
by the strange quark mass, the masses of $K^-$ and ${\bar K}^0$ are pushed up while 
$K^+$ and $K^0$ are lowered. As a result, the $K^+$ and $K^0$ become
massless if $m_s|_{crit} =3.03 ~ m_d^{1/3} \Delta^{2/3}$. For larger values
of $m_s$ the kaon modes are unstable, signaling the formation of a kaon condensate.
For review of the kaon condensate in the CFL phase, see Ref.~\cite{Schafer-r}. 
Recently, it was found that in the CFL phase,
there also may exist $\eta$ condensate \cite{CFLeta-Schafer}. 

{\bf g2SC and gCFL:} When the $\beta$-equilibrium and the charge neutrality condition 
are required for the two-flavor quark system, the Fermi surfaces of the pairing $u$ 
quark and $d$ quark differ by $\mu_e$, here $\mu_e$ is the chemical potential for 
electrons. It was found that when the gap parameter $\Delta < \mu_e/2$, the system
will be in a new ground state called the gapless 2SC (g2SC) phase \cite{g2SC-SH}. The 
g2SC phase has very unusual temperature properties \cite{g2SC-HS-T} and chromomagnetic 
properties \cite{g2SC-HS-M}. This phase will be introduced in more detail in Sec.~\ref{g2sc}. 

Similarly, for a charge neutral 3-flavor system with a nonzero strange quark mass $m_s$, with 
increasing $m_s$, the CFL phase transfers to a new gapless CFL (gCFL) phase 
when $m_s^2/\mu \simeq 2 \Delta$ \cite{gCFL-AKR}. The finite temperature property
of the charge neutral three-flavor quark matter was investigated in 
Ref.~\cite{udSC-Hatsuda,gCFL-Ruster,gCFL-Kenji}. Recently, it was shown that the kaon 
condensate shifts the critical strange quark mass to higher values for the 
appearance of the gCFL phase \cite{gCFL-condensate}.


\section{Competition between the chiral and diquark condensates}
\label{NG}

In this section, we are going to investigate the competition between the
chiral and diquark condensates, and the content is based on 
Refs.~\cite{NJL-Huang,NG-Huang}.

In the idealized case at asymptotically high baryon densities, the color 
superconductivity with two massless flavors and the color-flavor-locking 
(CFL) phase with three degenerate massless quarks have been widely discussed 
from first principle QCD calculations. Usually, the diagrammatic methods are 
used at the asymptotic densities. The Green-function of the eight-component 
field and the gap equation were discussed in details in \cite{weak-Dirk,NG-Dirk}. 
Neither current quark mass nor chiral condensate are necessary to be 
considered because they can be neglected compared to the very high 
Fermi surface. At less-than-asymptotic densities, the corrections of 
nonzero quark mass to the pure CFL phase can be treated perturbatively 
by expanding the current quark mass around the chiral limit 
\cite{smass-Schafer,smass-Rho}. 

For physical applications, we are more interested in the moderate baryon 
density regime, which may be related to the neutron stars and, in very 
optimistic cases even to heavy-ion collisions. To work out the phase structure 
from the hadron phase to the color superconducting phase, one should deal with 
the chiral condensate and the diquark condensate simultaneously. Because the chiral 
condensation contributes a dynamical quark mass, it is not reasonable any more 
to treat the quark mass term perturbatively in this density regime. 

Usually, at moderate baryon density regime, effective models such as the 
instanton model, as well as the Nambu--Jona-Lasinio (NJL) model, are used. 
The model parameters are fixed in the QCD vacuum. In this regime, 
the usual way is to use the variational methods, which are enough for 
constructing the thermodynamic potential and the gap equation to 
investigate the ground state. In order to investigate the 
diquark excitations, and to calculate the cross-section of physical 
process in the color superconducting phase, it is very 
helpful to use the Green-function approach. 

One of our main aims in this section is to apply the Green-function approach 
in the moderate baryon density regime. By using the energy projectors for 
massive quark, we can deal with the chiral and diquark condensates on an equal 
footing, and we will deduce the Nambu-Gorkov massive quark propagator \cite{NG-Huang}
as well as the thermodynamic potential \cite{NJL-Huang}. 

In the normal phase, the quarks of different colors are degenerate. While in 
the color broken phase, it is natural to assume that the quarks involved 
in the diquark condensate are different from the free quarks. It is difficult 
to obtain the mass expression for the quarks participating in the diquark 
condensate, because the particles and holes mix with each other, and the 
elementary excitations are quasi-particles and quasi-holes near the Fermi 
surface. The difference between quarks in different colors has 
been reflected by their propagators, and the difference can be read 
through calculating the quarks chiral condensate. In the chiral limit, the 
chiral condensate disappears entirely in the color superconducting phase, 
and it is not possible to investigate the influence of the color breaking on 
quarks in different colors, so we keep the small current quark mass in this section.
   
In the moderate baryon density regime, people are interested in whether there 
exists a regime where both the chiral symmetry and color symmetry are broken 
\cite{Berges-Rajagopal,Carter-Diakonov,NJL-Klevansky,Mishra-H,Koide-vector,Hatsuda-coex,Vanderheyden,Kerbikov}.  
Note that the presence of a small current quark mass induces that the chiral symmetry 
only restores partially and there always exists a small chiral condensate 
in the color superconducting phase. This phenomenon had been called 
coexistence of the chiral and diquark condensates in Ref.~\cite{Berges-Rajagopal}. 
In the coexistence regime resulted by current quark mass, the chiral condensate
is small comparing with the diquark condensate, and the role of the chiral 
condensate can be prescribed by the Anderson theorem \cite{Kerbikov}, i.e.,
in this phase, the contribution of the chiral condensate to 
thermodynamic quantities becomes strongly suppressed, and one can 
calculate the diquark condensate neglecting the influence of 
the chiral condensate.

In order to differ from the coexistence of the diquark condensate 
and the small chiral condensate in the chiral symmetry restoration phase, 
the coexistence of the diquark condensate and the dynamical chiral condensate 
in the chiral symmetry breaking phase is called the {\it double broken phase}. 
The existence of the double broken phase depends on the coupling constants 
$G_S$ and $G_D$ in the quark-antiquark and diquark channels \cite{NJL-Klevansky,Vanderheyden}.  
In the case of small ratio of $G_D/G_S<1$, the calculations in the
instanton model \cite{Berges-Rajagopal}, NJL model \cite{NJL-Klevansky}
and random matrix model \cite{Vanderheyden} show that there is no double broken phase.
While for $G_D/G_S>1$ the calculations in the the random matrix model 
\cite{Vanderheyden} and the NJL model \cite{NJL-Klevansky} show that there exists
a narrow regime where both chiral (dynamical) symmetry and color symmetry 
are broken, and the chiral and diquark condensate coexist.
The larger value of $G_D/G_S$ is, the wider regime of the double broken
phase has been found in the random matrix model \cite{Vanderheyden}. 
 
In this section, based on Ref.~\cite{NJL-Huang}, we explain the existence of the 
double broken phase by analyzing the influence of the diquark condensate on 
the Fermi surface of the constituent quark. In the mean-field approximation 
of the NJL model, the thermal system of the constituent quarks is nearly an 
ideal Fermi gas, and there is a sharp Fermi surface for the constituent quark. 
When a diquark condensate is formed, the Cooper pair smoothes the Fermi surface, 
and this induces the chiral symmetry restoring at a smaller chemical potential. 
The stronger the coupling constant in the diquark channel, the larger the diquark gap 
will be, and the smaller critical chemical potential will be for the color 
superconductivity phase transition. However, it is found in Ref.~\cite{NJL-Huang} 
that, if $G_D/G_S>1$, the diquark condensate starts to appear even in the vacuum.
This may suggest that in the NJL model, the diquark coupling strength $G_D$ 
should not be larger than the quark-antiquark coupling strength $G_S$.  
It is also possible that the appearance of the diquark condensate in 
vacuum is an artificial result due to the lack of deconfinement in the NJL model. 

\subsection{The extended NJL model}

To describe quark matter in the intermediate baryon density regime, we
use the extended NJL model. The choice of the NJL model \cite{Nambu} is 
motivated by the fact that this model displays the same symmetries as QCD 
and that it describes well the spontaneous breakdown of chiral symmetry 
in the vacuum and its restoration at high temperature and density. For the review
of the NJL model, please see \cite{review-NJL}. The model we used in this paper 
is an extended version of the two-flavor NJL model including interactions in 
the color singlet quark-antiquark channel as well as in the color anti-triplet 
diquark channel, which is not directly extended from the NJL model, but from the 
QCD Lagrangian \cite{dnjl1,dnjl2,dnjl3}.

The importance of color ${\bar 3}$ diquark degree of freedom is related to 
the fact that one can construct a color-singlet nucleon current based on it. 
Because the gluon exchange between two quarks in the color ${\bar 3}$ channel 
is attractive, one can view a color singlet baryon as a quark-diquark bound state.
And experimental data from $ep$ collisions indicate the existence of this 
quark-diquark component in nucleons \cite{quark-diquark}. The discovery of the
pentaquark also reminds us the importance of the diquark degree of freedom 
\cite{Jaffe-Wilczek,Shuryak:2003zi,Wilczek-D}.

The first attempt to investigate the diquark properties in the NJL model was 
taken in Ref.~\cite{Vogl1}. Starting from an NJL model for scalar, pseudoscalar, 
vector and axial-vector interactions of the $({\bar q} q) \times ({\bar q} q)$ type  
and Fierz-transforming away the vector and axial-vector interactions, 
the scalar and pseudoscalar mesons, and diquarks can be obtained.  
However, this method could not get a consistent treatment of vector 
and axial-vector particles. 

The extended NJL model we use is derived from the QCD Lagrangian
\cite{dnjl1,dnjl2}. Integrating out the gluon degrees of freedom from the
QCD Lagrangian, and performing a local approximation for the (nonperturbative)
gluon propagator, one obtains a contact current-current interaction.
By using a special Fierz-rearrangement \cite{cahill} 
(see Appendix \ref{Fierz-Appendix}), one can completely 
decompose the two-quark-current interaction term into ``attractive" color 
singlet (${\bar q}q$) and color antitriplet ($qq$) channels. 
In this way, a complete simultaneous description of scalar, 
pseudo-scalar, vector, and axial-vector mesons and diquarks is possible, 
thus the extended NJL model including $({\bar q} q) \times ({\bar q} q)$ 
interactions is completed by a corresponding 
$({\bar q} {\bar q}) \times (q q)$ interaction part.

It is worth to mention that, in the mean-field approximation, there
is a famous ambiguity connected with performing the Fierz transformation for 
the pointlike four-fermion interaction. For more detail discussion about
the Fierz ambiguity, see Ref.~\cite{Fierz-ambi}. Here, in this paper, 
we only consider the scalar, pseudoscalar mesons and 
scalar diquark, and the Lagrangian density has the form
\begin{eqnarray}
\label{lagr}
{\cal L} = {\bar q}(i\gamma^{\mu}\partial_{\mu}-m_0)q + 
   G_S[({\bar q}q)^2 + ({\bar q}i\gamma_5{\bf {\bf \tau}}q)^2 ]
 +G_D[(i {\bar q}^C  \varepsilon  \epsilon^{b} \gamma_5 q )
   (i {\bar q} \varepsilon \epsilon^{b} \gamma_5 q^C)].
\end{eqnarray}
Where $q^C=C {\bar q}^T$, ${\bar q}^C=q^T C$ are charge-conjugate spinors, 
$C=i \gamma^2 \gamma^0$ is the charge conjugation matrix (the superscript $T$ 
denotes the transposition operation). For more details about the charge conjugate, 
see Appendix \ref{ChargeConj}. $m_0$ is the current quark mass, the quark field 
$q \equiv q_{i\alpha}$ with $i=1,2$ and $\alpha=1,2,3$ is a flavor 
doublet and color triplet, as well as a four-component Dirac spinor, 
${\bf \tau}=(\tau^1,\tau^2,\tau^3)$ are Pauli matrices in the flavor 
space, where $\tau^2$ is antisymmetric, and $(\varepsilon)^{ik} \equiv \varepsilon^{ik}$,
$(\epsilon^b)^{\alpha \beta} \equiv \epsilon^{\alpha \beta b}$ are totally 
antisymmetric tensors in the flavor and color spaces.  

Though, from the Fierz transformation, the quark-antiquark coupling constant $G_S$ 
and the diquark coupling constant $G_D$ have the relation $G_D=G_S~ N_c/(2N_c-2)$,
in this article, $G_S$ and $G_D$ are treated independently. The former is responsible 
for the meson excitations, and the latter for the diquark excitations, which 
in principle can be determined by fitting meson and baryon properties 
in the vacuum. In Ref.~\cite{dnjl2}, the ratio of coupling constants $G_D/G_S \simeq 2.26/3$
was obtained by fitting the scalar diquark mass of $600~{\rm MeV}$ to 
get a realistic vacuum baryon mass. It is noticed
that this ratio is quite close to the value $3/4$ obtained from Fierz transformation.

The attractive interaction in different channels in the Lagrangian  
gives rise to a very rich structure of the phase diagram. At zero temperature and density, 
the attractive interaction in the color singlet channel is responsible for the 
appearance of a quark-antiquark condensate and for the spontaneous breakdown 
of the chiral symmetry, and the interaction in the $qq$ channel 
binds quarks into diquarks (and baryons), but is not strong enough
to induce diquark condensation. As the density increases, Pauli blocking 
suppresses the ${\bar q}q$ interaction, while the attractive interaction in the 
color anti-triplet diquark channel will induce the quark-quark condensate around 
the Fermi surface which can be identified as a superconducting phase.  

After bosonization \cite{dnjl1,dnjl2,bosonization}, one obtains the linearized 
version of the model, i.e., in the mean-field approximation, 
\begin{eqnarray}
\label{lagr2}
\tilde{\cal L} & =  & {\bar q}(i\gamma^{\mu}\partial_{\mu}-m_0)q - 
  {\bar q}(\sigma+i \gamma^5{\bf \tau}{\bf \pi}) q - 
  \frac{1}{2}\Delta^{*b} (i{\bar q}^C  \varepsilon \epsilon^{b}\gamma_5 q )
  -\frac{1}{2}\Delta^b (i {\bar q}  \varepsilon  \epsilon^{b} \gamma_5 q^C) \nonumber \\
  & & -\frac{\sigma^2+{\bf \pi}^2}{4G_S}
  -\frac{\Delta^{*b}\Delta^{b}}{4G_D},
\end{eqnarray}
with the bosonic fields 
\begin{eqnarray}
\Delta^b \sim i {\bar q}^C \varepsilon \epsilon^{b}\gamma_5 q, \ \ 
\Delta^{*b} \sim i {\bar q}  \varepsilon  \epsilon^{b} \gamma_5 q^C, \ \
\sigma \sim {\bar q} q , \ \ {\bf \pi} \sim i {\bar q}\gamma^5 {\bf \tau} q.
\end{eqnarray}
Clearly, the $\sigma$ and ${\bf \pi}$ fields are color singlets, and the 
diquark fields $\Delta^b$ and $\Delta^{*b}$ are color antitriplet and  
(isoscalar) singlet under the chiral $SU(2)_L \times SU(2)_R$ group.
$\sigma \neq 0$  and $\Delta^b \neq 0$ indicate that the chiral symmetry and 
the color symmetry are spontaneously broken. 
Here it has been regarded that only the red and green quarks participating in the 
condensate, while the blue quarks do not. The real ground state at any $T,\mu$ will 
be determined by the minimum of the thermodynamic potential.

The partition function of the grand canonical ensemble can be evaluated 
by using the standard method \cite{kapusta,bellac},
\begin{eqnarray}
\label{part-1}
{\cal Z}=N' \int [d {\bar q} ][d q] exp\{ \int_0^{\beta} d \tau \int d^3{\bf x} 
  ~ ( \tilde{\cal L} +\mu {\bar q} \gamma_0 q)\},
\end{eqnarray}
where $\mu$ is the chemical potential, and $\beta=1/T$ is the inverse of 
the temperature $T$.

In the mean field approximation, we can  
write the partition function as a product of three parts, 
\begin{eqnarray}
\label{part}
{\cal Z}={\cal Z}_{const}{\cal Z}_{b}{\cal Z}_{r,g}.
\end{eqnarray}
The constant part is
\begin{eqnarray}
{\cal Z}_{const}=N'{\rm exp} \{- \int_0^{\beta} d \tau \int d^3{\bf x} ~ 
[\frac{\sigma^2}{4 G_S}
      +\frac{\Delta^{*}\Delta}{4 G_D} \}.
\end{eqnarray}
The contribution from free blue quarks has the form 
\begin{eqnarray}
\label{z3}
{\cal Z}_{b}  & = & \int[d{\bar q}_b][d q_b]{\rm exp}\{ \int_0^{\beta} d \tau 
\int d^3{\bf x} ~[\frac{1}{2}{\bar q}_b ( i\gamma^{\mu}\partial_{\mu}-m
+\mu\gamma_0)q_b \nonumber \\
& & + \frac{1}{2}{\bar q}_b^C ( i\gamma^{\mu}\partial_{\mu}-m-\mu\gamma_0)q_b^C] \}.
\end{eqnarray}
For the quarks with  red and green color $Q=q_{r,g}$ participating in the quark 
condensate, their contribution is
\begin{eqnarray}
\label{z12}
{\cal Z}_{r,g} & = & \int[d{\bar Q}][d Q] {\rm exp} \{ \int_0^{\beta} d \tau \int d^3{\bf x} ~ 
[\frac{1}{2}{\bar Q} 
( i\gamma^{\mu}\partial_{\mu}-m+\mu\gamma_0)Q + \nonumber \\
 & & \frac{1}{2}{\bar Q}^C 
( i\gamma^{\mu}\partial_{\mu}-m-\mu\gamma_0)Q^C 
+\frac{1}{2}{\bar Q} \Delta^{-} Q^C + \frac{1}{2}{\bar Q}^C \Delta^{+} Q] \}.
\end{eqnarray}
Here we  have introduced the constituent quark mass 
\begin{eqnarray}
m=m_0+ \sigma,
\end{eqnarray}
and defined $\Delta^{\pm}$ as
\begin{eqnarray} 
\Delta^{-} = -i \Delta \varepsilon  \epsilon^{b} \gamma_5, \  \ 
\Delta^{+} = -i \Delta^{*} \varepsilon  \epsilon^{b} \gamma_5
\end{eqnarray}
with the relation $\Delta^{+}= \gamma^0 (\Delta^{-})^{\dagger} \gamma^0$.

\subsection{Nambu-Gorkov propagator with chiral and diquark condensates}

In the case of finite chemical potential, it is more convenient to use
the Nambu-Gorkov formalism. Introducing the 8-component spinors for the 
blue quarks and the quarks colored with red and green, respectively
\begin{equation}
\Psi_b =\left(\begin{array}{c} 
                    q_b \\
                    q_b^C
                   \end{array}
            \right), \ \ 
\bar{\Psi}_b =( \bar{q}_b \ \  \bar{q}_b^C ),
\end{equation}
\begin{equation}
\Psi =\left(\begin{array}{c} 
                    Q \\
                    Q^C 
                   \end{array}
            \right), \ \ 
\bar{\Psi} =( \bar{Q} \ \  \bar{Q}^C ),
\end{equation}
and using the Fourier transformation in the momentum space,
\begin{eqnarray}
q(x)=\frac{1}{\sqrt{V}}\sum_n\sum_{\bf p} 
e^{-i ( \omega_n\tau-{\bf p}\cdot{\bf x})}q({\bf p}),
\end{eqnarray}
where $V$ is the volume of the thermal system,
we can re-write the partition function Eqs.~(\ref{z3}) and (\ref{z12}) 
in the momentum space as
\begin{eqnarray}
\label{zq3}
 {\cal Z}_{b} & =  & \int[d \Psi_b]{\rm exp}\{\frac{1}{2}\sum_{n,{\bf p}}  
 ~ {\bar \Psi}_b\frac{G_0^ {-1}}{T}\Psi_b \} \nonumber \\
& = & {\rm Det} ^{1/2}(\beta G_0^{-1}),
\end{eqnarray}
and
\begin{eqnarray}
\label{zq12}
{\cal Z}_{r,g} & = & \int[d \Psi]{\rm exp} \{\frac{1}{2}  \sum_{n,{\bf p}} ~
{\bar \Psi}\frac{{\rm G}^{-1}}{T}  \Psi \}  \nonumber \\
& = & {\rm Det}^{1/2}(\beta {\rm G}^{-1}).
\end{eqnarray}
Where the determinantal operation ${\rm Det}$ is to be carried out over the Dirac, color, 
flavor and the momentum-frequency space.
In Eqs.~(\ref{zq3}) and (\ref{zq12}), we have defined the quark propagator 
in the normal phase
\begin{equation}
{\rm G_0}^{-1} = 
    \left( 
          \begin{array}{cc}
            \left[ G_0^{+} \right]^{-1}  &  0 \\  
            0 &  \left[ G_0^{-} \right]^{-1} 
              \end{array}  
             \right),
\end{equation}
with 
\begin{eqnarray}
[G_0^{\pm}]^{-1}=
(p_0 \pm \mu) \gamma_0 -{\mbox{\boldmath$\gamma$}}\cdot {\bf p} -m
\end{eqnarray}
and the quark propagator in the color broken phase
\begin{equation}
{\rm G}^{-1} = 
    \left( 
          \begin{array}{cc}
            \left[ G_0^{+} \right]^{-1}  &  \Delta^{-} \\  
            \Delta^{+}  &  \left[ G_0^{-} \right]^{-1} 
              \end{array}  
             \right). 
\end{equation}

The Nambu-Gorkov propagator ${\rm G}(P)$ is determined from solving
$1={\rm G}^{-1}\, {\rm G}$, 
\begin{equation} 
\label{full-S}
{\rm G} = \left( \begin{array}{cc}
  G^{+} &   \Xi^{-}  \\
 \Xi^{+}  & G^{-}
           \end{array}  \right) \,\, ,
\end{equation}
with the components
\begin{eqnarray}
G^{\pm}  \equiv \left\{ \left[ G_0^{\pm} \right]^{-1} - \Sigma^{\pm} 
\right\}^{-1} \,\,\, , \,\,\,\, \Sigma^{\pm} \equiv \Delta^{\mp}
\, G_0^{\mp}\, \Delta^{\pm} \,\, , \nonumber
\end{eqnarray}
\begin{eqnarray}
\Xi^{\pm} \equiv -  G^{\mp} \, \Delta^{\pm} \, G_0^{\pm} 
= -  G_0^{\mp} \, \Delta^{\pm} \, G^{\pm}.
\end{eqnarray}
Here all components depend on the 4-momentum $P$.  

In the case of the mass term $m=0$, the Nambu-Gorkov quark 
propogator has a simple form which could be derived from
the energy projectors for massless particles \cite{weak-Dirk,NG-Dirk}. 
If there is a small mass term, the quark propagator 
can be expanded perturbatively around $m=0$, but the form is very complicated 
\cite{smass-Rho}. In our case the quark mass term cannot be 
treated perturbatively, we have to find a general way to deal with the massive 
quark propagator. 

Fortunately, we can evaluate a simple form for the massive
quark propagator by using the energy projectors for massive particles.
The energy projectors onto states of positive and negative energy for free massive 
particles are defined as
\begin{eqnarray}
\Lambda_{\pm}({\bf p})=\frac{1}{2}(1\pm\frac{\gamma_0({\mbox{\boldmath$\gamma$}}
\cdot{\bf p}+m)}{E_p}),
\label{energy-project}
\end{eqnarray}
where the quark energy $E_p=\sqrt{{\bf p}^2+m^2}$.
Under the transformation of $\gamma_0$ and $\gamma_5$, we can get another 
two energy projectors $\tilde \Lambda_{\pm}$,
\begin{eqnarray}
\label{tilde}
\tilde \Lambda_{\pm}({\bf p})=
\frac{1}{2}(1\pm\frac{\gamma_0({\mbox{\boldmath$\gamma$}}\cdot{\bf p}-m)}{E_p})\ ,
\end{eqnarray}
which satisfy 
\begin{eqnarray}
\label{gamma}
 \gamma_0 \Lambda_{\pm}({\bf p}) \gamma_0=\tilde \Lambda_{\mp}({\bf p}), ~ ~
 \gamma_5 \Lambda_{\pm}({\bf p}) \gamma_5=\tilde \Lambda_{\pm}({\bf p}).
\end{eqnarray}

The normal quark propagator elements can be re-written as 
\begin{eqnarray}
\label{mass0}
G_0^{\pm} =   \frac{\gamma_0\tilde \Lambda_{+}}{p_0+E_p^{\pm}} + 
\frac{\gamma_0\tilde\Lambda_{-}}{p_0-E_p^{\mp}},
\end{eqnarray}
with $E_p^\pm = E_p \pm \mu$. The propagator has four poles, i.e., 
\begin{eqnarray}
p_0=\pm E_p^{-},  ~~~p_0=\mp E_p^{+},
\end{eqnarray}
where the former two correspond to the excitation energies of particles and holes, and the 
latter two are for antiparticles and antiholes, respectively.

The full quark propagator $G$, i.e., (\ref{full-S}) can be evaluated \cite{NG-Huang}.
The normal quark propagator has the form
\begin{eqnarray}
\label{gpmagain}
G^{\pm} = (\frac{p_0-E_p^{\pm}}{p_0^2-{E_{\Delta}^{\pm}}^2}\gamma_0\tilde\Lambda_{+}
+ \frac{p_0+E_p^{\mp}}{p_0^2-{E_{\Delta}^{\mp}}^2}\gamma_0\tilde\Lambda_{-})(\delta_{\alpha\beta}
-\delta_{\alpha b}\delta_{\beta b})\delta^{ij}, 
\end{eqnarray}
and the abnormal quark propagator reads
\begin{eqnarray}
\label{cosiagain}
\Xi^{\pm}=(\frac{\Delta^{\pm}}{p_0^2-{E_{\Delta}^{\pm}}^2}\tilde\Lambda_{+}
+\frac{\Delta^{\pm}}{p_0^2-{E_{\Delta}^{\mp}}^2}\tilde\Lambda_{-}),
\end{eqnarray}
with ${E_{\Delta}^{\pm}}^2={E_p^{\pm}}^2+\Delta^2$. This propagator is similar to
the massless propagator derived in \cite{NG-Dirk}.

The four poles of the Nambu-Gorkov propagator, i.e., 
\begin{eqnarray}
p_0=\pm E_{\Delta}^{-},  ~~~p_0=\mp E_{\Delta}^{+},
\end{eqnarray}
correspond to the excitation energies of quasi-particles (quasi-holes) and 
quasi-antiparticle (quasi-antiholes)
in the color broken phase. These quasi-particles are superpositions of
particles and holes, and are called ``Bogoliubons". 
In the normal phase, exciting a pair of a particle and a hole on the Fermi surface 
does not need energy, while in the superconducting phase, exciting a 
quasi-particle and a hole needs at least the energy $2\Delta$ at $E_p=\mu$. 

\subsection{Gap equations}

With the Nambu-Gorkov quark propagator Eqs. (\ref{gpmagain}) and (\ref{cosiagain}), 
one can investigate the meson excitations \cite{NG-Blascke}, and physical processes related
to the transport properties. The chiral and diquark condensates can also be generally 
expressed from the quasi-particle propagator. The diquark condensate has the form
\begin{eqnarray}
< {\bar q}^C \gamma_5 q>
= -iT\sum_n \int\frac{d^3{\bf p}}{(2\pi)^3}tr[\Xi^{-} \gamma_5].
\end{eqnarray}
From general consideration, there should be eight Dirac components of the diquark 
condensates \cite{Bailin-Love,NG-Dirk,Dirk-mass,Mass-Fugleberg}. In the case of the NJL 
type model, the diquark condensates related to momentum vanish, and there
is only one independent $ 0^+$ diquark gap with Dirac structure 
$\Gamma=\gamma_5$ for massless quark, and there exists
another $0^+$ diquark condensate with Dirac structure
$\Gamma=\gamma_0\gamma_5$ at nonzero quark mass.
In this paper, we assume the contribution of the diquark condensate with 
$\Gamma=\gamma_0\gamma_5$ is small, and only consider the diquark condensate
with $\Gamma=\gamma_5$.  

Performing the Matsubara frequency summation and taking the limit $T \rightarrow 0$,
we get the diquark condensate at finite chemical potential
\begin{eqnarray}
\label{ddt0}
< {\bar q}^C \gamma_5 q> = -2 \Delta N_c N_f 
\int\frac{d^3{\bf p}}{(2\pi)^3}[\frac{1}{2E_{\Delta}^{-}}
+\frac{1}{2E_{\Delta}^{+}}].
\end{eqnarray}
Using the relation between the diquark gap $\Delta$ and the diquark condensate 
$ <{\bar q}^C\gamma_5 q> $, i.e., 
\begin{eqnarray}
\label{dgap}
\Delta  = -2G_D<{\bar q}^C\gamma_5 q>,
\end{eqnarray}
the gap equation for the diquark condensate
in the limit of $T \rightarrow 0$ can be written as
\begin{eqnarray}
1 = 8 N_f G_D \int\frac{d^3{\bf p}}{(2\pi)^3} [\frac{1}{2E_{\Delta}^{-}}
+\frac{1}{2E_{\Delta}^{+}}].
\label{gapd-NG}
\end{eqnarray}


The quark mass has the expression
\begin{eqnarray}
\label{mgap}
m & = & m_0 -2 G_S <{\bar q}q>, \nonumber \\
<{\bar q}q> & = & 2 <{\bar q_r}q_r> + <{\bar q_b}q_b>.
\end{eqnarray}
Where the chiral condensate for the free blue quarks is
\begin{eqnarray}
<{\bar q}_b q_b>  =   -iT \sum_n \int\frac{d^3{\bf p}}{(2\pi)^3}tr[G_0^{+}],
\end{eqnarray}
and the chiral condensate for the red and green quarks reads
\begin{eqnarray}
<{\bar q}_{r} q_{r}>  = -iT \sum_n \int\frac{d^3{\bf p}}{(2\pi)^3}tr[G^{+}].
\end{eqnarray} 
In the limit $T \rightarrow 0$, the gap equation for the quark mass has the form
\begin{eqnarray}
m & = & m_0 -8 m N_f G_S \int\frac{d^3{\bf p}}{(2\pi)^3} \frac{1}{2 E_p} 
   \left[ \theta(\mu-E_p)-1 + 2 (n_p^{+}-n_p^{-}) \right].
\label{gapm-NG}
\end{eqnarray}
Where
\begin{eqnarray}
n_p^{\pm}=\frac{1}{2}(1 \mp \frac{E_p^{\mp}}{E_{\Delta}^{\mp}})
\end{eqnarray}
are the occupation numbers for quasi-particles and quasi-antiparticles at $T=0$. 
Correspondingly, $1-n_p^{\pm}$ are the occupation numbers of quasi-holes and 
quasi-antiholes, respectively.  
 
The difference between the quarks participating in the diquark condensate and the free 
blue quark can be read from their chiral condensates,  
\begin{eqnarray}
\label{delta}
\delta = <{\bar q}_b q_b>^{1/3}-<{\bar q}_r q_r>^{1/3},
\end{eqnarray}
where $\delta$ has the dimension of energy.
In the case of chiral limit, the quark mass $m$ decreases to zero in the color superconducting 
phase, and the influence of the diquark condensate on quarks in different colors vanishes.

\subsection{Derivation of the thermodynamic potential}

Now let us derive the thermodynamic potential. For the blue quarks which do not 
participate in the diquark condensate, from Eq. (\ref{zq3}), we have 
\begin{eqnarray}
{\rm ln} {\cal Z}_{q_b} = \frac{1}{2} {\rm ln} \{  {\rm Det}(\beta [G_0]^{-1})
 = \frac{1}{2} {\rm ln}[ {\rm Det} (\beta [G_0^+]^{-1})  {\rm Det} (\beta [G_0^-]^{-1})].
\end{eqnarray}
 
Using the Dirac matrix, we first perform the determinant in the Dirac space,
\begin{eqnarray}
{\rm Det} ~ \beta ~[G_0^{+}]^{-1} & = & 
{\rm Det} ~ \beta~ [ (p_0+\mu) \gamma_0 - {\mbox{\boldmath$\gamma$}} \cdot {\bf p} -m]  \nonumber \\
   & = & {\rm Det} ~ \beta \left( 
          \begin{array}{cc}
            (p_0+\mu)-m  &  {\mbox{\boldmath$\sigma$}} \cdot {\bf p}  \\  
           -  {\mbox{\boldmath$\sigma$}} \cdot {\bf p} &  -(p_0+ \mu)-m 
              \end{array}  
             \right), \nonumber \\
 & = & - \beta^2 ~ [ (p_0+\mu)^2- E_p^2 ],
\end{eqnarray}
and in a similar way, we get 
\begin{eqnarray}
{\rm Det} ~ \beta ~[G_0^{-}]^{-1} & = & - \beta^2 ~ [ (p_0-\mu)^2- E_p^2 ].
\end{eqnarray}
After performing the determinant in the Dirac space, we have
\begin{eqnarray}
{\rm Det} ~ \beta ~[G_0^{+}]^{-1} {\rm Det} ~ \beta ~[G_0^{-}]^{-1} = 
  \beta^2 [ p_0^2- (E_p + \mu)^2 ]~\beta^2  [ p_0^2- (E_p -\mu)^2].
  \label{DetG0}
\end{eqnarray}

Considering the determinant in the flavor, color, spin spaces and momentum-frequency 
space, we get the standard expression 
\begin{eqnarray}
{\rm ln} {\cal Z}_{q_b}  =  N_f
   \sum_n \sum_{\bf p} \{ {\rm ln} ( \beta^2 [ p_0^2- (E_p + \mu)^2 ] ) 
   +  {\rm ln} ( \beta^2  [ p_0^2- (E_p -\mu)^2] ) \},
\end{eqnarray}
remembering that the color space for the blue quark is one-dimensional.  


It is more complicated to evaluate the thermodynamic potential for the quarks 
participating in the diquark condensate. From Eq. (\ref{zq12}), we have 
\begin{eqnarray}
\label{detg}
{\rm ln} {\cal Z}_{q_{r,g}}=\frac{1}{2} {\rm ln} {\rm Det} (\beta G^{-1}).
\end{eqnarray}

For a $2 \times 2$ matrix with elements $A,B,C$ and $D$, we have the identity
\begin{eqnarray}
\label{identity}
{\rm Det} \left(  \begin{array}{cc}
            A  &  B  \\  
           C & D
              \end{array}  
             \right) = {\rm Det} ( -CB + CAC^{-1}D ) =={\rm Det}(-BC+BDB^{-1}A).
\end{eqnarray}
Replacing $A,B,C$ and $D$ with the corresponding elements of $G^{-1}$,   we have 
\begin{eqnarray}
{\rm Det} (\beta {\rm G}^{-1}) 
& = &\beta^2 {\rm Det} D_1  =  \beta^2 {\rm Det} [- \Delta^{+} \Delta^{-} 
+ \Delta^{+}[G_0^{+}]^{-1} [\Delta^{+}]^{-1}[G_0^{-}]^{-1}] \nonumber \\
& = & \beta^2 {\rm Det} D_2 =  \beta^2 {\rm Det} [- \Delta^{+} \Delta^{-} 
+ [G_0^{-}]^{-1} [\Delta^{-}]^{-1}[G_0^{+}]^{-1}\Delta^{-}].
\end{eqnarray}
Using the energy projectors $\tilde\Lambda_{\pm}$, we can work out $D_1$ and $D_2$ as
\begin{eqnarray}
D_1 & = &  \Delta^2 +  \gamma_5 [\gamma_0 (p_0-E_p^{-}) \Lambda_{+}
+ \gamma_0 (p_0+E_p^{+})\Lambda_{-}] \gamma_5 
[\gamma_0 (p_0-E_p^{+})\Lambda_{+}
+ \gamma_0 (p_0+E_p^{-})\Lambda_{-}]   \nonumber \\
& = &  - [ (p_0^2-{E_p^{-}}^2-\Delta^2)\Lambda_{-}+
       (p_0^2-{E_p^{+}}^2-\Delta^2)\Lambda_{+}], \nonumber \\
D_2 & = & - [ (p_0^2-{E_p^{-}}^2-\Delta^2)\Lambda_{+}+
       (p_0^2-{E_p^{+}}^2-\Delta^2)\Lambda_{-}].
\end{eqnarray}
Using the properties of the energy projectors, one can get 
\begin{eqnarray}
D_1D_2=[p_0^2-{E_{\Delta}^{-}}^2][p_0^2-{E_{\Delta}^{+}}^2].
\end{eqnarray}
With the above equations, Eq. (\ref{detg}) can be expressed as
\begin{eqnarray}
{\rm ln} {\cal Z}_{q_{r,g}} & = &  \frac{1}{2} {\rm ln} [ {\rm Det} \beta G^{-1}] 
 =  \frac{1}{4} {\rm Tr} {\rm ln}[ \beta^2 D_1 \beta^2 D_2] \nonumber \\
& = &  \frac{1}{4} \{ {\rm Tr}{\rm ln} [\beta^2 (p_0^2-{E_{\Delta}^{-}}^2) ] + 
                 {\rm Tr}{\rm ln}[\beta^2  (p_0^2-{E_{\Delta}^{+}}^2) ] \} \nonumber \\
& = & 2 N_f \sum_n \sum_p \{ {\rm ln} [\beta^2 (p_0^2-{E_{\Delta}^{-}}^2 )]+ 
                      {\rm ln} [\beta^2 (p_0^2-{E_{\Delta}^{+}}^2 ) ] \}.
\end{eqnarray}
 
The frequency summation of the free-energy
\begin{eqnarray}
\label{lnzf}
{\rm ln}{\cal Z}_f  =  \sum_n{\rm ln}[\beta^2(p_0^2-E_p^2)]
\end{eqnarray}
can always be obtained by performing the frequency summation of the 
propagator $1/(p_0^2-E_p^2)$.
Differentiate Eq. (\ref{lnzf}) with respect to $E_p$:
\begin{eqnarray}
\frac{\partial {\rm ln}{\cal Z}_f}{\partial E_p}  =  -2 E_p \sum_n \frac{1}{p_0^2-E_p^2}
=\beta [1-2 {\tilde f}(E_p)], 
\end{eqnarray}
where $\tilde f(x) = 1/(e^{\beta x} +1 )$ is the usual 
Fermi-Dirac distribution function.
Then integrating with respect to $E_p$, one can get the free-energy 
\begin{eqnarray}
{\rm ln}{\cal Z}_f=\beta [E_p +2T {\rm ln}(1+e^{-\beta E_p})].
\end{eqnarray}

With the help of the above expression, and replacing 
\begin{eqnarray}
\sum_p \rightarrow V \int\frac{d^3{\bf p}}{(2\pi)^3},
\end{eqnarray}
one can have 
\begin{eqnarray}
{\rm ln}{\cal Z}_{q_b} &=&  N_f \beta V \int \frac{d^3 p}{(2\pi)^3} [E_p^{+} 
+2T{\rm ln}(1+e^{-\beta E_p^{+}}) + E_p^{-} +2T {\rm ln}(1+ e^{-\beta E_p^{-}})], \nonumber \\
{\rm ln}{\cal Z}_{q_{r,g}} &=&  2 N_f \beta V \int \frac{d^3 p}{(2\pi)^3} [E_{\Delta}^{+} 
+2T{\rm ln}(1+e^{-\beta E_{\Delta}^{+}}) + E_{\Delta}^{-} 
+2T {\rm ln}(1+ e^{-\beta E_{\Delta}^{-}})].
\end{eqnarray}

Finally, we obtain the familiar expression of the thermodynamic potential  
\begin{eqnarray}
\label{potential}
\Omega & = & -T\frac{ {\rm ln} Z}{ V } =
  \frac{\sigma^2}{4G_S}+\frac{\Delta^2}{4G_D} 
-2N_f \int\frac{d^3 p}{(2\pi)^3} [ E_p + T{\rm ln}(1+e^{-\beta E_p^{+}}) 
+ T {\rm ln}(1+ e^{-\beta E_p^{-}})  \nonumber \\
& & + E_{\Delta}^{+} 
+2T{\rm ln}(1+e^{-\beta E_{\Delta}^{+}}) + E_{\Delta}^{-} 
+2T {\rm ln}(1+ e^{-\beta E_{\Delta}^{-}}) ].
\end{eqnarray}

The two gap equations $\Delta$ and $m$ in Eqs. (\ref{gapd-NG}) and (\ref{gapm-NG})
can be equivalently derived by minimizing the 
thermodynamic potential Eq. (\ref{potential}) with respect to $m$ and $\Delta$,
respectively, i.e.,
\begin{eqnarray}
\frac{\partial \Omega}{\partial \Delta} =0, ~~ \frac{\partial \Omega}{\partial m} = 0.
\end{eqnarray}

\subsection{Competition between the chiral and diquark condensates}

In this subsection, through numerical calculations, we will
investigate the phase structure along the 
chemical potential direction, analyze the competition mechanism between the 
chiral condensate and diquark condensate, and discuss the influence of the color
breaking on the quarks with different colors.

Before the numerical calculations, we should fix the model parameters. 
The current quark mass $m_0=5. 5 {\rm MeV}$, the Fermion momentum 
cut-off $\Lambda_f=0.637 {\rm GeV}$, and the coupling constant in color
singlet channel $G_S=5.32 {\rm GeV}^{-2}$ are determined by fitting vacuum properties.
The corresponding constituent quark mass in the vacuum is taken to be
$m(\mu=0)=330 {\rm MeV}$. The coupling constant in the color anti-triplet channel 
$G_D$ can in principle be determined by fitting the nucleon properties, 
here, we choose $G_D/G_S$ as a free parameter.

\vskip 0.2cm
{\bf The double broken phase}
\vskip 0.2cm
 
\begin{figure}[ht]
\centerline{\epsfxsize=10cm\epsffile{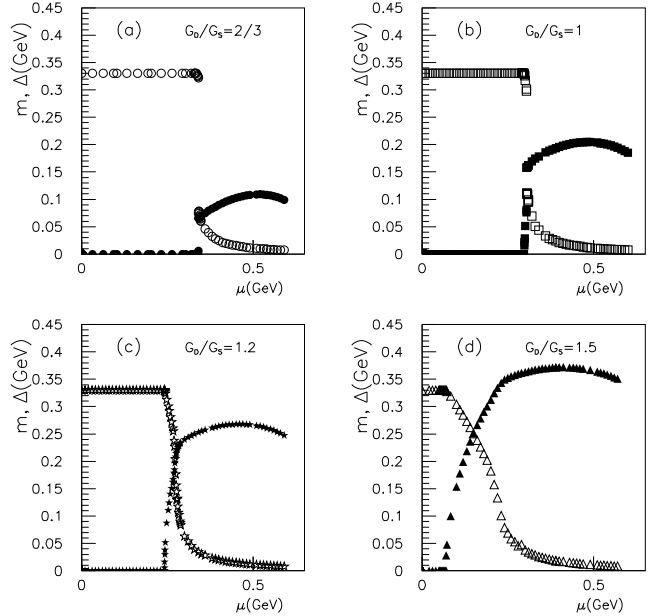}}
\vspace{-1cm}
\caption{The two gaps $m$ (white points) and $\Delta$ (black points)
as functions of chemical potential $\mu$ for $G_D/G_S=2/3, 1, 1.2, 1.5$,
respectively.}
\label{mqdd}
\end{figure}

Firstly, we investigate the phase structure along the chemical potential 
direction with respect to different magnitude of $G_D/G_S$. 
In the explicit chiral symmetry breaking case, we define the point at which
the chiral condensate has maximum change as the critical chemical potential
$\mu_{\chi}$ for the chiral phase transition, and
the point at which the diquark condensate starts to appear as the 
critical chemical potential $\mu_{\Delta}$ for the color 
superconductivity phase transition. 
The two gaps $m$ (white points) and $\Delta$ (black points)
determined by Eqs.~(\ref{gapd-NG}) and (\ref{gapm-NG}) are plotted in 
Fig.~\ref{mqdd} as functions of $\mu$ with respect to different 
$G_D/G_S=2/3, 1, 1.2, 1.5$ in $(a), (b), (c)$ and $(d)$, respectively.

In Fig.~\ref{mqdd} $(a)$, i.e., in the case of $G_D/G_S=2/3$, 
we see that the chiral phase transition and the color superconductivity phase 
transition occur nearly at the same chemical potential 
$\mu_{\chi} \simeq \mu_{\Delta} = 340~{\rm MeV}$. The two phase
transitions are of first order. In the explicit chiral symmetry 
breaking case, there is a small chiral condensate in the 
color superconductivity phase $\mu>\mu_{\chi}$. This phenomena 
has been called the coexistence of the chiral and diquark condensate 
in Ref.~\cite{Berges-Rajagopal}. In this coexistence regime, 
the chiral condensate is small and can be described by the 
Anderson theorem \cite{Kerbikov}.  

In Fig.~\ref{mqdd} $(b)$, with $G_D/G_S=1$, the diquark condensate starts 
to appear at $\mu_{\Delta}=298~{\rm MeV}$, and the chiral symmetry restores at 
$\mu_{\chi}=304.8~{\rm MeV}$. Both $\mu_{\Delta}$ and $\mu_{\chi}$ are smaller 
than those in the case of $G_D/G_S=2/3$. In the regime from $\mu_{\Delta}$ to 
$\mu_{\chi}$, we also see the coexistence of the chiral and diquark condensates. 
However, the chiral condensate in this regime is not due to the explicit 
symmetry breaking but due to the dynamical symmetry breaking. In order to differ the
coexistence of the diquark condensate and the small quark-antiquark condensate in the
chiral symmetric phase as in Fig.~\ref{mqdd} $(a)$, we call the coexistence 
of the diquark condensate and the large quark-antiquark condensate in the chiral 
symmetry broken phase as the ``double broken phase". The diquark gap increases 
continuously from zero to $82~{\rm MeV}$ in this double broken phase, and jumps up to 
$152~{\rm MeV}$ at the critical point $\mu_{\chi}$. 

In Fig.~\ref{mqdd} $(c)$ and $(d)$, we see that with increasing of $G_D/G_S$, 
the diquark condensate starts to appear at smaller $\mu_{\Delta}$, and 
chiral symmetry restores at smaller $\mu_{\chi}$, while the width of the regime 
of double broken phase, $\mu_{\chi}-\mu_{\Delta}$, becomes larger. 
This phenomena has also been found in Ref.~\cite{Vanderheyden} in the random 
matrix model. 

\begin{figure}[ht]
\centerline{\epsfxsize=10cm\epsffile{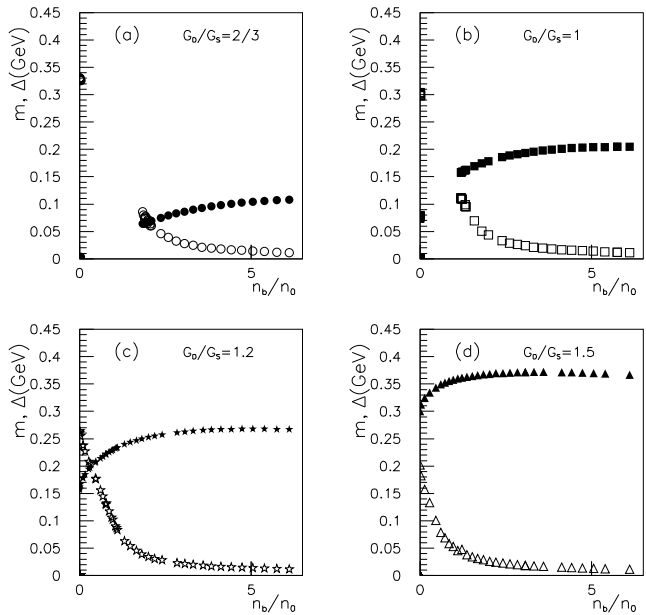}}
\vspace{-1cm}
\caption{The two gaps $m$ (white points) and $\Delta$ (black points)
as functions of scaled baron density $n_b/n_0$ for $G_D/G_S=2/3, 1, 1.2, 1.5$,
respectively. Where $n_0$ is the normal nuclear matter density.}
\label{nbmqdd}
\end{figure}

However, as shown in Fig.~\ref{nbmqdd} $(b), (c)$ and $(d)$,
in the three cases $G_D/G_S=1, 1.2, 1.5$, the diquark condensate starts to 
appear even in the vacuum where $n_b=0$ with $n_b$ the baryon number density. 
This, of course, is not physical, because a nonzero diquark condensate
means the appearance of a nonzero Majorana mass for quarks in vacuum. This 
would be in conflict with the standard definition of the bayron number in vacuum.
The appearance of the diquark condensate in the vacuum may indicate that,
in the NJL model, the diquark coupling strength $G_D$ cannot be very large,
and it should be smaller than the coupling strength $G_S$ in the 
quark-antiquark channel. However, it is worth
to mention that the appearance of the diquark condensate in the vacuum may be a 
consequence of the lack of deconfinement in the NJL model. 

We summarize the phase structure along the chemical potential direction:
1) when $\mu< \mu_{\Delta}$, the chiral symmetry is broken;
2) in the regime from $ \mu_{\Delta}$ to $\mu_{\chi}$, both chiral and 
color symmetries are broken, and 3) when $\mu > \mu_{\chi}$, 
the chiral symmetry restores partially and the phase 
is dominated by the color superconductivity. 
The phase structure depends on the magnitude of $G_D/G_S$. If $G_D/G_S$ is very 
small, the diquark condensate will never appear; If $G_D/G_S<1$ but not too small,
there will be no double broken phase, the chiral phase transition 
and the color superconductivity phase transition
occur at the same critical point $\mu_{\chi}=\mu_{\Delta}$.    
With increasing of $G_D/G_S$, the diquark condensate starts 
to appear at smaller $\mu_{\Delta}$ and the chiral symmetry restores at smaller 
$\mu_{\chi}$, and the width of the double broken phase, $\mu_{\chi}-\mu_{\Delta}$, 
increases. 

\vskip 0.2cm
{\bf Competition between the chiral and diquark condensates}
\vskip 0.2cm
 
In order to understand the competition mechanism and 
explicitly show how the diquark condensate influences the chiral phase
transition, let us analyze Fig.~\ref{mqdd} in more detail.
In the case of $G_D/G_S=0$, only chiral phase transition occurs,
the thermal system in the mean-field approximation is nearly a free 
Fermi gas made of constituent quarks. In the limit of $T=0$, there 
is a very sharp Fermi surface of the constituent quark.  When the chemical 
potential is larger than the constituent quark mass in the vacuum, the chiral 
symmetry restores, and the system of constituent quarks becomes a system of 
current quarks. When a diquark gap $\Delta$ forms in the case of $G_D/G_S \neq 0$,  
it will smooth the sharp Fermi surface of the constituent quark.  
In other words, the diquark pair lowers the sharp Fermi surface, and 
induces a smaller critical chemical potential of chiral restoration.

In TABLE I, we list the chemical potentials $\mu_{\Delta}$, at which the 
diquark gap starts appearing, and $\mu_{\chi}$, at which 
the chiral symmetry restores, for different values of $G_D/G_S$.
$\mu_F^0=345.3~{\rm MeV}$ is the critical chemical potential
in the case of $G_D/G_S=0$. $\Delta_{\chi}$ is the value of diquark gap 
at $\mu_{\chi}$,  if there is a jump, it is the lower value. 

\begin{table}[ht]
\begin{center}
\begin{tabular}{|c|c|c|c|c|c|c|} 
$G_D/G_S$  &  $\mu_{\Delta} ({\rm MeV}) $ & $\mu_{\chi}  ({\rm MeV}) $  & 
$\mu_{\chi}-\mu_{\Delta} ({\rm MeV}) $ & $\Delta_{\chi}  ({\rm MeV})  $
& $ (\mu_F^0-\mu_{\chi})/\Delta_{\chi}  $  
& $ (\mu_F^0-\mu_{\Delta})/\Delta_{\chi}  $
\\ \hline
 1 & 298 & 304.8 & 6 & 82  &  0.49  &  0.57   \\  
\hline
 1.2 & 242 & 266 & 24 &   162   &  0.49   & 0.64   \\   
 \hline
 1.5 & 70 & 190 & 120 & 310  & 0.50  &  0.89  \\   
\end{tabular}
\vskip 0.5cm
\caption{ The $G_D$ dependence of chemical potentials $\mu_{\Delta}$ 
and $\mu_{\chi}$,  
$\mu_F^0=345. 3 {\rm MeV}$.}
\label{final_ex}
\end{center}
\end{table}
\vspace{-0.3cm}

We see that for larger $G_D/G_S$, the diquark condensate
appears at a smaller chemical potential $\mu_{\Delta}$, and the chiral phase
transition occurs at a smaller critical chemical potential $\mu_{\chi}$,
the gap of the diquark condensate $\Delta_{\chi}$ at $\mu_{\chi}$ becomes larger, 
and the regime of the double broken phase becomes wider.

We assume the relation between $\Delta_{\chi}$ and  
$\mu_{\chi}$ as
\begin{eqnarray}
\mu_{\chi}=\mu_F^0 - x \Delta_{\chi}, 
\label{muchi}
\end{eqnarray}
and the relation between $\Delta_{\chi}$ and $\mu_{\Delta}$ as 
\begin{eqnarray}
\mu_{\Delta}=\mu_F^0 - y \Delta_{\chi}.
\label{mudelta}  
\end{eqnarray}

In TABEL I, we listed the values of $x=(\mu_F^0-\mu_{\chi})/\Delta_{\chi}$ and
$y=(\mu_F^0-\mu_{\Delta})/\Delta_{\chi}$ for different $G_D/G_S$.
It is found that $x$ is almost $G_D/G_S$ independent and equal to $1/2$.
As for $y$, it is larger than $1/2$
and increases with increasing of $G_D/G_S$. 
From the Eqs. (\ref{muchi}) and (\ref{mudelta}), we have the relation
\begin{eqnarray}
\mu_{\chi}-\mu_{\Delta}= (y-1/2) \Delta_{\chi}
\end{eqnarray}
for the double broken phase. 
With increasing of $G_D/G_S$, $y$ and $\Delta_{\chi}$ increase,
and then the width of the double broken phase, $\mu_{\chi}-\mu_{\Delta}$,
becomes larger. 

\begin{figure}[ht]
\centerline{\epsfxsize=10cm\epsffile{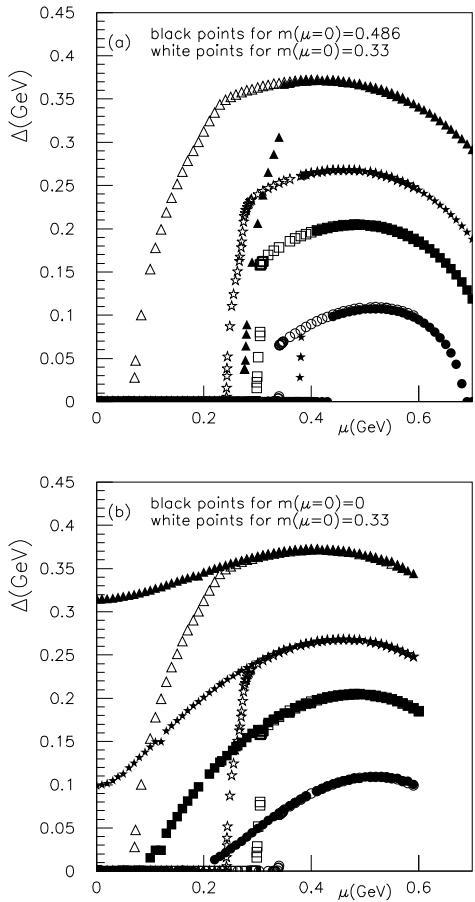}}
\caption{The influence of chiral gap on the 
color superconductivity phase transition in the case of 
$m(\mu=0)= 486 {\rm MeV}$ in $(a)$ and $m(\mu=0)= 0$
in $(b)$. }
\label{gsingp}
\end{figure}

Now we turn to study how the chiral gap influences the color 
superconductivity phase transition. 
Firstly, we change the constituent quark mass in the vacuum from
$330~{\rm MeV}$ to $486 {\rm MeV}$.
To fit the pion properties, the coupling constant in the quark-antiquark channel
is correspondingly increased from $G_S$ to $1.2 G_S$.
We plot the diquark gap as a function of $\mu$ in Fig.~\ref{gsingp}$a$ for
the two vacuum masses and for $G_D/G_S=2/3, 1, 1.2, 1.5$. 
We find that for the same $G_D/G_S$, the diquark gap starts to appear at
much larger chemical potential $\mu_{\Delta}$ when the vacuum mass
increases from $330~{\rm MeV}$ to $486~{\rm MeV}$.

Then we withdraw the quark mass, i.e., taking $m=0$ even in the vacuum (this,
of course, is artificial).
We plot the diquark gap as a function of $\mu$ in Fig.~\ref{gsingp} $b$ 
for $m(\mu=0)=0$ and $ 330~{\rm MeV}$ and for $G_D/G_S=2/3, 1, 1.2, 1.5$. 
We find that for any $G_D/G_S$, the diquark condensate starts 
to appear at a much smaller  
chemical potential $\mu_{\Delta}$ for $m(\mu=0)=0$ compared with
$m(\mu=0)=330~{\rm MeV}$.

From Fig.~\ref{gsingp} $a$ and $b$, we can see that the vacuum quark mass
only changes the critical point of color superconductivity $\mu_{\Delta}$, 
the diquark gaps for different vacuum quark masses 
coincide in the overlap regime of the color superconductivity phase,
where chiral symmetry restores partially.  

From the influence of the diquark gap on the chiral phase transition
and the influence of the chiral gap on the color superconductivity phase 
transition, it is found that there does exist a strong competition between 
the two phases. The competition starts at $\mu_{\Delta}$ and ends at $\mu_{\chi}$.
We call the double broken phase as the competition regime, which becomes wider 
with increasing of $G_D/G_S$. This competition regime is the result of the  
diquark gap smoothing the sharp Fermi surface of the constituent quark.     
If the attractive interaction in the diquark channel is too small, 
there will be no diquark pairs, the system
will be in the chiral breaking phase before $\mu_{\chi}$, and in the chiral
symmetry restoration phase after $\mu_{\chi}$.
If the attractive interaction in the diquark channel is strong enough, 
diquark pairs can be formed and smooth the sharp Fermi surface,
and induce a smaller critical chemical potential of chiral
symmetry restoration. 

\vskip 0.2cm
{\bf The influence of color breaking on quark properties}
\vskip 0.2cm

Finally, we study how the diquark condensate influences the quark properties.
In the normal phase, the quarks in different colors are degenerate. 
However, in the color 
breaking phase, the red and green quarks are involved in the diquark condensate, 
while the blue quarks do not.

\vspace{-1.5cm}
\begin{figure}[ht]
\centerline{\epsfxsize=11cm\epsffile{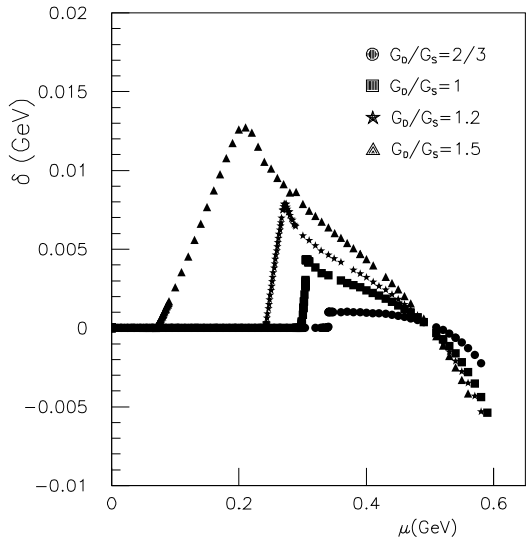}}
\vspace{-1.5cm}
\caption{The difference of the 
chiral condensates for quarks in different colors $\delta$ as 
a function of the chemical potential $\mu$ with respect to
$G_D/G_S=2/3, 1, 1.2, 1.5$, respectively.}
\label{qq13}
\end{figure}

The quark mass $m$ appeared in the formulae of this section is the mass for 
the third quark which does not participate in the diquark condensate. We have
seen that the diquark condensate influences much the quark mass $m$ in the 
competition regime $\mu_{\Delta} < \mu < \mu_{\chi}$. In the color breaking 
phase, i.e., when $\mu > \mu_{\Delta}$, the quark mass $m$ in different cases 
of $G_D$ decreases slowly with increasing $\mu$, and reaches the same value 
at about $\mu=500 {\rm MeV}$. 

The difference of the chiral condensates for quarks in different colors $\delta$ 
defined in Eq. (\ref{delta}) is shown in Fig.~\ref{qq13} as a function of the 
chemical potential $\mu$ with respect to $G_D/G_S=2/3, 1, 1.2, 1.5$. 
It is found that in any case 
$\delta$ is zero before $\mu_{\Delta}$, then begins to increase 
at $\mu_{\Delta}$ and reaches
its maximum at $\mu_{\chi}$, and starts to decrease after $\mu > \mu_{\chi}$, and
approaches to zero at about $\mu=500~{\rm MeV}$,
when $\mu > 500~{\rm MeV}$, $\delta$ becomes negative.
With increasing $G_D$, $\delta(\mu_{\chi})$ increases from $1~{\rm MeV}$ 
for $G_D/G_S=2/3$ to $13~{\rm MeV}$ for $G_D/G_S=1.5$.
Comparing with the magnitude of the diquark condensate, $\delta$ 
is relatively small in the color superconductivity phase. 

\vskip 0.4cm
{\bf Summary of this section}
\vskip 0.4cm

In the mean-field approximation of the extended NJL model, considering only the 
attractive interactions in the $0^{+}$ color singlet quark-antiquark channel and 
in the color anti-triplet diquark channel, the Nambu-Gorkov form of the quark 
propogator has been evaluated with a dynamical quark mass. The Nambu-Gorkov massive 
propagator makes it possible to extend the Green-function approach to the moderate 
baryon density regime. The familiar expression of the thermodynamic potential 
has been re-evaluated by using the massive quark propagator. 

The phase structure along the chemical potential direction has been 
investigated. The system is in the chiral symmetry breaking phase 
below $\mu_{\Delta}$, in the color superconducting phase above $\mu_{\chi}$, 
and the two phases compete with each other in the double broken phase with 
width $\mu_{\chi}-\mu_{\Delta}$. The existence of the double broken phase 
depends on the magnitude of $G_D/G_S$. If $G_D<G_S$, there is no
double broken phase, the chiral phase transition and the color superconductivity 
phase transition occur at the same chemical potential. The double broken phase 
starts to appear when $G_D/G_S \simeq 1$, and the regime of this phase becomes
wider when the diquark coupling strength increases. 

The double broken phase is a consequence of the competition between the 
chiral condensate and the diquark condensate. The competition mechanism 
has been analyzed by investigating the influence of the diquark condensate 
on the sharp Fermi surface of the constituent quark. 
The diquark condensate smoothes the sharp Fermi surface, and induces the 
chiral phase transition and the color superconducting phase transition
occurring at smaller critical chemical potentials. The stronger the diquark coupling 
strength is, the smaller critical chemical potential (or the critical baryon
number density) will be for the appearance of the diquark condensate.  

However, it is found that when $G_D/G_S \ge 1$, the diquark condensate appears 
even in the vacuum. It indicates a nonzero Majorana mass for quarks in the
vacuum, which would be in conflict with the standard definition of the bayron 
number in the vacuum. This unphysical result may suggest that in the NJL model, 
the diquark coupling strength $G_D$ cannot be very large. For the parameters 
of the NJL model used in this section, our results suggest that the diquark 
coupling strength should be smaller than the coupling strength $G_S$ in the 
quark-antiquark channel. Correspondingly, the upper limit of the diquark gap 
parameter $\Delta$ is about $200~{\rm MeV}$. It is also worth to mention that 
the appearance of the diquark condensate in the vacuum may be a result of
the lack of deconfinement in the NJL model.

\section{The gapless 2SC phase}
\label{g2sc}

In the previous section, we discussed the phase transition from the hadronic phase to
the color superconducting phase. Now we are going to discuss the 
two-flavor color superconducting phase under the local charge neutrality condition 
based on Refs.~\cite{g2SC-SH,g2SC-HS-T,g2SC-HS-M,N2SC-HZC}.
 
It is very likely that the color superconducting phase may exist in the core of 
compact stars, where bulk matter should satisfy the charge neutrality condition. 
This is because bulk matter inside the neutron star is bound by the gravitation 
force, which is much weaker than the electromagnetic and the strong color forces. 
Any electric charges or color charges will forbid the formation of bulk matter. 
In addition, matter inside neutron star also needs to satisfy the $\beta$-equilibrium.

In the ideal two-flavor color superconducting (2SC) phase, the pairing
$u$ and $d$ quarks have the same Fermi momenta. Because $u$ quark carries 
electric charge $2/3$, and $d$ quark carries electric charge $-1/3$, 
it is easy to check that quark matter in the ideal 2SC phase is positively 
charged. To satisfy the electric charge neutrality condition, 
roughly speaking, twice as many $d$ quarks as $u$ quarks are needed. This 
induces a large difference between the Fermi surfaces of the two pairing 
quarks, i.e., $\mu_d - \mu_u = \mu_e \approx \mu/4$, where $\mu,\mu_e$ are 
chemical potentials for quarks and electrons, respectively. Naively, one would 
expect that the requirement of the charge neutrality condition will destroy 
the $ud$ Cooper pairing in the 2SC phase. 

Indeed, the interest in the charge neutral 2SC phase was stirred by the paper 
Ref.~\cite{absence2SC}. It was claimed in this paper that there will be no 2SC 
phase inside neutron star under the requirement of the charge neutrality 
condition. In fact, the authors meant that for a charge neutral three flavor 
system, the 2SC+s phase is not favorable compared to the CFL phase.  
This is a natural result under the assumption of a {\it small} 
strange quark mass, even without the requirement of the charge neutrality condition. 
In the framework of the bag model, in which the strange quark mass is very small, 
the CFL phase is always the ground state for cold dense quark matter, and there 
is no space for the existence of two-flavor quark matter. 

However, there is another scenario about the hadron-quark phase transition in 
the framework of the SU(3) NJL model. In the vacuum, quarks obtain their dynamical 
masses induced by the chiral condensate. $u,d$ quarks have constituent mass around 
$330 {\rm MeV}$, while the $s$ quark has heavier constituent mass, which is 
around $500 {\rm MeV}$. With the increasing of the bayron density, the constituent 
quark mass starts to decrease when the chemical potential becomes larger than
its vacuum constituent mass. In this scenario, $s$ quark restores chiral symmetry 
at a larger critical chemical potential than that of $u,d$ quarks. 
If the deconfinement phase transition happens sequentially, there will exist
some baryon density regime for only $u,d$ quark matter and $s$ quark 
is still too heavy to appear in the system. 

It is worth to mention that the effect of the electric charge neutrality condition 
on a three-flavor quark system is very different from that on a two-flavor quark system. 
Because $s$ quark carries $-1/3$ electric charges, it is much easier to neutralize 
the electric charges in a three-flavor quark system than that in a two-flavor quark 
system. However, the color charge neutrality condition is nontrivial 
in a three-flavor quark system, when the strange quark mass is not very small. 
For a  detailed consideration of the charge neutral three-flavor system, see 
recent papers Refs.~\cite{gCFL-AKR,udSC-Hatsuda,gCFL-Ruster,gCFL-Kenji}.   
 
In the following, we focus on the charge neutral two flavor quark system. 
Motivated by the sequential deconfinement scenario, the authors of Ref.~\cite{N2SC-Steiner} 
investigated charge neutral quark matter based on the SU(3) NJL model. To large extent, 
their results agree with those in Ref.~\cite{absence2SC}, i.e., the CFL phase 
is more favorable than the 2SC+s phase in charge neutral three-flavor cold dense 
quark matter, and they did not find the charge neutral 2SC phase. 

However, it was found in Ref.~\cite{N2SC-HZC} that a charge neutral two-flavor color 
superconducting (N2SC) phase does exist, which was confirmed in 
Refs.~\cite{Diploma-Ruster,Mishra-Mishra}. Comparing with the ideal 2SC phase, the 
N2SC phase found in Ref.~\cite{N2SC-HZC} has a largely reduced diquark gap parameter, 
and the pairing quarks have different number densities. The latter contradicts
the paring ansatz in Ref.~\cite{pairansatz}. Therefore, one could suggest that this
phase is an unstable Sarma state \cite{Sarma}. In Ref.~\cite{g2SC-SH}, it was 
shown that the N2SC phase is a stable state under the restriction of the charge 
neutrality condition. As a by-product, which comes out as a very important feature, 
it was found that the quasi-particle spectrum has zero-energy excitation in this 
charge neutral two-flaovr color superconducting phase. Thus this phase was named 
the ``gapless 2SC(g2SC)" phase. 
 
In the following of this section, firstly, using the method introduced in 
the previous section, I will derive the thermodynamic potential for the charge 
neutral two-flavor quark system based on Ref.~\cite{N2SC-HZC}. Then I discuss the 
ground state of the charge neutral two-flavor quark system and introduce the g2SC 
phase \cite{g2SC-SH}. At last, I show the properties of the g2SC phase 
at zero as well as at nonzero temperatures \cite{g2SC-HS-T}, and the color screening 
properties \cite{g2SC-HS-M} in the g2SC phase.

\subsection{Thermodynamic potential of the neutral two-flavor quark system}

We consider the SU(2) NJL model by assuming that the strange 
quark does not appear in the system, and we only consider scalar, 
pseudoscalar mesons and scalar diquark. The Lagrangian density has the 
same form as Eq.~(\ref{lagr}). As it was shown in 
Ref.~\cite{Berges-Rajagopal,Kerbikov}, in the color superconducting phase, 
the small quark mass due to the explicit chiral symmetry breaking has little 
effect on the diquark condensate. Therefore, for simplicity, in this section, 
we only consider the chiral limit case $m_0=0$, and the parameters $G_S$ 
and $\Lambda$ in the chiral limit are fixed as
\begin{eqnarray}
G_S=5.0163 {\rm GeV}^{-2}, \Lambda=0.6533 {\rm GeV}.
\end{eqnarray}
The corresponding effective mass $m=0.314 {\rm GeV}$, and we still choose $G_D$ 
as a free parameter with the ratio $\eta=G_D/G_S$. 

The partition function of the grand canonical ensemble can be evaluated 
by using standard method \cite{kapusta,bellac},
\begin{eqnarray}
\label{Z-CN}
{\cal Z}=N' \int [d {\bar q} ][d q] exp\{ \int_0^{\beta} d \tau \int d^3{\bf x} 
  ~ ( \tilde{\cal L} +\mu {\bar q} \gamma_0 q)\},
\end{eqnarray}
where $\beta=1/T$ is the inverse of the temperature $T$, and 
$\mu$ is the chemical potential. When the electric and color charge 
neutrality conditions are considered, the chemical
potential $\mu$ is a diagonal $6 \times 6$ matrix in flavor and color space,
and can be expressed as
\begin{eqnarray}
\mu=diag(\mu_{ur},\mu_{ug},\mu_{ub},\mu_{dr},\mu_{dg},\mu_{db}),
\end{eqnarray}
the chemical potential for each color and flavor quark is specified by its 
electric and color charges
\begin{eqnarray}
\mu_{ij,\alpha\beta}=(\mu\delta_{ij}-\mu_e Q_{ij})\delta_{\alpha\beta}+ 
\frac{2}{\sqrt{3}}\mu_{8} \delta_{ij}(T_8)_{\alpha\beta},
\end{eqnarray} 
where $Q$ and $T_8$ are generators of $U(1)_Q$ and $U(1)_8$.
Here we assumed $\mu_{3}=0$, because the diquark condenses in the anti-blue 
direction, and the red and green colored quarks are degenerate. 
For the same flavor, the difference of chemical potentials between the red 
and green quarks and the blue quark is induced by $\mu_{8}$.
While for the same color, the difference of chemical potentials between 
$u$ and $d$ is induced by $\mu_e$.

The explicit expressions for each color and flavor quark chemical potential are given as:
\begin{eqnarray}
\mu_{ur}=\mu_{ug}=\mu-\frac{2}{3}\mu_e+\frac{1}{3}\mu_{8}, \nonumber \\
\mu_{dr}=\mu_{dg}=\mu+\frac{1}{3}\mu_e+\frac{1}{3}\mu_{8}, \nonumber \\
\mu_{ub}=\mu-\frac{2}{3}\mu_e-\frac{2}{3}\mu_{8}, \nonumber \\
\mu_{db}=\mu+\frac{1}{3}\mu_e-\frac{2}{3}\mu_{8}.
\end{eqnarray}

For the convenience of calculations, we define 
the mean chemical potential ${\bar \mu}$ for the pairing quarks 
$q_{ur}, q_{dg}$, and $q_{ug}, q_{dr}$
\begin{eqnarray}
{\bar \mu}=\frac{\mu_{ur}+\mu_{dg}}{2}=\frac{\mu_{ug}+\mu_{dr}}{2}
=\mu-\frac{1}{6}\mu_e+\frac{1}{3}\mu_{8},
\end{eqnarray}
and the difference of the chemical potential $\delta\mu$  
\begin{eqnarray}
\delta\mu = \frac{\mu_{dg}-\mu_{ur}}{2}=\frac{\mu_{dr}-\mu_{ug}}{2}=\mu_e /2.
\end{eqnarray}

Because the blue quarks do not participate in the diquark condensate, the partition function 
can be written as a product of three parts, 
\begin{eqnarray}
\label{part-CN}
{\cal Z}={\cal Z}_{const} {\cal Z}_{b} {\cal Z}_{rg} .
\end{eqnarray}
Where the constant part is
\begin{eqnarray}
{\cal Z}_{const}=N'{\rm exp} \{- \int_0^{\beta} d \tau \int d^3{\bf x} ~ 
[\frac{\sigma^2}{4 G_S}
      +\frac{\Delta^{*}\Delta}{4 G_D} \},
\end{eqnarray}
${\cal Z}_{b}$ part is for the unpairing blue quarks, and 
${\cal Z}_{rg}$ part is for the quarks participating in pairing. In the following,
we will derive the contributions of ${\cal Z}_{b}$ and ${\cal Z}_{rg}$.

\vskip 0.2cm 
{\bf Calculation of ${\cal Z}_{b}$}
\vskip 0.2cm

Introducing the  8-spinors for $q_{ub}$ and $q_{db}$,
\begin{equation}
{\bar \Psi_{b}} = ( {\bar q}_{ub}, {\bar q}_{db}; {\bar q}_{ub}^C,{\bar q}_{db}^C ),
\end{equation}
we can express ${\cal Z}_{b}$ as
\begin{eqnarray}
\label{zq3-CN}
 {\cal Z}_{b} & =  & \int[d \Psi_{b}]{\rm exp}\{\frac{1}{2}\sum_{n,{\bf p}}  
 ~ {\bar \Psi}_{b}\frac{[G_0^ {-1}]_{b}}{T}\Psi_{b} \} \nonumber \\
& = & {\rm Det} ^{1/2}(\beta [G_0^{-1}]_{b}),
\end{eqnarray}
where the determinantal operation ${\rm Det}$ is to be carried out over the Dirac, color, 
flavor and the momentum-frequency space, and $[G_0^{-1}]_{b}$ has the form of
\begin{equation}
[G_0^{-1}]_{b} = 
    \left( 
          \begin{array}{cccc}
             [G_0^{+}]_{ub}^{-1}  &  0 & 0 & 0 \\  
            0 &   [G_0^{+}]_{db}^{-1}  & 0 & 0 \\
             0 &  0 &  [G_0^{-}]_{ub}^{-1}  & 0  \\
              0 &  0 & 0 &  [G_0^{-}]_{db}^{-1} 
              \end{array}  
             \right),
\end{equation}
with 
\begin{eqnarray}
[G_0^{\pm}]^{-1}_{i\alpha}=  {\fslash p} \pm {\fslash \mu_{i\alpha}} -m.
\end{eqnarray}
Here we have used ${\fslash p}=p_{\mu}\gamma^{\mu}$ and 
${\fslash \mu_{i\alpha}} = \mu_{i\alpha} \gamma_0$.

For the two blue quarks not participating in the diquark condensate, 
from Eq. (\ref{zq3-CN}), one can have 
\begin{eqnarray}
{\rm ln} {\cal Z}_{b} & = & \frac{1}{2} {\rm ln}  {\rm Det}
(\beta [G_0^{-1}]_{b}) \nonumber \\
 & = & \frac{1}{2} {\rm ln}[{\rm Det} (\beta [G_0^+]_{ub}^{-1}) 
 {\rm Det} (\beta [G_0^-]_{ub}^{-1})] [{\rm Det} (\beta [G_0^+]_{db}^{-1}) 
 {\rm Det} (\beta [G_0^-]_{db}^{-1})].
\end{eqnarray}
Using the results of Eq.~(\ref{DetG0}) in the previous section, one has 
\begin{eqnarray}
 & [{\rm Det} (\beta [G_0^+]_{ub}^{-1})  {\rm Det} (\beta [G_0^-]_{ub}^{-1}) ] & =  \beta^4
  [p_0^2-E_{ub}^{+^2}] [p_0^2-E_{ub}^{-^2}],  \nonumber \\
 &  [{\rm Det} (\beta [G_0^+]_{db}^{-1}) {\rm Det} (\beta [G_0^-]_{db}^{-1})] & =  \beta^4 
 [p_0^2-E_{db}^{+^2}][p_0^2-E_{db}^{-^2}], 
\end{eqnarray}
with $E_{ub}^{\pm}=E \pm \mu_{ub}$ and $ E_{db}^{\pm}=E \pm \mu_{db}$ 
where $E=\sqrt{{\bf p}^2+m^2}$.
Considering the determinant in the flavor, color, spin spaces and the 
momentum-frequency space, we get the expression 
\begin{eqnarray}
{\rm ln} {\cal Z}_{b}   & = &  
   \sum_n \sum_{\bf p} \{ {\rm ln} ( \beta^2 [ p_0^2- (E_{ub}^{+})^2 ] ) 
   +  {\rm ln} ( \beta^2  [ p_0^2- (E_{ub}^{-})^2 ] )  \nonumber \\
   & &  + {\rm ln} ( \beta^2 [ p_0^2- (E_{db}^{+})^2 ] ) 
   +  {\rm ln} ( \beta^2  [ p_0^2- (E_{db}^{-})^2 ] ) \}.
\end{eqnarray}

\vskip 0.2cm
{\bf Calculation of ${\cal Z}_{rg}$}
\vskip 0.2cm

The calculation of ${\cal Z}_{rg}$ here is much more complicated than that
when the two pairing quarks have the same Fermi momonta \cite{NJL-Huang}.
Also, we introduce the Nambu-Gokov formalism for $q_{ur}, q_{ug}, q_{dr}$ and $q_{dg}$, i.e.,
\begin{equation}
{\bar \Psi} = ( {\bar q}_{ur}, {\bar q}_{ug}, {\bar q}_{dr}, {\bar q}_{dg}; {\bar q}_{ur}^C,
{\bar q}_{ug}^C,{\bar q}_{dr}^C,{\bar q}_{dg}^C ).
\end{equation} 
The ${\cal Z}_{rg}$ can have the simple form as
\begin{eqnarray}
\label{zq12-CN}
{\cal Z}_{rg} & = & \int[d \Psi]{\rm exp} \{\frac{1}{2}  \sum_{n,{\bf p}} ~
{\bar \Psi}\frac{G^{-1}_{rg}}{T}  \Psi \}  \nonumber \\
& = & {\rm Det}^{1/2}(\beta G^{-1}_{rg}),
\end{eqnarray}
where
\begin{equation}
{\rm G}^{-1}_{rg} = 
    \left( 
          \begin{array}{cc}
             [G_0^{+}]_{rg}^{-1}  &  \Delta^{-} \\  
            \Delta^{+}  &  [ G_0^{-}]_{rg}^{-1} 
              \end{array}  
             \right),
\end{equation}
with
\begin{eqnarray}
[G_0^{\pm}]_{rg}^{-1}=
\left( 
          \begin{array}{cccc}
            [G_0^{\pm}]_{ur}^{-1}  &  0  & 0 & 0\\  
            0 &  [G_0^{\pm}]_{ug}^{-1}  & 0 & 0\\
            0  & 0 & [G_0^{\pm}]_{dr}^{-1}  &  0 \\  
            0 & 0  & 0 &  [G_0^{\pm}]_{dg}^{-1}
              \end{array}  
             \right),
\end{eqnarray}
and the matrix form for $\Delta^{\pm}$  is
\begin{eqnarray} 
\Delta^{-} = -i \Delta \gamma_5 \left( 
          \begin{array}{cccc}
            0  &  0  & 0 & 1\\  
            0 & 0 & -1 & 0\\
            0  & -1 & 0  &  0 \\  
            1 & 0  & 0 & 0 
              \end{array}  
             \right),  \  \ 
\Delta^{+} = \gamma^0 (\Delta^{-})^{\dagger} \gamma^0.
\end{eqnarray}

From Eq. (\ref{zq12-CN}), one obtains 
\begin{eqnarray}
\label{detg-CN}
{\rm ln} {\cal Z}_{rg}=\frac{1}{2} {\rm ln} {\rm Det} (\beta G^{-1}_{rg}).
\end{eqnarray}
Using the identity Eq.~(\ref{identity}), one can have 
\begin{eqnarray}
{\rm Det} ({\rm G}_{rg}^{-1}) 
& = &{\rm Det} D_1  =  {\rm Det} (- \Delta^{+} \Delta^{-} 
+ \Delta^{+}[G_0^{+}]_{rg}^{-1} 
[\Delta^{-}]^{-1}[G_0^{-}]_{rg}^{-1} ) \nonumber \\
& = & {\rm Det} D_2 = {\rm Det} ( - \Delta^{-} \Delta^{+} 
+ \Delta^{-} [G_0^{-}]_{rg}^{-1} [\Delta^{+}]^{-1}  [G_0^{+}]_{rg}^{-1} ).
\end{eqnarray}
Using the massive energy projectors $\Lambda_{\pm}$ in Eq.~(\ref{energy-project})
for each flavor and color quark, we can work out the diagonal matrix 
$D_1$ and $D_2$ as
\begin{eqnarray}
(D_1)_{11}= (D_1)_{22} = [(p_0 + \delta \mu)^2 - [{ E}_{\Delta}^-]^2] \Lambda_{-}
+ [(p_0 + \delta \mu)^2 - [{E}_{\Delta}^+]^2] \Lambda_{+} \nonumber \\
(D_1)_{33}= (D_1)_{44} = [(p_0 - \delta \mu)^2 - [{E}_{\Delta}^-]^2] \Lambda_{-}
+ [(p_0 - \delta \mu)^2 - [{E}_{\Delta}^+]^2] \Lambda_{+} \nonumber \\
(D_2)_{11}= (D_2)_{22} = [(p_0 - \delta \mu)^2 - [{E}_{\Delta}^+]^2] \Lambda_{-}
+ [(p_0 - \delta \mu)^2 - [ {E}_{\Delta}^-]^2] \Lambda_{+} \nonumber \\ 
(D_2)_{33}= (D_2)_{44} = [(p_0 + \delta \mu)^2 - [{E}_{\Delta}^+]^2] \Lambda_{-}
+ [(p_0 + \delta \mu)^2 - [{E}_{\Delta}^-]^2] \Lambda_{+}, 
\end{eqnarray}
where ${E}_{\Delta}^{\pm} = \sqrt{ (E \pm {\bar \mu})^2+\Delta^2}$.

With the above equations, Eq. (\ref{detg-CN}) can be expressed as
\begin{eqnarray}
{\rm ln} {\cal Z}_{rg} & = &  
2  \sum_n \sum_p \{ {\rm ln}[\beta^2 (p_0^2-(E_{\Delta}^-+\delta\mu))^2]
 + {\rm ln}[\beta^2 (p_0^2-(E_{\Delta}^- - \delta\mu))^2 ] \nonumber \\
  & + & {\rm ln}[\beta^2 (p_0^2-(E_{\Delta}^+ + \delta\mu))^2 ]
 + {\rm ln}[\beta^2 (p_0^2-(E_{\Delta}^+ -\delta\mu))^2] \}.
\end{eqnarray}

\vskip 0.2cm
{\bf The thermodynamic potential}
\vskip 0.2cm

Using the helpful relation Eq.~(\ref{lnzf}),
one can evaluate the thermodynamic potential of the quark system
in the mean-field approximation. The total thermodynamic potential for $u, d$ quarks in 
$\beta$-equilibrium with electrons takes the form \cite{N2SC-HZC,g2SC-SH,g2SC-HS-T}:
\begin{eqnarray} 
\Omega_{u,d,e} &=& 
-\frac{1}{12\pi^2}\left(\mu_{e}^{4}+2\pi^{2}T^{2}\mu_{e}^{2}
+\frac{7\pi^{4}}{15} T^{4} \right) + \frac{m^2}{4G_S} \nonumber\\
&+&  \frac{\Delta^2}{4G_D}
-\sum_{a} \int\frac{d^3 p}{(2\pi)^3} \left[E_{a}
+2 T\ln\left(1+e^{-E_{a}/T}\right)\right] ,
\label{pot}
\end{eqnarray} 
where the electron mass was taken to be zero, which is sufficient
for the purposes of the current study. The sum in the second line of
Eq.~(\ref{pot}) runs over all (6 quark and 6 antiquark)
quasi-particles. The explicit dispersion relations and the degeneracy 
factors of the quasi-particles read
\begin{eqnarray}
E_{ub}^{\pm} &=& E(p) \pm \mu_{ub} , \hspace{26.6mm} [\times 1]
\label{disp-ub} \\
E_{db}^{\pm} &=& E(p) \pm \mu_{db} , \hspace{26.8mm} [\times 1]
\label{disp-db}\\
E_{\Delta^{\pm}}^{\pm} &=& E_{\Delta}^{\pm}(p) \pm  \delta \mu .
\hspace{25.5mm} [\times 2]
\label{2-degenerate}
\end{eqnarray} 

\subsection{Gap equations and number densities}

From the thermodynamic potential Eq.~(\ref{pot}), we can
derive the gap equations of the order parameters $m$ and  $\Delta$ 
for the chiral and color superconducting phase transitions, as well
as the number densities for quarks with different color and flavor.

\vskip 0.2cm
{\bf Gap equation for the quark mass} 
\vskip 0.2cm

The gap equation for the quark mass can be 
derived by using

\begin{eqnarray}
\frac{\partial \Omega_{u,d,e}}{\partial m}=0.
\end{eqnarray}
The explicit expression for the above equation is 
\begin{eqnarray}
 m &=& 4G_S 
\int\frac{d^3{\bf p}}{(2\pi)^3}\frac{m}{E} 
\Bigg[ 2\frac{E-\bar\mu}{E_{\Delta}^-}[1- {\tilde f}(E_{\Delta^+}^-) 
- {\tilde f}(E_{\Delta^-}^- ) ] \nonumber \\
&  &  + 2 \frac{E+\bar\mu}{E_{\Delta}^+}
[1- {\tilde f}(E_{\Delta^+}^+)
- {\tilde f}(E_{\Delta^-}^+ ) ]  \nonumber \\
&  & + 
2- {\tilde f}(E_{ub}^+) - {\tilde f}(E_{ub}^-) 
- {\tilde f}(E_{db}^+) - {\tilde f}(E_{db}^-)
\Bigg].
\label{gap-m}
\end{eqnarray}
In the chiral limit, the trivial solution $m=0$ corresponds to a 
chirally symmetric phase of quark matter, while a nontrivial solution 
$m \neq 0$ corresponds to a phase with spontaneously broken chiral symmetry.

\vskip 0.2cm
{\bf Gap equation for the diquark gap}
\vskip 0.2cm

Similarly, the gap equation for the diquark 
condensate can be derived using
\begin{eqnarray}
\frac{\partial \Omega_{u,d,e}}{\partial \Delta}=0,
\end{eqnarray}
the explicit expression has the form 
\begin{eqnarray}
 \Delta [1- & & 4G_D\int\frac{d^3{\bf p}}{(2\pi)^3} 
[ 2 \frac{1}{E_{\Delta}^-}(1- {\tilde f}(E_{\Delta^+}^-) 
- {\tilde f}(E_{\Delta^-}^- ) ) \nonumber \\
&  & + 2 \frac{1}{E_{\Delta}^+}(1- {\tilde f}(E_{\Delta^+}^+)
- {\tilde f}(E_{\Delta^-}^+ ) ) ] =0.
\label{gap-T}
\end{eqnarray}
This equation can also have trivial as well as nontrivial solutions for $\Delta$,
and the nontrivial solution $\Delta \neq 0$ corresponds to a color superconducting phase.

\vskip 0.2cm
{\bf Number densities of quarks}
\vskip 0.2cm

According to one of the criteria of the g2SC phase, the densities 
of the quark species that participate in pairing dynamics are not 
equal at zero temperature \cite{g2SC-SH,N2SC-HZC}. This is in contrast to regular 
pairing in the conventional gapped color superconductors \cite{pairansatz}. 

The quark densities are obtained by taking the partial 
derivatives of the potential $\Omega_{u,d,e}$ in Eq.~(\ref{pot}) with 
respect to the chemical potentials for each flavor and color. 
Because of the residual $SU(2)_c$ symmetry in the ground state of 
quark matter, the densities of red and green quarks for the same flavor,
are equal in the g2SC ground state. For example, the densities of 
the up quarks participating in the Cooper pairing read
\begin{eqnarray}
 n_{ur} = n_{ug} & = & \int\frac{d^3{\bf p}}{(2\pi)^3} 
\Bigg[\frac{E+\bar\mu}{E_{\Delta}^+}
[1- {\tilde f}(E_{\Delta^+}^+)
    -{\tilde f}(E_{\Delta^-}^+ ) ] \nonumber \\
 &  &- \frac{E-\bar\mu}{E_{\Delta}^-}
[1- {\tilde f}(E_{\Delta^+}^-)
- {\tilde f}(E_{\Delta^-}^- ) ]\nonumber \\
&  & + {\tilde f}(E_{\Delta^+}^-) + {\tilde f}(E_{\Delta^+}^+)
 - {\tilde f}(E_{\Delta^-}^+) - {\tilde f}(E_{\Delta^-}^-)\Bigg].
\end{eqnarray}
Similarly, for the densities of the down quarks, we get
\begin{eqnarray}
 n_{dg} = n_{dr} & = & \int\frac{d^3{\bf p}}{(2\pi)^3} 
   \Bigg[\frac{E+\bar\mu}{E_{\Delta}^+}
[1- {\tilde f}(E_{\Delta^+}^+)
    -{\tilde f}(E_{\Delta^-}^+ )] \nonumber \\
&&-\frac{E-\bar\mu}{E_{\Delta}^-}[1- {\tilde f}(E_{\Delta^+}^-)
- {\tilde f}(E_{\Delta^-}^- ) ]\nonumber \\
&  & - {\tilde f}(E_{\Delta^+}^-) - {\tilde f}(E_{\Delta^+}^+)
 + {\tilde f}(E_{\Delta^-}^+) + {\tilde f}(E_{\Delta^-}^-)\Bigg]. 
\end{eqnarray}
As we see, the densities of the up and down quarks participating in 
the Cooper pairing are not equal. In fact, the difference of the 
densities is given by
\begin{eqnarray}
n_{dg} - n_{ur} &=& n_{dr} - n_{ug} \nonumber \\
 & = & 2 \int\frac{d^3{\bf p}}{(2\pi)^3}
\left[{\tilde f}(E_{\Delta^-}^-)-{\tilde f}(E_{\Delta^+}^-) 
+{\tilde f}(E_{\Delta^-}^+)-{\tilde f}(E_{\Delta^+}^+)\right],
\label{n_dg-n_ur}
\end{eqnarray}
which is always nonzero at finite temperature provided the mismatch 
parameter $\delta\mu$ is nonzero. It is even more important for us here 
that this difference is nonzero at zero temperature in the g2SC phase 
of quark matter,
\begin{eqnarray}
\left.\left(
n_{dg} - n_{ur} \right)\right|_{T=0} &=& \left.\left(
n_{dr} - n_{ug} \right)\right|_{T=0} \nonumber \\
& = & 
\theta\left(\delta\mu-\Delta\right)
\frac{2}{3\pi^2} \sqrt{(\delta\mu)^2-\Delta^2}
\left(3\bar\mu^2+(\delta\mu)^2-\Delta^2\right).
\label{n_dg-n_ur-T0}
\end{eqnarray}
This is in contrast to the 2SC phase ($\delta\mu <\Delta$) where
this difference is zero in agreement with the arguments of 
Ref.~\cite{pairansatz}. 

The unpaired blue quarks, $u_b$ and $d_b$, are singlet states with 
respect to SU(2)$_c$ symmetry of the ground state. The densities of
these quarks are 
\begin{eqnarray}
n_{ub}= 2 \int\frac{d^3{\bf p}}{(2\pi)^3}
\left[ {\tilde f}(E_{ub}^-)-{\tilde f}(E_{ub}^+) \right]
\simeq \frac{\mu_{ub}}{3}\left(\frac{\mu_{ub}^2}{\pi^2} + T^2\right),
\end{eqnarray}
for the up quarks, and
\begin{eqnarray}
n_{db}= 2 \int\frac{d^3{\bf p}}{(2\pi)^3}
\left[ {\tilde f}(E_{db}^-)-{\tilde f}(E_{db}^+) \right]
\simeq \frac{\mu_{db}}{3}\left(\frac{\mu_{db}^2}{\pi^2} + T^2\right),
\end{eqnarray}
for the down quarks, respectively.

\subsection{The charge neutral ground state and the g2SC phase}

If a macroscopic chunk of quark matter exists inside compact stars, it must 
be neutral with respect to electric as well as color charges.

\vskip 0.2cm
{\bf electric and color charge neutrality}
\vskip 0.2cm

It has been shown that, in QCD case, the two-flavor color superconductivity is 
automatically color neutral, because of a non-vanishing expectation value of the 
8th gluon field, for more details, see Refs. \cite{Color-neutral-Gerhold,Color-neutral-Dirk}. 
However, in the framework of the NJL model which lacks gluons, the color 
charge neutrality condition is to choose $\mu_{8}$ such that the system has zero 
net charge $n_8$, i.e.,
\begin{eqnarray} 
n_8 &=& \frac{4}{3}\int\frac{d^3{\bf p}}{(2\pi)^3}
\Bigg[ - \frac{E-\bar\mu}{E_{\Delta}^-}
[1- {\tilde f}(E_{\Delta^+}^-) 
- {\tilde f}(E_{\Delta^-}^- ) ] \nonumber \\
&  & +  \frac{E+\bar\mu}{E_{\Delta}^+}
[1- {\tilde f}(E_{\Delta^+}^+)
- {\tilde f}(E_{\Delta^-}^+ ) ] \nonumber \\
& & + {\tilde f}(E_{ub}^+) - {\tilde f}(E_{ub}^-)
+ {\tilde f}(E_{db}^+) - {\tilde f}(E_{db}^-) \Bigg].
\label{n8}
\end{eqnarray}
The detailed numerical calculation in Ref.~\cite{N2SC-HZC} shows that it is very easy to
satisfy the color neutrality condition, $\mu_8$ is around several ${\rm MeV}$, which is 
very small comparing with the quark chemical potential $\mu$. In the following discussions, 
color charge neutrality condition is always satisfied. 

In the two-flavor quark system, the electric neutrality plays an essential role.
Similar to the color charge, the electric charge neutrality condition is to choose 
$\mu_e$ such that the system has zero net electric charge $n_Q$, i.e.,
\begin{eqnarray}
n_Q &= & 
-\frac{1}{2} n_8 +\frac{\mu_e^3}{3\pi^2} + \frac{1}{3}\mu_e T^2
+2 \int\frac{d^3{\bf p}}{(2\pi)^3} \Bigg[ 
{\tilde f}(E_{ub}^+) - {\tilde f}(E_{ub}^-) \nonumber \\
& &
-{\tilde f}(E_{\Delta^+}^-) - {\tilde f}(E_{\Delta^+}^+)
 + {\tilde f}(E_{\Delta^-}^+) + {\tilde f} (E_{\Delta^-}^-)
\Bigg] .
\label{nQ}
\end{eqnarray}
To neutralize the electric charge in the homogeneous dense $u, d$ quark matter, 
roughly speaking, twice as many $d$ quarks as $u$ quarks are needed, i.e., $n_d \simeq 2 n_u$, 
where $n_{u,d}$ are the number densities for $u$ and $d$ quarks. This induces a mismatch
between the Fermi surfaces of pairing quarks, i.e., $\mu_d - \mu_u = \mu_e = 2 \delta\mu$, where
$\mu_{e}$ is the electron chemical potential. 

\vskip 0.2cm
{\bf The proper way to find the charge neutral ground state}
\vskip 0.2cm

Now, we discuss the role of the electric charge neutrality condition. 
If a macroscopic chunk of quark matter has nonzero net electric charge density $n_Q$,
the total thermodynamic potential for the system should be given by
\begin{eqnarray}
\Omega &=&  \Omega_{Coulomb} + \Omega_{u,d,e}, 
\end{eqnarray}
where $\Omega_{Coulomb} \sim n_Q^2 V^{2/3}$ ($V$ is the volume of the 
system) is induced by the repulsive Coulomb interaction. The energy 
density grows with increasing the volume of the system, as a result, it is almost
impossible for matter inside stars to remain charged over macroscopic distances.
So bulk quark matter must satisfy electric neutrality condition
with $\Omega_{Coulomb}|_{n_Q=0}=0$, and $\Omega_{u,d,e}|_{n_Q=0}$ 
is on the neutrality line. Under the charge neutrality condition, the total 
thermodynamic potential of the system is $\Omega|_{n_Q=0}=\Omega_{u,d,e}|_{n_Q=0}$. 
 
\begin{figure}
\vskip 0.3cm
\centerline{\epsfxsize=10cm\epsffile{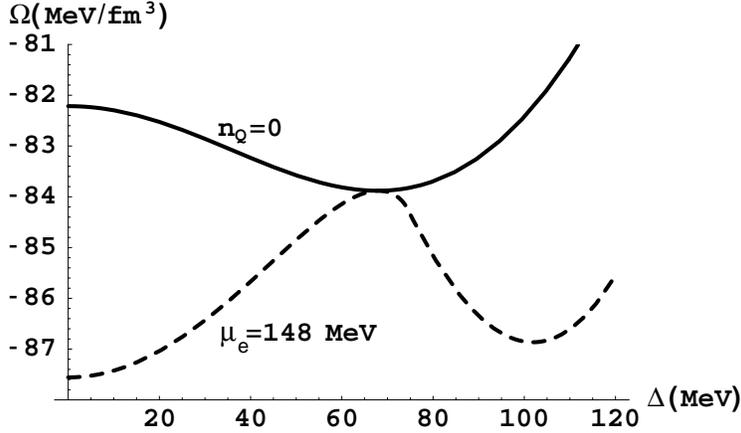}}
\caption{The effective potential as a function of the diquark gap $\Delta$
calculated at a fixed value of the electric chemical potential $\mu_e =
148 $ ~MeV (dashed line), and the effective potential defined along the
neutrality line (solid line). The results are plotted for $\mu=400$ MeV
with $\eta=0.75$.}
\label{V2D}
\end{figure}

Here, we want to emphasize that: {\it  The proper way to find the ground state of 
homogeneous neutral $u, d$ quark matter is to minimize the thermodynamic 
potential along the neutrality line $\Omega|_{n_Q=0} = \Omega_{u,d,e}|_{n_Q=0}$. This
is different from that in the flavor asymmetric quark system, where $\beta$-equilibrium 
is required but $\mu_e$ is a free parameter, and the ground state for flavor asymmetric 
quark matter is determined by minimizing the thermodynamic potential $\Omega_{u,d,e}$.}

From Fig.~\ref{V2D}, we can see the difference in determining the ground state for a charge 
neutral system and for a flavor asymmetric system when $\mu_e$ is a free parameter. In 
Fig.~\ref{V2D}, at a given chemical 
potential $\mu=400 ~ {\rm MeV}$ and $\eta=G_D/G_S=0.75$, the thermodynamic potential 
along the charge neutrality line $\Omega|_{n_Q=0}$ as a function of the diquark 
gap $\Delta$ is shown by the solid line. The minimum gives the ground state of the neutral 
system, and the corresponding values of the chemical potential and the diquark gap are 
$\mu_e=148 ~{\rm MeV}$ 
and $\Delta=68 ~{\rm MeV}$, respectively. If we switch off the charge neutrality conditions, 
and consider the flavor asymmetric $u, d$ quark matter in $\beta-$ equilibrium like in 
Refs.~\cite{Bed-asym,Kiriyama-asym}, the electric chemical potential $\mu_e$ becomes a 
free parameter. At a fixed $\mu_e=148 ~{\rm MeV}$ and with color charge neutrality, the 
thermodynamic potential is shown as a function of the diquark gap by the dashed line in 
Fig.~\ref{V2D}. The minimum gives the ground state of the flavor asymmetric system, and 
the corresponding diquark gap is $\Delta=0$, but this state has negative electric charge 
density, and cannot exist in the interior of compact stars.

\vskip 0.2cm
{\bf $G_D$ dependent charge neutral ground state and the g2SC phase}
\vskip 0.2cm

Equivalently, the neutral ground state can also be determined by solving the diquark gap 
equation Eq.~(\ref{gap-T}) together with the charge neutrality conditions Eqs.~(\ref{n8}) 
and (\ref{nQ}), here we regarded the quark mass is zero in the chiral symmetric phase. 

\begin{figure}
\centerline{\epsfxsize=10cm\epsffile{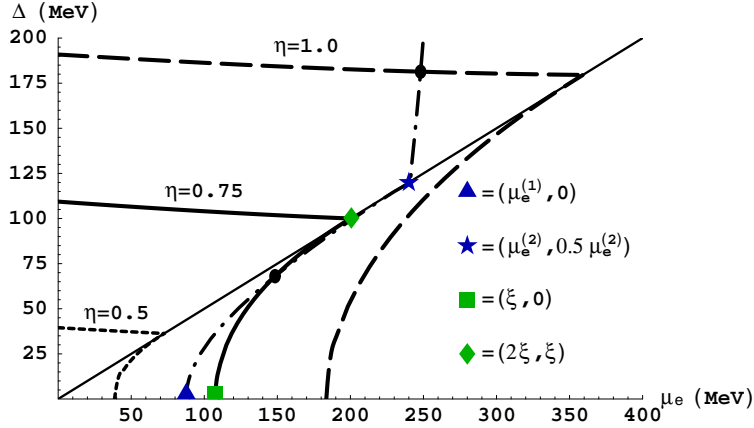}}
\vskip 0.3cm
\caption{The graphical representation of the solution to the charge 
neutrality conditions (thick dash-dotted line) and the solution to 
the gap equation for three different values of the diquark coupling 
constant (thick solid and dashed lines). The intersection points 
represent the solutions to both. The thin solid line divides two 
qualitatively different regimes, $\Delta<\delta\mu$ and $\Delta>
\delta\mu$. The results are plotted for $\mu=400$ MeV and three 
values of diquark coupling constant $G_{D} = \eta G_S$ with $\eta=0.5$, 
$\eta=0.75$, and $\eta=1.0$.}
\label{gapneutral}
\end{figure}

We visualize this method in Fig.~\ref{gapneutral}, with
color neutrality always satisfied, at a given chemical potential $\mu=400 ~{\rm MeV}$.
The nontrivial solutions to the diquark gap equation as functions of the electric 
chemical potential $\mu_e$ are shown by a thick-solid line ($\eta=G_D/G_S=0.75$), a long-dashed 
line ($\eta=1.0$), and a short-dashed line ($\eta=0.5$). It is found that for each 
$\eta$, the solution is divided into two branches by the thin-solid line $\Delta=\delta\mu$, 
and the solution is very sensitive to $\eta$. Also, there is always a trivial solution to 
the diquark gap equation, i.e., $\Delta =0$.  The solution of the charge neutrality 
conditions is shown by a thick dash-dotted line, which is also divided into two branches 
by the thin-solid line $\Delta=\delta\mu$, but the solution of the charge neutrality is 
independent of $\eta$.

The cross-point of the solutions to the charge neutrality conditions and 
the diquark gap gives the solution of the system. We find that the neutral 
ground state is sensitive to the coupling constant $G_D=\eta G_S$ in the 
diquark channel. In the case of a very strong coupling (e.g., $\eta=1.0$ case), 
the charge neutrality line crosses the upper branch of the solution to the 
diquark gap, the ground state is a charge neutral regular 2SC phase with 
$\Delta>\delta\mu$. In the case of weak coupling (e.g., $\eta=0.5$), the 
charge neutrality line crosses only the trivial solution of the diquark gap, 
i.e., the ground state is a charge neutral normal quark matter with $\Delta=0$. 

The regime of intermediate coupling (see, e.g., $\eta=0.75$ case) is most interesting,
the charge neutrality line crosses the lower branch of the solution of the diquark 
gap. It should be noticed that $\eta=0.75$ is from the Fierz transformation, and 
$\eta = 2.26/3 \simeq 0.75$ obtained in the SU(2) NJL model in Ref.~\cite{dnjl2} from 
fitting the vacuum baryon mass. 
We will see that in the regime of intermediate coupling, the phase with 
$\Delta<\delta\mu$ is a gapless 2SC (g2SC) phase.

\vskip 0.2cm
{\bf Quasi-particle spectrum} 
\vskip 0.2cm

As we already mentioned that the pairing quarks have different number densities
in the g2SC phase, which is different from 
the regular 2SC phase. It is the quasi-particle spectrum that makes the g2SC 
phase different from the gapped 2SC phase.

\begin{figure}
\hbox{
\centerline{\epsfxsize=6cm\epsffile{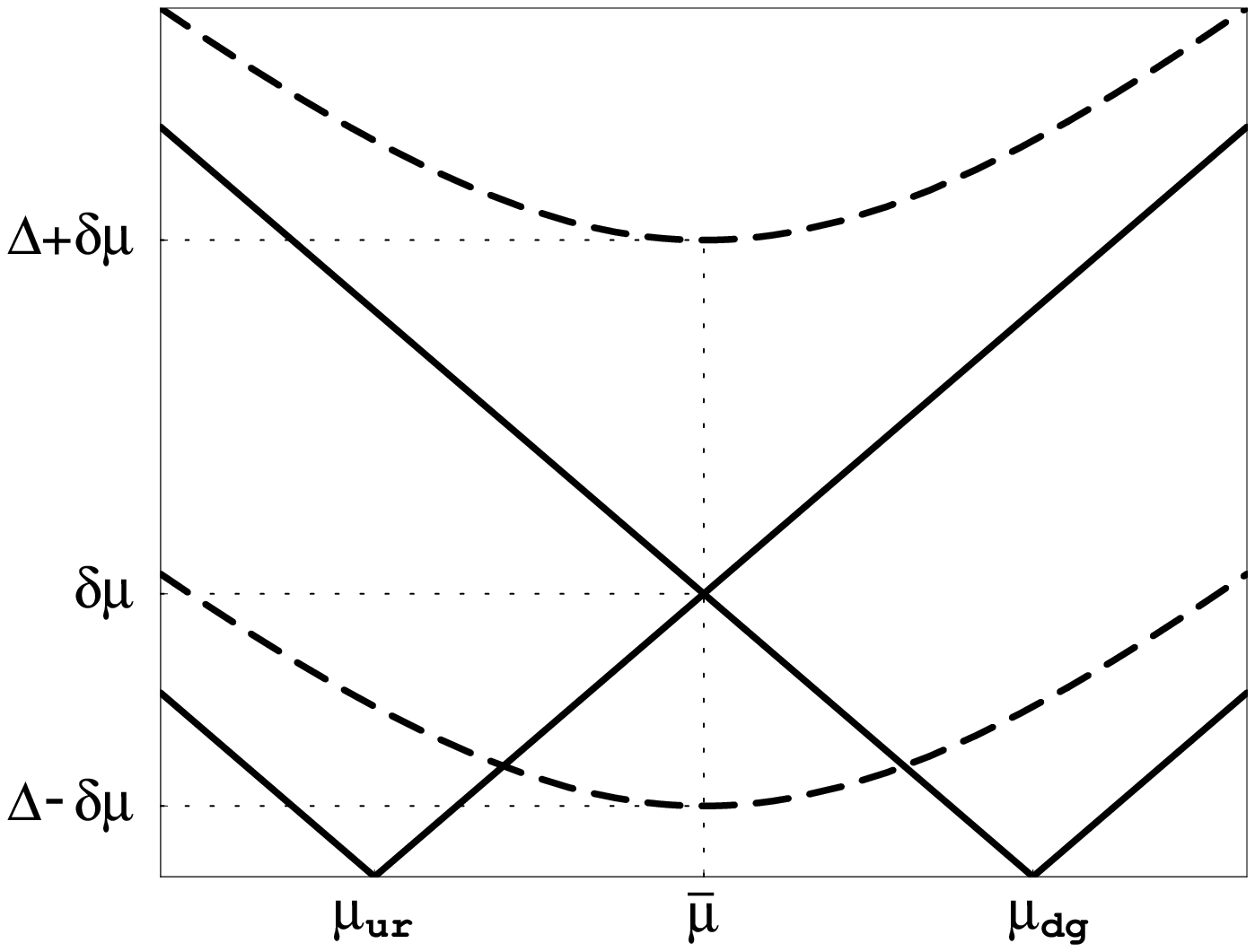}
\hskip 0.5cm
\epsfxsize=6cm\epsffile{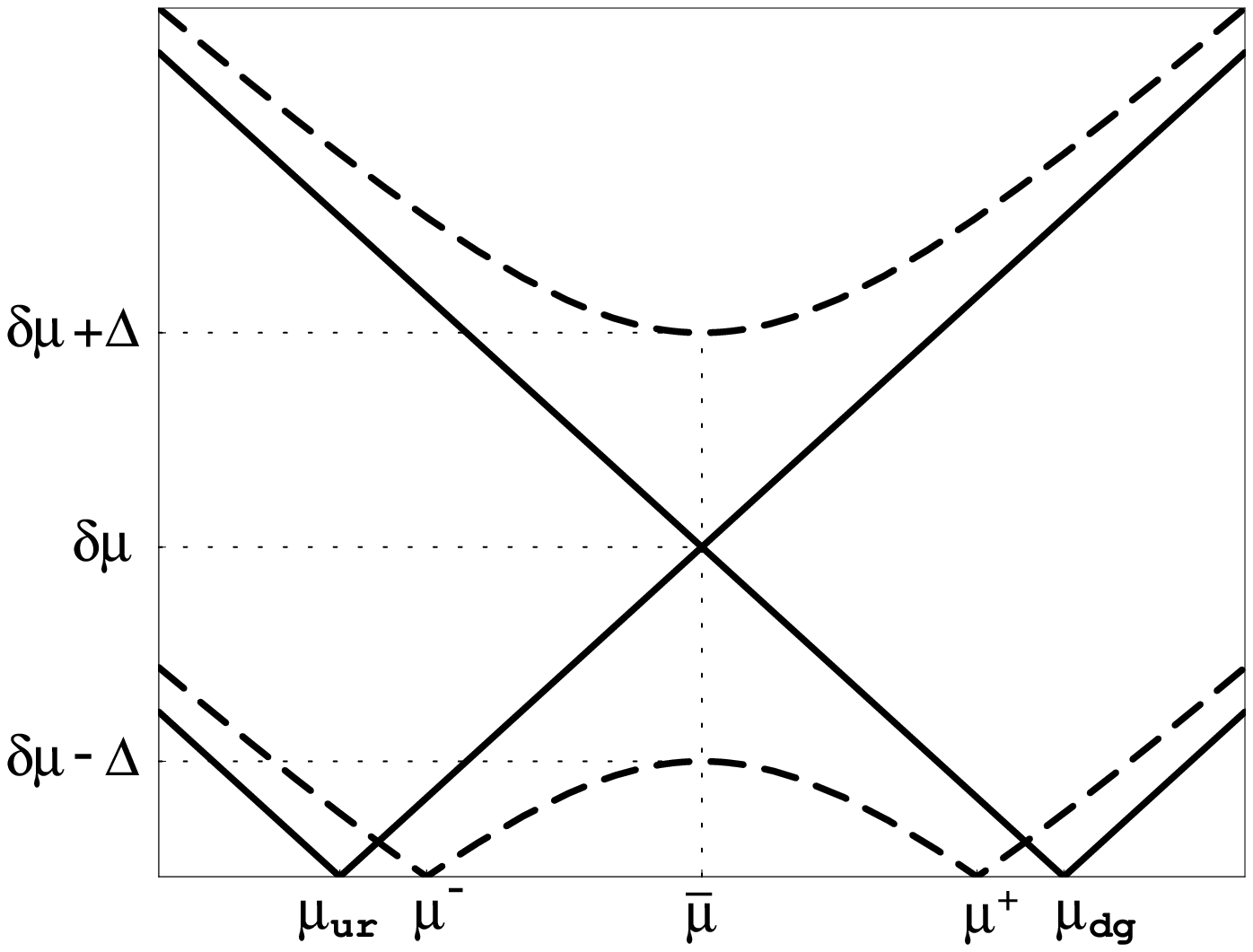}}}
\vskip 0.3cm
\caption{The quasi-particle dispersion relations at low energies in the 2SC phase 
(left panel) and in the g2SC phase (right panel).}
\label{quasi-spec}
\end{figure}

It is instructive to start with the excitation spectrum in the case of the 
ideal 2SC phase when $\delta\mu=0$. With the conventional choice of the 
gap pointing in the anti-blue direction in the color space, the blue quarks are 
not affected by the pairing dynamics, and the other four quarsi-particle 
excitations are linear superpositions of $u_{r,g}$ and $d_{r,g}$ quarks and 
holes. The quasi-particle is nearly identical with a quark at large momenta
and with a hole at small momenta. We represent the quasi-particle in the form 
of $Q(quark, hole)$, then the four quasi-particles can be represented explicitly 
as $Q(u_r, d_g)$, $Q(u_g, d_r)$, $Q(d_r, u_g)$ and $Q(d_g, u_r)$. When 
$\delta\mu=0$, the four quasi-particles are degenerate, and have a common 
gap $\Delta$.
If there is a small mismatch ($\delta\mu < \Delta$) between the Fermi surfaces 
of the pairing $u$ and $d$ quarks, the excitation spectrum will change. For 
example, we show the excitation spectrum of $Q(u_r, d_g)$ and $Q(d_g, u_r)$ 
in the left panel of Fig.~\ref{quasi-spec}. We can see that $\delta \mu$ induces 
two different dispersion relations, the quasi-particle $Q(d_g, u_r)$ has a smaller 
energy gap $\Delta - \delta\mu$, and the quasi-particle $Q(u_r, d_g)$ has a larger 
energy gap $\Delta + \delta\mu$. This is similar to the case when the mismatch is 
induced by the mass difference of the pairing quarks \cite{gapless-ABR}.

If the mismatch $\delta\mu$ is larger than the gap parameter $\Delta$, the 
lower dispersion relation for the quasi-particle $Q(d_g, u_r)$ will cross 
the zero-energy axis, as shown in the right panel of Fig.~\ref{quasi-spec}. 
The energy of the quasi-particle $Q(d_g, u_r)$ vanishes at two values of momenta 
$p=\mu^{-}$ and $p=\mu^{+}$ where 
$\mu^{\pm}\equiv \bar\mu\pm\sqrt{(\delta\mu)^2-\Delta^2}$. 
Thus this phase is called the gapless 2SC (g2SC) phase. 

An unstable gapless CFL phase has been found in Ref.~\cite{gapless-ABR}, 
and a similar stable gapless color superconductivity could also appear in 
a cold atomic gas \cite{Liu-Wilczek} or in $u, s$ or $d, s$ quark matter 
when the number densities are kept fixed \cite{gapless-GLW}. Also, some 
gapless phases may appear due to P-wave interactions in the cold atomic 
system \cite{P-wave}. 
 
As we shall see, the interplay of the neutrality condition
and the solution to the gap equation produces some unusual properties
of the g2SC phase that have no analogue in well known systems.

\subsection{The g2SC phase at nonzero temperatures}
\label{finite-T}

In a superconducting system, when one increases the temperature at a given 
chemical potential, thermal motion will eventually break up the quark 
Cooper pairs. In the weakly interacting Bardeen-Copper-Schrieffer (BCS) 
theory, the transition between the superconducting and normal phases is 
usually of second order. The ratio of the critical temperature 
$T_c^{\rm BCS}$ to the zero temperature value of the gap 
$\Delta_0^{\rm BCS}$ is a universal value \cite{ratio-in-BCS}
\begin{eqnarray}
r_{\rm BCS}=\frac{T_c^{\rm BCS}}{\Delta_0^{\rm BCS}} =
\frac{{\rm e}^{\gamma_E}}{\pi} \approx 0.567,
\label{r_BCS}
\end{eqnarray}
where $\gamma_E \approx 0.577$ is the Euler constant.
In the conventional 2SC phase of quark matter with equal densities of
the up and down quarks, the ratio of the critical temperature to the 
zero temperature value of the gap is also the same as in the BCS theory 
\cite{weak-Tc-Dirk}. In the spin-0 color flavor locked phase as well as 
in the spin-1 color spin locked phase, on the other hand, this ratio
is larger than BCS ratio by the factors $2^{1/3}$ and $2^{2/3}$, 
respectively. These deviations are related directly to the presence 
of two different types of quasiparticles with nonequal gaps 
\cite{Spin1-Tc-Schmitt}. 

This commonly accepted picture of the finite temperature effects in 
superconducting phases changes drastically in the case of dense quark 
matter when the $\beta$-equilibrium and the local neutrality conditions 
are enforced. Below, we study this in detail.

\vskip 0.2cm
{\bf Gap equation and charge neutrality condition}
\vskip 0.2cm

Here we consider the gap equation and charge neutrality conditions in 
the g2SC phase at finite temperature. Because of additional technical 
complications appearing at finite temperature, most of the results that 
follow will be obtained by numerical computation. It is important to 
keep in mind, however, that our approach remains conceptually the 
same as that at zero temperature.
  
Before studying the gap equation (\ref{gap-T}), we recall that the right 
hand side of this equation is a function of the three chemical potentials 
$\mu$, $\mu_e$ and $\mu_8$. The values of $\mu_e$ and $\mu_8$ are not the 
free parameters in neutral quark matter. They are determined by satisfying 
the two charge neutrality conditions. In our analysis, on the other hand, 
it is very convenient to enforce only the color charge neutrality by 
choosing the function $\mu_8(\mu,\mu_e,\Delta)$ properly. As for the 
electric chemical potential $\mu_e$, it is treated as an independent 
parameter at intermediate stages of calculation. Only at the very end its 
value is adjusted to make the quark matter also electricly neutral.

As in the case of zero temperature, the resulting color chemical potential 
$\mu_8$ is much smaller than the other two chemical potentials in neutral 
quark matter. This, of course, is not surprizing. Small nonzero values of 
$\mu_8$ [typically, $\mu_8 \sim \Delta^2/\mu$] are required only because 
an induced color charge of the diquark condensate should be compensated.
Therefore, the smallness of $\mu_8$ is protected by the smallness of the
order parameter. The numerical calculations show, in fact, that even 
the approximation $\mu_8=0$ does not modify considerably the exact solutions 
to the gap equation and to the electric charge neutrality condition. In 
practice, we either tabulate the function $\mu_8(\mu,\mu_e,\Delta)$ for a 
given set of parameters, or determine it numerically in the vicinity of 
the ground state solution.

\begin{figure}
\centerline{\epsfxsize=10.5cm\epsffile{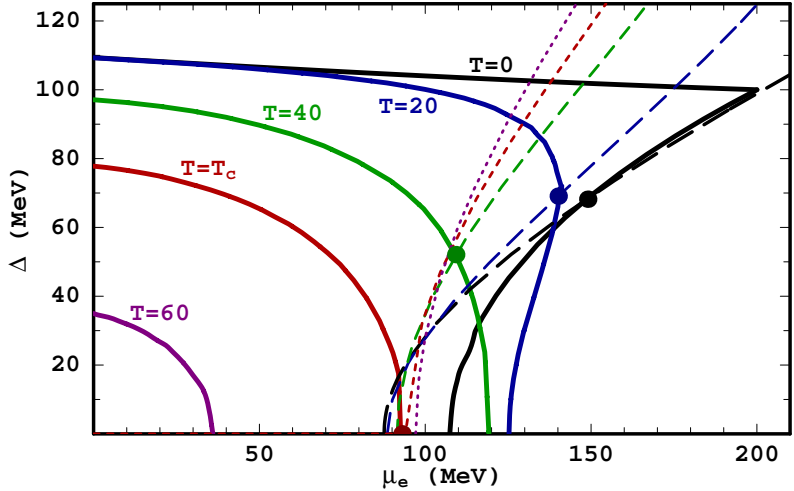}}
\vskip 0.3cm
\caption{The solutions to the gap equation (solid lines) and the 
neutrality condition (dashed lines) calculated for several values of 
temperature. The results are plotted for $\mu=400$ MeV and the 
diquark coupling constant $G_D=\eta G_S$ with $\eta=0.75$. In this 
case, $T_c\approx 50.2$ MeV.}
\label{gap-vs-mue-T.eps}
\end{figure}

The solutions to the gap equation (\ref{gap-T}) for several different 
values of temperature ($T=0$, $20$, $40$, $50.2$ and $60$ MeV) are shown 
graphically in Fig.~\ref{gap-vs-mue-T.eps} (solid lines). The values of 
the temperature are marked along the curves. The results are plotted in 
the ($\mu_e$,$\Delta$)-plane. In computation we kept the quark chemical 
potential fixed, $\mu=400$ MeV, and used the diquark coupling constant 
$G_D=\eta G_S$ with $\eta=0.75$. This is the same choice that we used at
zero temperature in Fig.~\ref{gapneutral}.
   
Now let us briefly discuss the temperature dependence of the solution. 
It is not very surprizing that the shape of the graphical solution is 
smoothed with increasing the temperature. The same applies to the 
disappearance of the double-branch structure of the solution at a 
finite value of the temperature, $T_s \simeq 30$ MeV. Indeed, in a 
model that treats the electric chemical potential as a free parameter, 
this value of the temperature $T_{s}$ marks the expected switch of two 
regimes. Namely, while the phase transition controlled by the $\mu_e$ 
parameter (i.e., neutrality is not required) is a first order phase 
transition at $T<T_s$, it becomes a second order phase transition at 
$T>T_s$. 

We also observe that the value of the electric chemical potential 
at the point where the solution to the gap equation intersects with 
the $\mu_e$-axis has a nonmonotonic dependence on the temperature. With
increasing temperature, this value increases first and, after reaching 
some maximum value, starts to decrease and goes down to zero eventually.
When the interplay with the neutrality condition is taken into account
later, this simple property of the solution would produce rather unusual 
physical results.

We note that the temperature dependence of the solutions to the gap 
equation is very similar to the dependence found by Sarma in a  
solid state physics analogue of the g2SC phase \cite{Sarma}. Of 
course, there was no analogue of the neutrality condition in the
system studied in Ref.~\cite{Sarma}. Therefore, the mentioned similarity 
does not extend to the complete analysis of the g2SC phase that 
follows.

In Fig.~\ref{gap-vs-mue-T.eps}, we also show the neutrality lines (dashed
lines) for the same values of temperature ($T=0$, $20$, $40$, $50.2$ and 
$60$ MeV). The convention is that the lengths of the dashes decreases with 
increasing the value of the temperature. As we see, with increasing 
temperature, the neutrality line gets steeper while its intersection 
point with the $\mu_e$-axis moves towards larger values of $\mu_e$. The 
points of intersection of these neutrality lines with the lines of 
solutions to the gap equation, when they exist, are shown as well. 

\begin{figure}
\centerline{\epsfxsize=11cm\epsffile{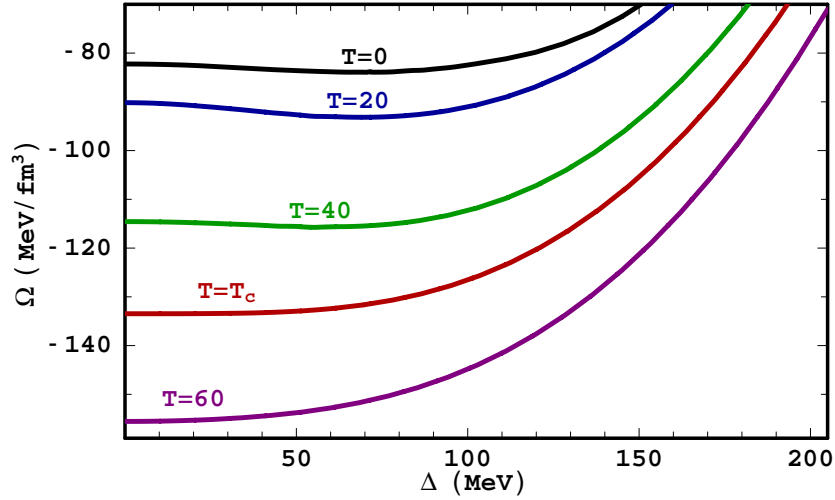}}
\vskip 0.3cm
\caption{The thermodynamic potential of neutral quark matter as a 
function of the diquark gap $\Delta$ calculated for several values of 
temperature.} 
\label{potdd}
\end{figure}

As in the case of zero temperature, see 
Fig.~\ref{V2D}, we need to show that the points of
intersections of the gap solutions with the corresponding neutrality 
lines in Fig.~\ref{gap-vs-mue-T.eps} represent the ground state of the 
neutral quark matter. To this end, we calculate the dependence of the 
thermodynamic potential on the value of the gap $\Delta$. Since the 
charge neutrality condition is satisfied only along the neutrality 
line, we restrict the thermodynamic potential only to this line. The 
numerical results for several values of the temperature are shown in 
Fig.~\ref{potdd}.

The results at $\mu=400$ MeV and $\eta=0.75$ show that, for any 
$T<T_c\approx 50.2$ MeV, the thermodynamic potential 
has the global minimum away from the origin, meaning that the 
corresponding ground state develops a nonzero expectation value of 
$\Delta(T)$. Needless to say that this is the same value that one 
extracts from the geometrical construction in 
Fig.~\ref{gap-vs-mue-T.eps}. The expectation value $\Delta(T)$
disappears gradually when the temperature approaches $T_{c}$ from
below. This is an indication of the second order phase transition. At 
temperature higher than $T_c$, the ground state of neutral quark 
matter is the normal phase with $\Delta(T)=0$. 

\vskip 0.3cm
{\bf Temperature dependence of the gap}
\vskip 0.3cm

The temperature dependence of the gap is obtained by numerical solution
of the gap equation (\ref{gap-T}) together with the two charge neutrality 
conditions. Of course, this is equivalent to the 
geometrical construction used in Fig.~\ref{gap-vs-mue-T.eps}, where the
intersection points of two types of lines determine the values of the
gaps in the ground state. 

\begin{figure}
\centerline{\epsfxsize=11cm\epsffile{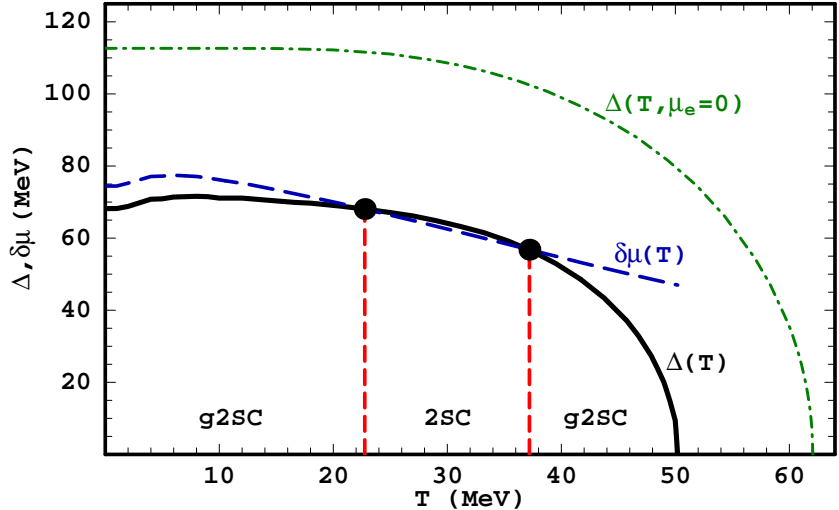}}
\vskip 0.3cm
\caption{The temperature dependence of the diquark gap (solid line) 
and the value of $\delta\mu\equiv \mu_e/2$ (dashed line) in neutral 
quark matter. For comparison, the diquark gap in the model with 
$\mu_e=0$ and $\mu_8=0$ is also shown (dash-dotted line). The results 
are plotted for $\mu=400$ MeV and $\eta=0.75$.}
\label{gap-mue0-vs-T.eps}
\end{figure}

The typical results for the default choice of parameters $\mu=400$ MeV 
and $\eta=0.75$ are shown in Fig.~\ref{gap-mue0-vs-T.eps}. Both the 
values of the diquark gap (solid line) and the mismatch parameter 
$\delta\mu=\mu_e/2$ (dashed line) are plotted. One very unusual property 
of the shown temperature dependence of the gap is the nonmonotonic behavior.
Only at sufficiently high temperatures, the gap is a decreasing function. 
In the low temperature regime, $T\lesssim 10$ MeV, however, it increases 
with temperature. For comparison, in the same figure, the diquark gap 
in the model with $\mu_e=0$ and $\mu_8=0$ is also shown (dash-dotted 
line). This latter has the standard BCS shape.

Another interesting thing regarding the temperature dependences in
Fig.~\ref{gap-mue0-vs-T.eps} appears in the intermediate temperature 
regime, $22.5 \lesssim T \lesssim 37$ MeV. By comparing the values of 
$\Delta(T)$ and $\delta\mu$ in this regime, we see that the g2SC phase 
is replaced by a ``transitional'' 2SC phase there. Indeed, the 
energy spectrum of the
quasiparticles even at finite temperature is determined by the same 
relations in Eqs.~(\ref{disp-ub}) and (\ref{2-degenerate}) that we 
used at zero temperature. When $\Delta > \delta \mu$, the modes 
determined by Eq.~(\ref{2-degenerate}) are gapped. Then, according 
to our standard classification, the ground state is the 2SC phase.

It is fair to say, of course, that the qualitative difference of the 
g2SC and 2SC phases is not so striking at finite temperature as it 
is at zero temperature. This difference is particularly negligible 
in the regime of interest where temperatures $22.5 \lesssim T 
\lesssim 37$ MeV are considerably larger than the actual value of 
the smaller gap, $\Delta - \delta \mu$. By increasing the value of 
the coupling constant slightly, however, the transitional 2SC phase 
can be made much stronger and the window of intermediate temperatures 
can become considerably wider. In either case, we find it 
rather unusual that the g2SC phase of neutral quark matter is 
replaced by a transitional 2SC phase at intermediate temperatures 
which, at higher temperatures, is replaced by the g2SC phase again. 

\begin{figure}
\centerline{\epsfxsize=11cm\epsffile{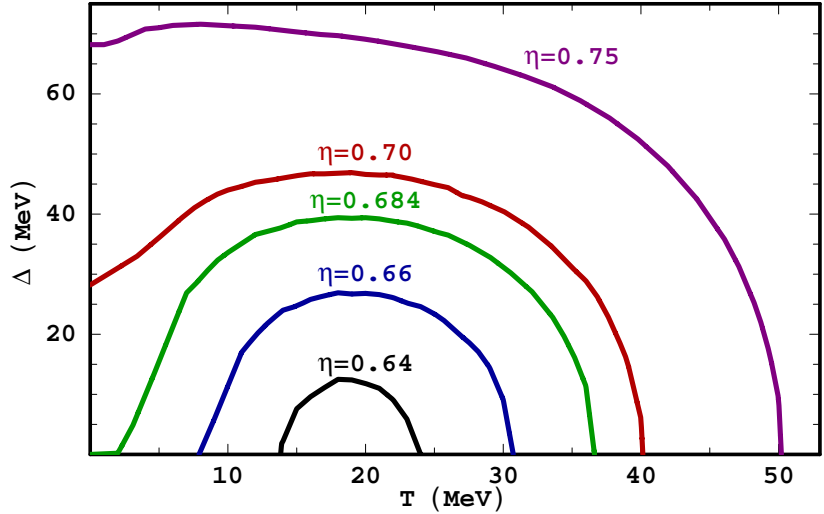}}
\vskip 0.3cm
\caption{The temperature dependence of the diquark gap in neutral 
quark matter calculated for several values of the diquark coupling 
strength $\eta=G_D/G_S$.}
\label{gap-eta}
\end{figure}

It appears that the temperature dependence of the diquark gap is 
very sensitive to the choice of the diquark coupling strength 
$\eta=G_D/G_S$ in the model at hand. This is not surprising because 
the solution to the gap equation is very sensitive to this choice. 
The resulting interplay of the solution for $\Delta$ with the 
condition of charge neutrality, however, is very interesting.
This is demonstrated by the plot of the temperature dependence 
of the diquark gap calculated for several values of the diquark 
coupling in Fig.~\ref{gap-eta}.

The most amazing are the results for weak coupling. It appears that
the gap function could have sizable values at finite temperature even
if it is exactly zero at zero temperature. This possibility comes
about only because of the strong influence of the neutrality condition
on the ground state preference in quark matter. Because of the thermal
effects, the positive electric charge of the diquark condensate is
easier to accommodate at finite temperature. This opens a possibility 
of the Cooper pairing that is forbidden at zero temperature. 

We should mention that somewhat similar results for the temperature 
dependence of the gap were also obtained in Ref.~\cite{SedLom} in
a study of the asymmetric nuclear matter and in Ref.~\cite{Liao-T} in
superfluidity. This behavior can be easily understood: When the
coupling strength is relatively weak, at zero temperature, it is 
very difficult to form a pair for two fermions at largely separating sharp 
Fermi surfaces. However, when the termperature increases, the thermal
excitations smooth the Fermi surfaces, thus offer the possibility for
forming Cooper pairs. Of course, when the temperature
is very high, the thermal excitations will eventually destroy the Cooper 
pairs.  

\vskip 0.2cm
{\bf Nonuniversal ratio $T_c/\Delta_0$}
\vskip 0.2cm

In the preceding subsection, we saw that the temperature dependence of
the gap in the g2SC phase is very different from the standard benchmark 
result in the BCS theory. One of the main properties 
of the BCS temperature dependence is a universal value of the ratio of 
the critical temperature $T_c$ to the value of the gap at zero 
temperature $\Delta_0$, see Eq.~(\ref{r_BCS}). It is instructive, 
therefore, to calculate the same quantity in the g2SC phase of neutral
quark matter. 

Let us start from a simple exercise, and consider the results plotted
in Fig.~\ref{gap-mue0-vs-T.eps} more carefully. First of all, we find 
that $T_c^{2SC} \approx 62.06$ MeV and $\Delta_0^{2SC} = 109.4$ 
MeV when there is no mismatch parameters in the model (i.e., $\mu_e=0$ 
and $\mu_8=0$ represented by dash-dotted line). It is not very 
surprizing that the ratio of interest $r_{2SC}\approx 0.567$ is in 
agreement with the BCS result. Now, if we check the results for the 
gap function in the g2SC phase (solid line), we find that $T_c^{g2SC}
\approx 50.2$ MeV and $\Delta_0^{g2SC} \approx 68.2$ MeV. Thus, the
ratio is $r_{g2SC}\approx0.7357$ which is considerably larger than the 2SC 
result. 

It appears that the real situation is even more interesting. The result 
for the ratio $r_{g2SC}\approx 0.7357$ is not universal. In fact, its 
value depends very much on the diquark coupling constant and, moreover, 
it can even be arbitrarily large. This last statement may not be so 
unexpected if we recall the temperature dependences of the gap shown in 
Fig.~\ref{gap-eta}. 

\begin{figure}
\centerline{\epsfxsize=11cm\epsffile{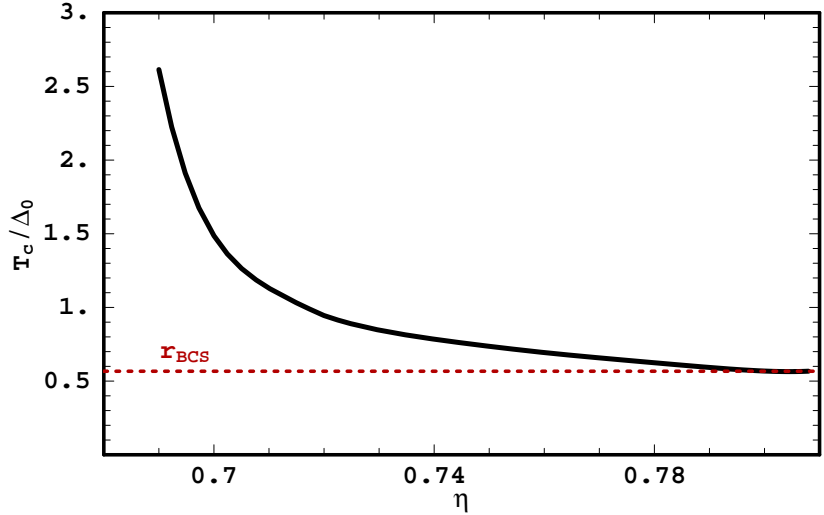}}
\vskip 0.3cm
\caption{The ratio of the critical temperature to the zero temperature 
gap in neutral quark matter as a function of the coupling strength
$\eta=G_D/G_S$.}
\label{ratio}
\end{figure}

The numerical results for the ratio of the critical temperature to 
the zero temperature gap in the g2SC case as a function of the diquark 
coupling strength $\eta=G_D/G_S$ are plotted in Fig.~\ref{ratio}. 
The dependence is shown for the most interesting range of values of 
$\eta=G_D/G_S$, $0.68 \lesssim \eta \lesssim 0.81$, which allows 
the g2SC stable ground state at zero temperature. 
When the coupling gets weaker in this range,
the zero temperature gap vanishes gradually. As we saw from 
Fig.~\ref{gap-eta}, however, this does not mean that the critical 
temperature vanishes too. Therefore, the ratio of a finite value of 
$T_c$ to the vanishing value of the gap can become arbitrarily 
large. In fact, it remains strictly infinite for a range of 
couplings.

\subsection{Chromomagnetic instability in the g2SC phase}

The g2SC phase has four gapless modes and two gapped modes, one may 
think that the low energy (large distance scale) properties in the g2SC phase 
should interpolate between those in the normal phase and those in 
the 2SC phase. However, its color screening properties do not fit 
this picture.  

As we know, one of the most important properties of the ordinary superconductor
is its Meissner effect, i.e., a superconductor expels the magnetic field, which
was discovered by Meissner and Ochsenfeld in 1933 \cite{Meissner}. From 
theoretical point of view, the Messiner effect can be explained using the linear 
response theory. The induced current $j^{ind}_{i}$ is related to the magnetic 
field $A_{j}$ by $j^{ind}_{i} = \Pi_{ij} A^{j} $, where the response function 
$\Pi_{ij}$ is the photon polarization tensor. 
The response function has two components, diamagnetic and paramagnetic part \cite{PD-Nam}. 
In the static and long-wavelength limit, for the normal metal, the paramagnetic 
component cancels exactly the diamagnetic component. While in the superconducting 
phase, the paramagnetic component is quenched by the energy gap and producing a 
net diamagnetic response. Thus the ordinary superconductor is a prefect diamagnet.

In color superconducting phases, the gluon self-energy (the response function to an 
external color field), has been investigated in the ideal 2SC phase \cite{Meissner2SC} 
and in the CFL phase \cite{Meissner-CFL}. The results show that the gauge bosons connected
with the broken generators obtain masses in these phases, which indicate the Meissner 
screening effect in these phases. 

It is very interesting to know the chromomagnetic property in the g2SC phase. 
We studied the g2SC phase in the framework of the SU(2) NJL model, and the NJL 
model lacks gluons. As reflection of this, it possesses the global instead of gauged 
color symmetry. In addition, there appear five Nambu-Goldstone (NG) bosons in the ground 
state of the model when the color symmetry is broken. In QCD, there is no room for such 
NG bosons. However, the NJL model can be thought of as the low energy theory 
of QCD in which the gluons, as independent degrees of freedom, are 
integrated out. The gluons could be reintroduced back by gauging the 
color symmetry in the Lagrangian density of the NJL model, providing
a semirigorous framework for studying the effect of the Cooper pairing 
on the physical properties of gluons. 

The existence of the g2SC phase can be regarded as a physical and model independent 
result under the restriction of local charge neutrality condition, the order parameter 
for this phase is $\Delta < \delta\mu$. In Ref.~\cite{g2SC-HS-M}, we calculated the gluon
self-energy in the g2SC phase. It is found that, in this phase, the symmetry broken
gauge bosons have imaginary Meissner screening masses, which is induced by the 
dominant paramagnetic contribution to the gluon self-energy. In condensed matter, 
this phenomenon is called the paramagnetic Meissner effect(PME) \cite{PME}, and has 
been observed in some high temperature superconductors and small superconductors.  

Unavoidably, the imaginary Meissner screening mass indicates a chromomagnetic 
instability of the g2SC phase.  There are several possibilities to resolve
this instability. One is through a gluon condensate to stabilize the system, 
which may not change the structure of the g2SC phase. 
It is also possible that the instability drives a new stable ground state,
which may have a rotational symmetry breaking like in Refs.~\cite{DFS,DFS-1},
or even have an inhomogeneous phase structure, like a crystal \cite{LO,FF,LOFF-1} 
or a vortex \cite{votex} structure. In the future, one has to resolve this problem.

\section{The mixed phase and nonstrange hybrid star}
\label{mixed}

We have discussed the homogeneous two-flavor quark matter when charge neutrality
conditions are satisfied locally, and found that the local charge neutrality 
conditions impose very strong constraints on determining the ground state of 
the system. In this section, we are going to discuss the mixed phase and the 
nonstrange hybrid star based on Ref.~\cite{Mixed-SHH}.

\subsection{The mixed phase}

When charge neutrality conditions are satisfied globally, one can construct a mixed 
phase \cite{Mixed-SHH,Mixed-Blaschke,Mixed-Reddy}. Inside mixed phases, the charge 
neutrality is satisfied ``on average'' rather than locally. This means that different 
components of mixed phases may have nonzero densities of conserved charges, but the total
charge of all components still vanishes. In this case, one says that the local
charge neutrality condition is replaced by a global one. There are three possible 
components: (i) normal phase, (ii) 2SC phase, and (iii) g2SC phase. 

\begin{figure}
\centerline{\epsfxsize=7cm\epsffile{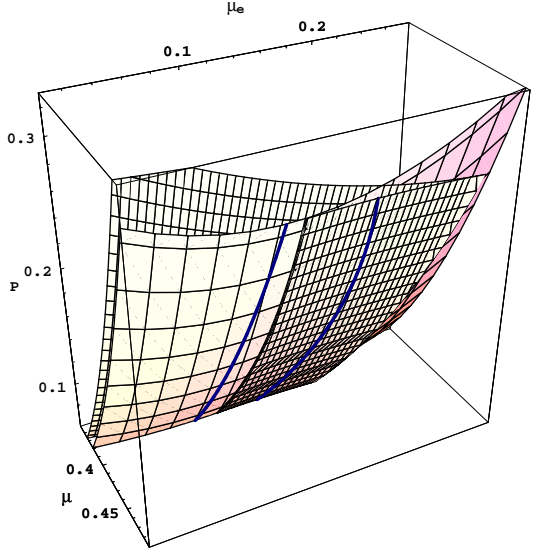}}
\vskip 0.3cm
\caption{\label{fig-back}
At $\eta=0.75$, pressure as a function of $\mu\equiv\mu_B/3$ and $\mu_e$ for 
the normal and color superconducting quark phases. The dark solid lines 
represent two locally neutral phases: (i) the neutral normal quark phase on the
left, and (ii) the neutral gapless 2SC phase on the right. The appearance
of the swallowtail structure is related to the first order type of the
phase transition in quark matter.}
\end{figure}

The pressure of the main three phases of two-flavor quark matter 
as a function of the baryon and electric chemical potentials is shown in 
Fig.~\ref{fig-back} at $\eta=0.75$. In this figure, we also show the pressure of the 
neutral normal quark and gapless 2SC phases (two dark solid lines). 
The surface of the g2SC phase extends only over a finite range of the values of 
$\mu_{e}$. It merges with the pressure surfaces of the normal quark phase 
(on the left) and with the ordinary 2SC phase (on the right).

It is interesting to notice that the three pressure surfaces in Fig.~ 
\ref{fig-back} form a characteristic swallowtail structure. As one
could see, the appearance of this structure is directly related to the
fact that the phase transition between color superconducting and 
normal quark matter, which is driven by changing parameter $\mu_{e}$, is
of first order. In fact, one should expect the appearance of a similar
swallowtail structure also in a self-consistent description of the
hadron-quark phase transition. Such a description, however, is possible.

From Fig.~\ref{fig-back}, one could see that the surfaces of normal  
and 2SC quark phases intersect along a common line. This means that the      
two phases have the same pressure along this line, and therefore could
potentially co-exist. Moreover, as is easy to check, normal quark
matter is negatively charged, while 2SC quark matter is positively
charged on this line.  This observation suggests that the appearance of   
the corresponding mixed phase is almost inevitable. 

Let us start by giving a brief introduction into the general method of
constructing mixed phases by imposing the Gibbs conditions of equilibrium
\cite{glen92,Weber}. From the physical point of view, the Gibbs
conditions enforce the mechanical as well as chemical equilibrium between
different components of a mixed phase. This is achieved by requiring that
the pressure of different components inside the mixed phase are equal,
and that the chemical potentials ($\mu$ and $\mu_{e}$) are the same
across the whole mixed phase. For example, in relation to the mixed phase
of normal and 2SC quark matter, these conditions read
\begin{eqnarray}
P^{(NQ)}(\mu,\mu_{e}) &=& P^{(2SC)}(\mu,\mu_{e}), 
\label{P=P}\\ 
\mu &=& \mu^{(NQ)}=\mu^{(2SC)}, 
\label{mu=mu}\\ 
\mu_{e} &=& \mu^{(NQ)}_{e}=\mu^{(2SC)}_{e}.  
\label{mue=mue}
\end{eqnarray}
 
\begin{figure}
\centerline{\epsfxsize=7cm\epsffile{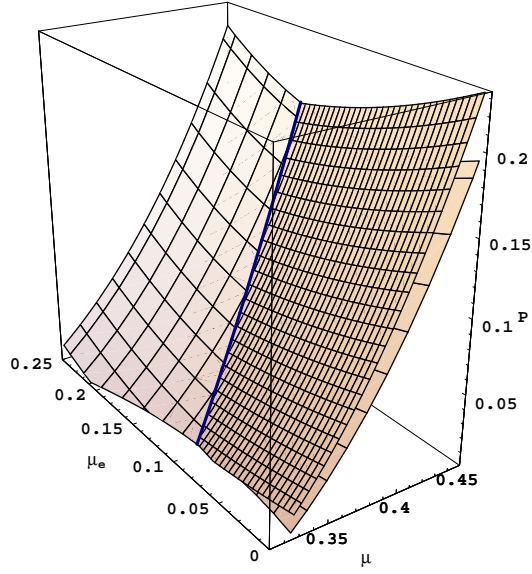}}
\vskip 0.3cm
\caption{\label{fig-front-quark}
At $\eta=0.75$, pressure as a function of $\mu\equiv\mu_B/3$ and $\mu_e$ for 
the normal and color superconducting quark phases (the same as in 
Fig.~\ref{fig-back}, but from a different viewpoint). The dark solid line
represents the mixed phase of negatively charged normal quark matter and
positively charged 2SC matter.}
\end{figure}

It is easy to visualize these conditions by plotting the pressure as a
function of chemical potentials ($\mu$ and $\mu_{e}$) for both components
of the mixed phase. This is shown in Fig.~\ref{fig-front-quark}. As
should be clear, the above Gibbs conditions are automatically satisfied
along the intersection line of two pressure surfaces (dark solid line in
Fig.~\ref{fig-front-quark}).

Different components of the mixed phase occupy different volumes of
space. To describe this quantitatively, we introduce the volume fraction
of normal quark matter as follows: $\chi^{NQ}_{2SC}\equiv V_{NQ}/V$
(notation $\chi^{A}_{B}$ means volume fraction of phase A in a mixture
with phase B). Then, the volume fraction of the 2SC phase is given by
$\chi^{2SC}_{NQ}=(1-\chi^{NQ}_{2SC})$. From the definition, it is clear
that $0\leq \chi^{NQ}_{2SC} \leq 1$.

The average electric charge density of the mixed phase is determined by
the charge densities of its components taken in the proportion of the
corresponding volume fractions. Thus,
\begin{equation}
n^{(MP)}_{e} = \chi^{NQ}_{2SC} n^{(NQ)}_{e}(\mu,\mu_e) 
+(1-\chi^{NQ}_{2SC}) n^{(2SC)}_{e}(\mu,\mu_e).
\end{equation}
If the charge densities of the two components have opposite signs, one
can impose the global charge neutrality condition, $n^{(MP)}_{e}=0$.
Otherwise, a neutral mixed phase could not exist. In the case of 
quark matter, the charge density of the normal quark phase is negative,
while the charge density of the 2SC phase is positive along the line of
the Gibbs construction (dark solid line in Fig.~\ref{fig-front-quark}).
Therefore, a neutral mixed phase may exist. The volume fractions of its
components are
\begin{eqnarray}
\chi^{NQ}_{2SC} &=& \frac{n^{(2SC)}_{e}}{n^{(2SC)}_{e}-n^{(NQ)}_{e}}, \\
\chi^{2SC}_{NQ} &\equiv& 1-\chi^{NQ}_{2SC}=
\frac{n^{(NQ)}_{e}}{n^{(NQ)}_{e}-n^{(2SC)}_{e}}.
\end{eqnarray}

After the volume fractions have been determined from the condition of the
global charge neutrality, we could also calculate the energy density of
the corresponding mixed phase,
\begin{equation}
\varepsilon^{(MP)} = \chi^{NQ}_{2SC} \varepsilon^{(NQ)}(\mu,\mu_e)
+(1-\chi^{NQ}_{2SC}) \varepsilon^{(2SC)}(\mu,\mu_e).
\end{equation}
This is essentially all that we need in order to construct the equation of
state of the mixed phase. 

So far, we were neglecting the effects of the Coulomb forces and the
surface tension between different components of the mixed phase. In a real
system, however, these are important. In particular, the balance
between the Coulomb forces and the surface tension determines the
size and geometry of different components inside the mixed phase. 

In our case, nearly equal volume fractions of the two quark phases are
likely to form alternating layers (slabs) of matter. The energy cost per
unit volume to produce such layers scales as $\sigma^{2/3}
(n_{e}^{(2SC)}-n_{e}^{(NQ)})^{2/3}$ where $\sigma$ is the surface tension
\cite{geometry}. Therefore, the quark mixed phase is a favorable phase of
matter only if the surface tension is not too large. Our simple estimates
show that $\sigma_{max} \leq 20$ MeV/fm$^{2}$. However, even for slightly
larger values, $20 \leq \sigma \leq 50$ MeV/fm$^{2}$, the mixed phase is
still possible, but its first appearance would occur at larger densities,
$3\rho_0 \leq \rho_B \leq 5\rho_0$. The value of the maximum surface
tension obtained here is comparable to the estimate in the case of the
hadronic-CFL mixed phase obtained in Ref.~\cite{interface}. The thickness
of the layers scales as $\sigma^{1/3} (n_{e}^{(2SC)}-n_{e}^{(NQ)})^{-2/3}$
\cite{geometry}, and its typical value is of order $10$ fm in the quark
mixed phase. This is similar to the estimates in various hadron-quark and
hadron-hadron mixed phases \cite{geometry,interface}. While the actual
value of the surface tension in quark matter is not known, in this study
we assume that it is not very large. Otherwise, the homogeneous gapless
2SC phase should be the most favorable phase of nonstrange quark matter
\cite{g2SC-SH}.

Under the assumptions that the effect of Coulomb forces and the surface tension is 
small, the mixed phase of normal and 2SC
quark matter is the most favorable neutral phase of matter in the model
at hand with $\eta=0.75$. This should be clear from observing the pressure surfaces in
Figs. \ref{fig-back} and \ref{fig-front-quark}. For a given value of the
baryon chemical potential $\mu=\mu_{B}/3$, the mixed phase is more
favorable than the gapless 2SC phase, while the gapless 2SC phase is more
favorable than the neutral normal quark phase.

\subsection{Nonstrange hybrid star} 

In order to construct a neutron star, we need to know the equation of state from 
very low baryon density to very high baryon density. Unfortunately, currently there is 
no single model that can describe the whole bayron density regime very successfully. In the 
following, we use a QCD-motivated hadronic chiral SU(3)$_L \times $SU(3)$_R$ model in 
normal baryon density regime, and use the NJL model in the quark phase.

\vskip 0.2cm
{\bf Hadronic matter}
\vskip 0.2cm
 
At densities around normal nuclear matter density $\rho_0\approx 0.15$
fm$^{-3}$, the description of baryonic matter in terms of quarks could
hardly be adequate. At such low densities, quarks are confined inside
hadrons. Thus, it is more natural to use an effective hadronic model.
 
We use a QCD-motivated hadronic chiral SU(3)$_L \times
$SU(3)$_R$ model as an effective theory of strong interactions to
describe the low density regime of the baryonic matter
\cite{papa98,papa99,han2000}. This model was found to describe reasonably
well the hadronic masses of various SU(3) multiplets, finite nuclei,
hypernuclei, excited nuclear matter and neutron star properties
\cite{papa98,papa99,han2000}. The basic features of the chiral model 
are:
(i) Lagrangian of the model is constructed in accordance with the
nonlinear realization of the chiral SU(3)$_L \times $SU(3)$_R$ symmetry; 
(ii) heavy baryons and mesons get their masses as a result of
spontaneous symmetry breaking;
(iii) the masses of pseudoscalar mesons (pseudo-Nambu-Goldstone
bosons) result from an explicit symmetry breaking;
(iv) a QCD-motivated field that describes the gluon condensate 
(dilaton field) is introduced.

\vskip 0.2cm
{\bf The mixed phase of hadronic matter and quark matter}
\vskip 0.2cm

It is expected that the phase transition between the hadronic phase and
the normal quark phase is a first order phase transition at zero
temperature and finite baryon chemical potential. Then, the hadronic and
quark phases could co-exist in a mixed phase \cite{glen92,Weber}. This
mixed phase should satisfy the Gibbs conditions of equilibrium which 
are similar to those discussed in the previous section, see Eqs.
(\ref{P=P})--(\ref{mue=mue}). The total energy density in the
hadron-quark mixed phase is given by
\begin{eqnarray}
\varepsilon^{(MP)} = \chi^{NQ}_{H} \epsilon^{(NQ)}(\mu,\mu_e)
+(1-\chi^{NQ}_{H}) \epsilon^{(H)}(\mu,\mu_e) ,
\label{eq:epsilon_MP}
\end{eqnarray}
where $\chi^{NQ}_{H}$ denotes the volume fraction of normal quark 
matter inside a mixture with hadronic matter.

\begin{figure}
\centerline{\epsfxsize=7cm\epsffile{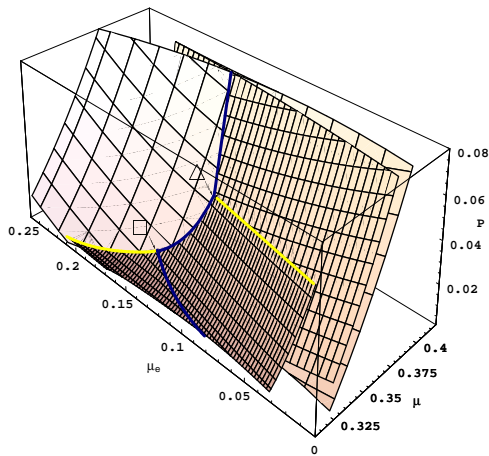}}
\caption{\label{fig:3D_HadQu}
Pressure as a function of $\mu\equiv\mu_B/3$ and $\mu_e$ for the
hadronic phase (at the bottom), for the two-flavor color superconducting   
phase (on the right in front) and the normal phase of quark matter (on
the left, and back on the right). The dark thick line follows the
neutrality line in hadronic matter, and two mixed phases: (i) the
mixed phase of hadronic and normal quark matter; and (ii) the mixed
phase of normal and color superconducting quark matter.}
\end{figure}

To visualize the Gibbs construction of the hadron-quark mixed phase, we
plot the hadronic surface of the pressure along with the quark surfaces
discussed in the previous section. Thus, Fig.~\ref{fig-front-quark} is
replaced by Fig.~\ref{fig:3D_HadQu}. The new figure shows the surfaces of
the pressure of the pure hadronic and quark phases as a function of their
chemical potentials $\mu$ and $\mu_e$. The intersection lines of different
surfaces indicate all potentially viable mixed phase constructions.
Although the Gibbs conditions are satisfied along all these lines, not all
of them could produce globally neutral phases (e.g., there are no neutral
constructions along the light shaded solid lines in
Fig.~\ref{fig:3D_HadQu}).

The dark solid line gives the complete, most favorable construction of
globally neutral matter in general $\beta$-equilibrium. This line consists
of three pieces. The lowest part of the curve (up to the point denoted by
$\square$ and $P\alt 10$ MeV$/$fm$^{3}$) corresponds to the pure confined
hadronic phase. Within this regime matter is mostly composed of neutrons
with little fractions of protons and electrons to realize the charge
neutrality and $\beta$-equilibrium. The hyperonic particles ($\Lambda,
\Sigma$ and $\Xi$) are not present in this lowest density regime. Such
particles would appear in the hadronic phase at considerably higher values
of pressure and density.

The mixed phase of hadronic and normal quark matter starts at the baryonic
density $\rho_B\approx 1.49 \rho_0$ which corresponds to the
$\square$-point in Fig.~\ref{fig:3D_HadQu}. At this point the first
bubbles of deconfined quark matter appear in the system. At the beginning
of this hadron-quark mixed phase, the deconfined bubbles are small but
highly negatively charged, whereas hadronic matter, in which the bubbles
are embedded, is slightly positively charged. The global charge neutrality
condition reads
\begin{equation} 
n^{(MP)}_{e} \equiv \chi^{NQ}_{H} n^{(NQ)}_{e}(\mu,\mu_e)
+(1-\chi^{NQ}_{H}) n^{(H)}_{e}(\mu,\mu_e) =0.
\label{eq:q_MP} 
\end{equation}
where $n^{(H)}_{e}$ and $n^{(NQ)}_{e}$ are the charge densities of
hadro\-nic and normal quark matter, respectively. This condition should be
satisfied at each point along the middle part of the dark solid line
(i.e., between the points denoted by $\square$ and $\bigtriangleup$).
With increasing density (from about $1.49 \rho_0$ up to $2.56 \rho_0$), 
the volume fraction of hadronic matter decreases (down to about
$0.59$), while the fraction of normal quark matter increases (up to
about $0.41$). 

\vskip 0.2cm
{\bf Equation of state}
\vskip 0.2cm

\begin{figure}
\centerline{\epsfxsize=8cm\epsffile{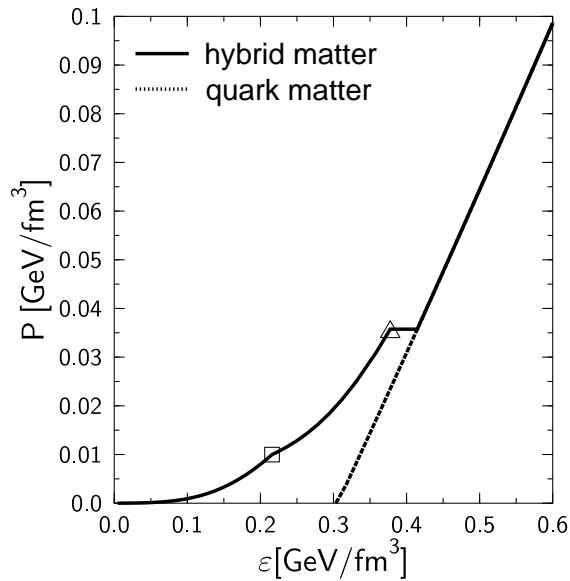}} 
\vskip 0.3cm
\caption{\label{fig:eos} 
The equation of state for globally neutral hybrid matter (solid line)  
and globally neutral quark matter (dashed line). The points of the
beginning of the two mixed phases are denoted by $\square$ and
$\bigtriangleup$.}
\end{figure}

The equations of state for quark and hybrid matter are shown in
Fig.~\ref{fig:eos}. The first equation of state corresponds to globally
neutral quark matter which is a mixture of the normal quark and 2SC
phases.  As for the equation of state of hybrid
matter, it is constructed out of the equation of state of neutral
hadronic matter and two Gibbs constructions in accordance with
Fig.~\ref{fig:3D_HadQu}. As before, the points that indicate the
beginning of two different mixed phases are denoted by $\square$ and
$\bigtriangleup$ in Fig.~\ref{fig:eos}.

\vskip 0.2cm
{\bf Mass-radius relation for hybrid star}
\vskip 0.2cm

\begin{figure}
\centerline{\epsfxsize=8cm\epsffile{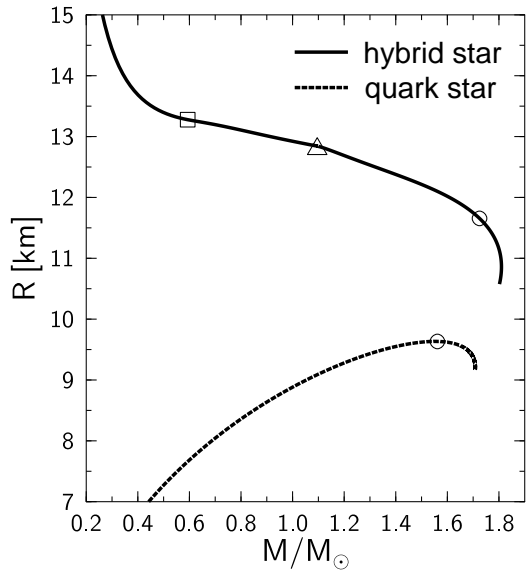}}
\vskip 0.3cm
\caption{\label{fig:mr}
The mass-radius relations for hybrid stars (solid line) and quark stars
(dashed line).}
\end{figure}

The mass-radius relations for hybrid and pure quark stars are shown in
Fig.~\ref{fig:mr}. As one would expect, the pure quark stars composed of
the mixed phase have much smaller radii and the value of their maximum 
mass is slightly smaller, see Fig.~\ref{fig:mr}. Our results for pure quark 
stars are comparable to those in Refs.~\cite{Diploma-Ruster,M-R}.
The difference between hybrid and pure quark stars is
mostly due to the low density part of the equation of state. This is also
evident from the qualitative difference in the dependence of the radius
as a function of mass for the hybrid and quark stars with low masses. The
corresponding hybrid stars are large because of a sizable low density
hadronic layer, while the quark stars are small because that have no such
layers. 

Here we use a two-flavor version of the NJL model to describe the
deconfined phase. However, the three-flavor extension of the NJL model in
Ref.~\cite{N2SC-Steiner} suggests that strange quarks might be present
in matter above a critical density of about $5\rho_0$. 
For the consideration of strange quark star in the framework of the 
NJL model, please see Ref.~\cite{Strange-star}.

\section{Conclusion and outlook}
\label{outlook}

This article focuses on the two-flavor color superconducting phase at moderate 
baryon density. 
   
In order to simultaneously investigate the chiral phase transition and the color 
superconducting phase transition, the Nambu-Gorkov formalism has been extended to 
treat the quark-antiquark and diquark condensates on an equal footing, and the 
competition between the chiral and diquark condensates has been investigated.  

Dense $u,d$ quark matter under local and global charge neutrality conditions 
in $\beta$-equilibrium has been discussed in detail. Under the local charge neutrality 
condition, the homogeneous neutral ground state is sensitive to the coupling constant 
in the diquark channel, it will be in the regular 2SC phase when the diquark coupling is 
strong, in the normal phase when the diquark coupling is weak, and in the g2SC phase in 
the case of intermediate diquark coupling. The low energy quasi-particle spectrum of the 
g2SC phase contains four gapless modes and only two gapped modes. 
In the intermediate diquark coupling regime, under global charge neutrality condition, 
assuming that the effect of Coulomb forces and the surface tension is small, one 
can construct a mixed phase composed of a positive charged 2SC phase and a negative 
charged normal quark phase. It will be valuable to compare the homogeneous g2SC 
phase and the mixed phase, and give a clear answer to which phase is preferable.

Even though the regime of the diquark coupling strength for the existence of the g2SC phase 
is rather narrow, i.e., $0.7\lesssim\eta\lesssim0.8$, it is noticed that the value 
$\eta=0.75$ from Fierz transformation and also the value $\eta\simeq 2.26/3$ from fitting 
the baryon vacuum mass, are inside this regime. The g2SC has rather 
unusual properties at zero as well as at finite temperature, and its chromomagnetic 
properties are extremely surprising. The paramagnetic Meissner effect indicates a 
chromomagnetic instability in the g2SC phase. In the near future, we need to  
resolve the problem of the chromomagnetic instability in the g2SC phase, and to 
investigate whether the chromomagnetic instability will drive a new inhomogeneous 
structure. 

In this article, I only discussed the two-flavor quark system. In the moderate baryon
density regime $\mu \simeq 500 {\rm MeV}$, the strange quark may appear in the system.
The cold dense three-flavor quark system has a more complicated and more interesting
phase structure. In different conditions, the three-flavor quark system can be in one 
of the phases of CFL, CFL+K, CFL+$\eta$, gCFL, and dSC
\cite{CFLK-Schafer,CFLK-Reddy,CFLeta-Schafer,gCFL-AKR,udSC-Hatsuda,gCFL-Ruster,gCFL-Kenji,gCFL-condensate}. 
It is, of course, very interesting and also challenging to investigate all these 
phases in one self-consistent framework. For the charge neutral gCFL phase, it is 
very attractive to know whether there is chromomagnetic instability. 

In our calculations, we used the mean-field approximation. It is very 
interesting to see how the diquark fluctuation will affect the results. It should
be noticed that till now, we are using the BCS scenario even in the intermediate
baryon density regime. From the experience in condensed matter, the BCS scenario
only works when the attractive interaction is very weak, where the Cooper pairs
are loosely bound states. While with the increasing of the coupling strength, the
Cooper pairs become tightly bounded, the fluctuation from the Cooper pairs cannot
be neglected. In the strong coupling case, the Cooper pairs become real bound sates 
in the coordinates space, and a Fermion system transfers to a boson system. The boson
system may experience the Bose-Einstein condensation (BEC). This is the basic
idea of the crossover from BCS to BEC \cite{BCS-BEC,strongBCS}. The fluctuation of 
the gauge fields as well as of the diquark were investigated in 
Refs.~\cite{fluctuation-gauge,pseudogap}, and it will be very
interesting to see how the Cooper pair structure \cite{Hatsuda-Cooper} 
changes when fluctuations are considered. In addition to the diquark fluctuation, 
the quark-antiquark fluctuation may also need to be considered in the intermediate
baryon density regime. It was found in Ref.~\cite{Beyond-Huang-rho} that the 
quark-antiquark fluctuation has significant influence on the chiral phase transition. 
When the quark-antiquark fluctuation is 
considered self-consistently \cite{Beyond-Huang}, the critical baryon density for 
the chiral phase transition becomes much larger than that in the mean-field approximation. 

The most interesting thing in the field of color superconductivity, of course, is to 
look for this phase in the nature or to create this phase in the laboratory. There are 
already some work to find the possibility of the existence of the color superconducting 
phase inside the neutron star, like the mass-radius relation 
\cite{Diploma-Ruster,Mixed-SHH,M-R,Strange-star} and the cooling history \cite{cooling}. The 
energy release from a hadronic star to a color superconducting star may relate to the 
supernovae or the gamma-ray burst \cite{GRB}. About the 
possibility of creation color superconductivity in heavy-ion collisions, the work is in 
progress \cite{CS-GSI}. Nevertheless, we are just at the very beginning of this field, and 
there is still a long way to go.

\section{Acknowledgements} 

I thank Prof. Wei-qin Chao and Prof. Peng-fei Zhuang for their continuous 
encouragement during my research work in this field.  
Most part of the research work in Sec. \ref{g2sc} and Sec. \ref{mixed} of 
this paper have been done in Frankfurt. I am grateful to my host professor, 
Prof. Horst Stoecker, for his hospitality and for giving me freedom to do 
what I like. I thank Prof. Dirk Rischke, who motivated our study on the thermal 
stability of the g2SC phase. I thank all the members in the ``Frankfurt-Darmstadt 
Color Superconductivity Group". I give my special thanks to Dr. Igor Shovkovy 
for reading this manuscript very carefully and for our fruitful and happy 
collaboration. During the two-year collaboration, we shared all the exciting 
discoveries, especially the g2SC phase and its unusual properties.  

Finally, I acknowledge the financial support from the Alexander von 
Humboldt-Foundation, and from the NSFC under Grant Nos. 10105005, 10135030.

\newpage
\appendix

\section{Fierz transformation}
\label{Fierz-Appendix}

Color-current interaction has the form of 
\begin{eqnarray}
{\cal L}_{int}^c = -G_c \sum_{a=1}^8[{\bar \psi}_{\alpha r}^i (\gamma_{\mu})_{rs} 
(t^a)_{\alpha \beta} \delta^{ij} \psi_{\beta s}^j][{\bar \psi}_{\gamma t}^k 
(\gamma^{\mu})_{tu} 
(t^a)_{\gamma \delta} \delta^{kl} \psi_{\delta u}^l],
\label{current}
\end{eqnarray}
where $t^a=(\lambda^a)_C/2$ are the SU(3) color generators with 
$tr(t^a t^b)=\delta_{ab}/2$, $\alpha, \beta, \gamma,
\delta$ are color indices acting on color space, 
$i,j,k,l$ are flavor indices acting on flavor space, 
and $r,s, t, u$ are spinor indices acting on Dirac space.  

\vskip 0.2cm
{\bf Fierz transformation in the Dirac space}
\vskip 0.2cm

Defining $S=1 \otimes 1$, $V=\gamma_{\mu} \otimes \gamma^{\mu}$,
$ T= \sigma_{\mu\nu} \otimes \sigma^{\mu\nu}$,
$A= \gamma_{\mu}\gamma_5 \otimes  \gamma^{\mu}\gamma_5$,
$P= \gamma_5 \otimes \gamma_5$,
the exchange of the spinor indices $s$ and $u$ or $r$ and $t$, can be represented as  
\begin{eqnarray}
 \Gamma^i_{rs, tu} =   \sum_j C_{ij} \Gamma^j_{ru, ts}, 
\end{eqnarray}
where $\Gamma^i \in (S,V,T,A,P)$, and 
\begin{eqnarray}
 \sum_j C_{ij} C_{jk}=\delta_{ik}.
\end{eqnarray}
The explicit expression for the above equation reads  
\begin{equation}
    \left( \begin{array}{c} 
           S \\  
            V \\
             T  \\
              A\\
              P 
              \end{array}  
             \right)^{\prime}  = 
           \left( \begin{array}{ccccc} 
           -1/4 & -1/4 & -1/8 & 1/4 & -1/4 \\  
            -1 & 1/2 & 0 & 1/2 & 1 \\
             -3 & 0 & 1/2 & 0 & -3  \\
              1 & 1/2 & 0 & 1/2 & -1\\
              -1/4 & 1/4 & -1/8 & -1/4 & -1/4 
              \end{array}
              \right)  
             \left( \begin{array}{c} 
           S \\  
            V \\
             T  \\
              A\\
              P 
              \end{array}  
             \right) , 
\end{equation}

In the following, we only need the transformation of $V$, i.e.,
\begin{eqnarray}
\label{dirac1}
(\gamma_{\mu} \otimes \gamma^{\mu})_{rs,tu} & = &
- ({\bf 1} \otimes {\bf 1} + i\gamma_5 \otimes i \gamma_5)_{ru,ts} \nonumber \\
 & + & \frac{1}{2}(\gamma_{\mu} \otimes \gamma^{\mu} + 
  \gamma_{\mu} \gamma_5 \otimes \gamma^{\mu} \gamma_5)_{ru,ts} \nonumber \\
& = & - (K_a \otimes K^a)_{ru,ts},   
\end{eqnarray}
where $K_a=\{{\bf 1}, i\gamma_5, \frac{i}{\sqrt{2}}\gamma_{\mu}, 
\frac{i}{\sqrt{2}}\gamma_{\mu}\gamma_5 \}$,
from which we can see that the contribution of the tensor part is zero, and thus
the chiral symmetry is preserved.

For diquarks, we have
\begin{eqnarray}
(\gamma_{\mu})_{rs} (\gamma^{\mu})_{tu} & =  &
(\gamma_{\mu})_{rs} ((\gamma^{\mu})^T)_{ut} \nonumber \\
& = & (\gamma_{\mu})_{rs} (C \gamma^{\mu} C)_{ut} \nonumber \\
& = & C_{ua} C_{bt} (\gamma_{\mu})_{rs} ( \gamma^{\mu} )_{ab} \nonumber \\
& = & - (K_a C \otimes C K^a )_{rt,us},
\label{dirac2}
\end{eqnarray}
where $C=i \gamma^2 \gamma^0$ is the charge conjugation matrix.

\subsection{Fierz transformation in the quark-antiquark sector}

We will use the Fierz transformation in the color, flavor and Dirac spaces
respectively.

\vskip 0.2cm
{\bf Fierz identity in the color space}
\vskip 0.2cm

The standard identity
\begin{eqnarray}
t^c_{\alpha \beta} t^c_{\gamma \delta} = \frac{1}{2}(1-\frac{1}{N_c^2})
      \delta_{\alpha \delta} \delta_{\gamma \beta} 
      -\frac{1}{N_c}t^c_{\alpha \delta} t^c_{\gamma \beta},
\label{color1}
\end{eqnarray}
gives both the color singlet and color octet terms.

\vskip 0.2cm
{\bf Fierz identity in the flavor space}
\vskip 0.2cm

Rewrite 
\begin{eqnarray}
\delta_{ij} \delta_{kl} = \sum_e (G^e)_{il} (G^e)_{kj},
\label{flavor1}
\end{eqnarray}
where $G^e$ has been ordered in the way flavor singlet + octet,
with
\begin{eqnarray}
\{G^e \}=(\frac{1}{\sqrt{3}}{\bf 1} \equiv \frac{(\lambda^0)_F}{\sqrt{2}}, 
     \frac{(\lambda^a)_F}{\sqrt{2}}), a=1, \cdots, 8.
\end{eqnarray}

\vskip 0.2cm
{\bf Color singlet part}
\vskip 0.2cm

Using Eqs. (\ref{dirac1}), (\ref{color1}) and (\ref{flavor1}), we can get 
the color singlet part of the Lagrange Eq. (\ref{current}), 
\begin{eqnarray}
{\cal L}_{int}^{1c}& = &\frac{8}{9} G_c \sum_{a=0}^8[ ({\bar \psi} \frac{\lambda_F^a}{2}\psi)^2
        +     ({\bar \psi} i \gamma_5 \frac{\lambda_F^a}{2}\psi)^2 ] \nonumber \\
     & - &    \frac{4}{9} G_c \sum_{a=0}^8 [ ({\bar \psi} \gamma_{\mu} \frac{\lambda_F^a}{2}\psi)^2
        +     ({\bar \psi} \gamma_{\mu} \gamma_5 \frac{\lambda_F^a}{2}\psi)^2 ].
\end{eqnarray}
        
\subsection{Fierz transformation in the diquark sector}

\vskip 0.2cm
{\bf Fierz identity in the color space}
\vskip 0.2cm

\begin{eqnarray}
t^c_{\alpha \beta} t^c_{\gamma \delta} = \frac{2}{N_c} \sum_{S} 
         t^S_{\alpha \gamma} t^S_{\delta \beta} - \frac{4}{N_c} \sum_A 
         t^A_{\alpha \gamma} t^A_{\delta \beta}, 
\label{colord1}
\end{eqnarray}
where $S \in \{0,1,3,4,6, 8 \}$ denotes the symmetric and $A \in \{2,5,7 \}$
the antisymmetric Gell-Mann matrices. 

The above identity is decomposed in the way of $6_c +{\bar 3}_c$, which shows
that the interaction in the $6_c$ channel is oppositive to that in the ${\bar 3}_c$
channel. There is another decomposition in the way of $1_c + {\bar 3}_c$, i.e.,
\begin{eqnarray}
t^c_{\alpha \beta} t^c_{\gamma \delta} = \frac{1}{2}(1-\frac{1}{N_c}) \delta_{\alpha \delta}
\delta_{\gamma\beta} + \frac{1}{2N_c} \epsilon_{\alpha \gamma}^{\rho} 
\epsilon_{\delta \beta}^{\rho}, 
\label{colord2}
\end{eqnarray}
which shows that the interaction in the color antitriplet channel has the same symbol as 
that in the singlet channel. The interaction in the color singlet channel $1_c$ 
is attractive, so is the color antitriplet channel ${\bar 3}_c$. The former 
is responsible for the bound state of meson, the latter is responsible for the 
bound state of baryon, as well as the quark-quark Cooper pairing at high baryon density.
The ratio of the magnitude of the coupling constant in ${\bar 3}_c$ channel and $1_c$
channel is $1/2$ from Eq. (\ref{colord2}).

\vskip 0.2cm
{\bf Fierz identity in the flavor space}
\vskip 0.2cm

\begin{eqnarray}
\delta_{ij} \delta_{kl} = \sum_e (G^e)_{ik} (G^e)_{lj},
\end{eqnarray}
where 
\begin{eqnarray}
\{G^e \}=\{ G_A^e, G_S^e\}=\{ \frac{\lambda_F^2}{\sqrt{2}},\frac{\lambda_F^5}{\sqrt{2}},
\frac{\lambda_F^7}{\sqrt{2}}; \frac{1}{\sqrt{3}} {\bf 1} \equiv \frac{\lambda_F^0}{\sqrt{2}},
\frac{\lambda_F^1}{\sqrt{2}},\frac{\lambda_F^3}{\sqrt{2}},\frac{\lambda_F^4}{\sqrt{2}},
\frac{\lambda_F^6}{\sqrt{2}}, \frac{\lambda_F^8}{\sqrt{2}} \}.
\label{flavor2}
\end{eqnarray}
where $S \in \{0,1,3,4,6, 8 \}$ denotes the symmetric and $A \in \{2,5,7 \}$
the antisymmetric Gell-Mann matrices. 

\vskip 0.2cm
{\bf Color anti-triplet part} 
\vskip 0.2cm

First, using Eqs. (\ref{dirac2}), we have   
\begin{eqnarray}
& & [{\bar \psi}_{r} (\gamma_{\mu})_{rs}  \psi_{s}][{\bar \psi}_{t} 
(\gamma^{\mu})_{tu} \psi_{u}] \nonumber \\
& = &  [{\bar \psi}_{r} (K_a C)_{rt} {\bar \psi}_{t}][\psi_{s} 
(C K^a)_{us} \psi_{u}] \nonumber \\
& = & [{\bar \psi}_{r} (K_a C)_{rt} {\bar \psi}_{t}][\psi_{s}
(C K^a)_{us} \psi_{u}] \nonumber \\
& = & [{\bar \psi}  K_a C {\bar \psi}^T][\psi^T C K^a \psi].
\end{eqnarray}
 
Then, using Eq. (\ref{colord2}) 
the color anti-triplet part can be expressed as
\begin{eqnarray}
& & [{\bar \psi}_{\alpha r} (\gamma_{\mu})_{rs} 
(t^a)_{\alpha \beta} \psi_{\beta s}[{\bar \psi}_{\gamma t}
(\gamma^{\mu})_{tu} 
(t^a)_{\gamma \delta} \psi_{\delta u}] \nonumber \\
& = & -\frac{1}{3} [{\bar \psi}  K_a C \frac{i\epsilon^{\rho}}{\sqrt{2}}  {\bar \psi}^T][\psi^T 
C K^a \frac{i\epsilon^{\rho}}{\sqrt{2}} \psi].
\end{eqnarray}

At last, the flavor structure is determined by the Pauli principle.
A color triplet diquark state is antisymmetric under the exchange 
of colors, due to the Pauli principle, the matrices $CK^a\lambda_F$ must be totally 
symmetric. Thus the axial-vector and scalar diquark currents transform according to
antisymmetric flavor ${\bar 3}$ representation, whereas the scalar-, pseudoscalar-
and vector-diquarks are necessarily flavor sextetes.

\section{Charge conjugation}
\label{ChargeConj}

Charge conjugation is conventionally defined to take a fermion with
a given spin orientation into an antifermion with the same spin orientation.

The charge-conjugated filed $\psi^C$ is defined by
\begin{eqnarray}
& & \psi^C(x)=C {\bar \psi}^T (x), {\bar \psi}^C(x) = \psi^T C, \nonumber \\
& & \psi(x)=C {\bar \psi}^{CT}(x), {\bar \psi}(x)=\psi^{CT}(x)C,
\end{eqnarray}
where $C=i \gamma^2 \gamma^0$ is the charge-conjugation matrix in the Dirac
representation, 
\begin{eqnarray}
C=-C^{-1}=-C^T=-C^{\dagger}, 
\end{eqnarray}
and 
\begin{eqnarray}
C \gamma^{\mu} C^{-1} = - (\gamma^{\mu})^T.
\end{eqnarray}

The charge conjugation of bilinears is a little tricky, and it helps to write 
in spinor indices. For the scalar,
\begin{eqnarray}
{\bar \psi}^C\psi^C & = & {\bar \psi}^C_{\alpha}\psi^C_{\alpha} \nonumber \\
& = & (\psi^T C)_{\alpha} (C {\bar \psi}^T)_{\alpha} \nonumber \\
& = & \psi_{\lambda} C_{\lambda \alpha} C_{\alpha \lambda^{\prime}} 
         {\bar \psi}_{ \lambda^{\prime}} \nonumber \\
& = & {\bar \psi} \psi.
\end{eqnarray}
In the last step, the fermion anticommunication and $C C =-1$ have been used.

For the vector,
\begin{eqnarray}
{\bar \psi}^C \gamma^{\mu} \psi^C & = & {\bar \psi}^C_{\alpha} 
 (\gamma^{\mu})_{\alpha \beta} \psi^C_{\beta} \nonumber \\
& = & (\psi^T C)_{\alpha}  (\gamma^{\mu})_{\alpha \beta}
          (C {\bar \psi}^T)_{\beta} \nonumber \\
& = & \psi_{\lambda} C_{\lambda \alpha}  (\gamma^{\mu})_{\alpha \beta}
          C_{\beta \lambda^{\prime}} {\bar \psi}_{ \lambda^{\prime}} \nonumber \\
& = & \psi_{\lambda}(\gamma^{\mu})_{\lambda \lambda^{\prime}}^T 
      {\bar \psi}_{ \lambda^{\prime}} \nonumber \\
& = & - {\bar \psi}_{ \lambda^{\prime}}  
      (\gamma^{\mu})_{\lambda^{\prime} \lambda} \psi_{\lambda} \nonumber \\
 & =  & -{\bar \psi} \gamma^{\mu} \psi.
\end{eqnarray}         

For the kinetic term,
\begin{eqnarray}
{\bar \psi}^C \gamma^{\mu} \partial_{ \mu} \psi^C 
 & =  & -(\partial_{\mu} {\bar \psi}) \gamma^{\mu} \psi,
\end{eqnarray}         
using
\begin{eqnarray}
\partial_{\mu} ({\bar \psi} \gamma^{\mu} \psi) = (\partial_{\mu} {\bar \psi})
 \gamma^{\mu} \psi + {\bar \psi} \gamma^{\mu} (\partial_{\mu} \psi),
\end{eqnarray}
and the vector current conservation, 
we have
\begin{eqnarray}
{\bar \psi}^C \gamma^{\mu} \partial_{ \mu} \psi^C 
 & =  & {\bar \psi} \gamma^{\mu} \partial_{\mu} \psi.
\end{eqnarray}

\end{document}